\documentclass[aps,prc,twocolumn,showpacs]{revtex4}
\usepackage{graphicx}
\usepackage{dcolumn}
\usepackage{color} 
\RequirePackage{booktabs}
\usepackage{longtable}

\usepackage{amsfonts}
\usepackage{amssymb}
\usepackage{amsmath}
\usepackage{amsxtra}

\begin{document}

\title{Effective interactions in the \boldmath$sd$ shell}

\author{N.~A.~Smirnova} 
\email[]{smirnova@cenbg.in2p3.fr}
\affiliation{CENBG (CNRS/IN2P3 - Universit\'e de Bordeaux), 33175 Gradignan cedex, France }
\author{B. R. Barrett}
\email[]{bbarrett@physics.arizona.edu}
\affiliation{Department of Physics, University of Arizona, Tucson, Arizona 85721}
\author{Y. Kim}
\email[]{ykim@ibs.re.kr}
\affiliation{Rare Isotope Science Project, Institute of Basis Science, Daejeon 34037, Republic of Korea}
\author{I. J. Shin}
\email[]{geniean@ibs.re.kr}
\affiliation{Rare Isotope Science Project, Institute of Basis Science, Daejeon 34037, Republic of Korea}
\author{A. M. Shirokov}
\email[]{shirokov@nucl-th.sinp.msu.ru}
\affiliation{Department of Physics and Astronomy, Iowa State University, Ames, Iowa 50011}
\affiliation{Skobeltsyn Institute of Nuclear Physics, Lomonosov Moscow State University, 
Moscow 119991, Russia}
\affiliation{Pacific National University, 136 Tikhookeanskaya st., Khabarovsk 680035, Russia}
\author{E. Dikmen}
\email[]{erdaldikmen@sdu.edu.tr}
\affiliation{Department of Physics, Suleyman Demirel University, Isparta, Turkey}
\author{P. Maris}
\email[]{pmaris@iastate.edu}
\affiliation{Department of Physics and Astronomy, Iowa State University, Ames,
Iowa 50011}
\author{J. P. Vary}
\email[]{jvary@iastate.edu}
\affiliation{Department of Physics and Astronomy, Iowa State University, Ames, Iowa 50011}

\date{\today}

\bibliographystyle{prsty}

\begin{abstract}
We perform a quantitative study of the microscopic effective shell-model interactions in the valence $sd$ shell,
obtained from modern nucleon-nucleon potentials, chiral N3LO, JISP16 and Daejeon16, 
using No-Core Shell-Model wave functions and the Okubo--Lee--Suzuki transformation. 
We investigate the monopole properties of those interactions in comparison with the phenomenological
universal $sd$-shell interaction, USDB. 
Theoretical binding energies and low-energy spectra of O isotopes 
and of selected  $sd$-shell nuclei, are presented. 
We conclude that there is a noticeable improvement in the quality of 
the effective interaction when it is derived from the Daejeon16 potential.
We show that its proton-neutron centroids are consistent with those from USDB. 
We then propose monopole modifications of the Daejeon16 centroids 
in order to provide an adjusted interaction yielding significantly improved agreement with the experiment.
A spin-tensor decomposition of two-body effective interactions is applied in order
to extract more information on the structure of the centroids and to understand the reason for 
deficiencies arising from our current theoretical approximations.
{The issue of the possible role of the three-nucleon forces is addressed.}

\end{abstract}


\bibliographystyle{prsty}

\maketitle

\section{Introduction}

Owing to growth in accessible computing power and to progress in the theory of nuclear forces,
{\it ab-initio} studies of light nuclei have achieved remarkable success in providing accurate 
binding energies, low-energy excitation spectra, transition probabilities and 
other measurable quantities~\cite{Barrett_2013,Hergert2013,Bogner2014,Jansen2014}.
These results, in turn, probe the recently derived nucleon-nucleon ($NN$) and three-nucleon ($3N$) potentials,
in particular, those obtained within the chiral perturbation theory~\cite{EpHaRMP2009,MaEnPR2011}.
In spite of such a significant progress in {\em ab-initio} calculations for light nuclei,
the structure of the open-shell medium-mass nuclei can still be described only 
by restricted valence-space calculations. 
Thus, the goal to derive effective valence-space interactions represents a major area of endeavor.

The present study is focused on the use of the No-Core Shell Model (NCSM)~\cite{Barrett_2013} in conjunction 
with additional theoretical treatment that leads to 
effective interactions for valence space shell-model calculations.
Within the NCSM, all $A$ nucleons, interacting via realistic forces, 
are treated as active within a model space,
consisting of a large number of shells (typically, shells of a harmonic-oscillator potential).
The eigenvalue problem is solved by diagonalization of the many-body Hamiltonian matrix
in a spherically symmetric harmonic-oscillator basis. 
The many-body eigenstates, represented as mixing of configurations expressed 
as Slater determinants of the proton and neutron single-particle wave functions,
preserve all fundamental symmetries of atomic nuclei and can be used directly 
to calculate matrix elements of various operators.
As a fully {\it ab-initio} approach, the NCSM 
provides a \mbox{reliable} description of nuclei up to $A=16$ 
and it underlies pathways to {\it ab-initio} reaction theory~\cite{Barrett_2013,SSHORSE,SSHORSE-C}.

For practical reasons, the traditional shell-model approach for heavier nuclei consists in considering only 
valence nucleons, interacting via effective forces in a truncated model space.
The success of such a valence-space shell model with phenomenological forces is well-confirmed
for $sd$ and $pf$ shell and for heavier nuclei~\cite{CaurierRMP}.  
Those interactions are traditionally obtained from a fit of two-body matrix elements (TBMEs) 
to selected experimental spectra.
Well-established interactions include the Cohen--Kurath~\cite{CohenKurath} interaction for the $p$ shell,
the USD family of interactions for $sd$ shell~\cite{USD,USDab}, 
the KB3G~\cite{KB3G} and GXPF1A~\cite{GXPF1A} interactions for $pf$ shell and many others.
One challenge is the large number of TBMEs to be determined and a need for experimental information on key states
in nuclei with closed (sub)shells plus or minus one nucleon.

Microscopic approaches to construct a reliable and accurate effective interaction
for valence space calculations from a bare $NN$ potential
via an appropriate renormalization procedure is a long-standing challenge.
In the 1960's, Kuo and Brown~\cite{KuoBrown66,KB}, building on the earlier work of Arima and Horie~\cite{ArHo54a,ArHo54b}
and of Bertsch~\cite{Bertsch65}, constructed the first microscopic effective shell-model interactions
in a truncated model space starting from the Brueckner $G$ matrix and accounting for the core polarization effects.
More advanced effective interactions based on the $G$-matrices from high-precision potentials 
(Bonn-type potentials) and obtained within the many-body perturbation theory can be found in Ref.~\cite{MHJ95}.
It has been determined~\cite{PoZu81} that the main deficiency of such microscopic interactions 
is their monopole part, representing a spherical mean field. 
It is this monopole part which is responsible for sub-shell closures within a given shell and for the correct saturation properties.
The main reason is thought to be the absence of $3N$ forces~\cite{Zuker3N,Otsuka3N,Holt12}, 
while other correlations beyond the model space may be missing as well.

In the last decade, other promising approaches to the problem of microscopic valence effective interactions have been developed,
such as  
SRG-approaches~\cite{Bogner2014,Stroberg2016,StrPRL118}, and
methods based on the coupled-cluster theory~\cite{Jansen2014} 
as well as on the many-body perturbation theory~\cite{Fukui18} and the NCSM~\cite{Dikmen2015}.
As an important advantage, some of these newer approaches incorporated
$3N$ forces~\cite{Simonis2016,Stroberg2016,StrPRL118,Fukui18}, yielding apparent progress in the description
of the nuclear binding energies and spectroscopy.

We aim to investigate new microscopic effective interactions for the $sd$ shell
obtained from the NCSM wave functions 
by an 
Okubo--Lee--Suzuki (OLS) transformation as proposed in Ref.~\cite{Dikmen2015}.
We study the effective interactions obtained from three different $NN$ potentials, namely, 
the effective-field theory (EFT) inspired N3LO potential (from Ref.~\cite{EnMa03}), the $J$-matrix Inverse Scattering Potential
JISP16~\cite{JISP16} and the Daejeon16 potential~\cite{DJ16}.
The two latter potentials provide high quality descriptions of the $NN$ data and have been tuned off-shell to fit selected
properties of light nuclei up to $A=16$ avoiding the use of $3N$ forces. 
This means that these interactions partially incorporate many-body effects and, as shown by recent NCSM calculations,
they provide a very good description of $p$-shell nuclei~\mbox{\cite{Maris2009,NN3Nbook,DJ16,KimNTSE2018,MarisNTSE2018}}.
The resulting effective $sd$-shell Hamiltonians (single-particle energies and TBMEs) 
derived from N3LO and JISP16 potentials can be found in Ref.~\cite{Dikmen2015}, 
while the effective valence Hamiltonian obtained from Daejeon16 is presented in this work.
\enlargethispage{.7\baselineskip}

We note that the study of Ref.~\cite{Dikmen2015} was of a proof-of-concept: 
it demonstrated that the derived effective valence $sd$-shell interaction reproduces exactly the NCSM energy of $^{18}$F 
in calculations within the conventional shell model with the $^{16}$O core using realistic $NN$ potentials. 
This work also showed that the spectrum of $^{19}$F generated by the same effective interaction 
is very close to the NCSM predictions for this nucleus. 
Here, we apply this approach to heavier $sd$-shell nuclei, compare with experimental data and 
with successful phenomenological interactions.

After a short discussion of deriving the effective valence-space interactions and 
introducing the new effective $sd$-shell interaction ``DJ16'' obtained from the Daejeon16 $NN$ potential,
we discuss the dependence of theoretical single-particle energies and TBMEs on atomic mass number $A$ and 
on the other parameters of the NCSM calculations. 
The theoretical single-particle energies derived from the N3LO EFT  interaction and from JISP16 and
Daejeon16 potentials are found to be deficient. The origin of this deficiency in the case of the EFT
interaction is that,   in the NCSM calculations, we neglected the $3N$ force which is required to
reproduce the properties of O and F isotopes; in the case of JISP16 and Daejeon16 interactions, which
were designed for the use without the $3N$ forces, we note that these interactions were fitted only
to nuclei with $A\leq16$, so the deficiencies in describing $sd$-shell nuclei signal the
need to further tune these interactions at least to the lightest $sd$-shell nuclei. Nevertheless,
surprisingly, DJ16 and the effective interaction derived from JISP16 
are able to reproduce reasonably well binding energies in the chain of O isotopes. 
However, the deficiency of their derived single-particle energies yield an inadequate description
of spectra of the $sd$-shell nuclei. 
As a result of these current deficiencies, we adopt guidance from successful phenomenology and we 
employ empirical single-particle energies.
We also scale TBMEs with~$A$ in the same manner as in the case of empirical effective interactions.

In order to understand the properties of the newly derived microscopic TBMEs we will compare them with 
the USDB phenomenological TBMEs~\cite{USDab} and as well as with the effective interaction obtained
within the folded-diagram approach with a $\hat Q$-box of third order in the $G$-matrix computed from the BonnC $NN$ potential 
(see column 'C' of Table~20 in Ref.~\cite{MHJ95}).
We also compare our results 
with the latest, very successful, IMSRG results~\cite{StrPRL118}, obtained from an $NN$ plus $3N$ potential.
We begin by investigating the monopole properties of these interactions, followed by the comparison of
the low-lying spectra and binding energies of the O isotopes. We discuss how the subshell closures affect the 
spectra of the odd-$A$ F isotopes and $^{39}$K. We also study odd-odd $^{26}$F and $^{22}$Na to test some specific
proton-neutron TBMEs. Next, we turn to more collective spectra of mid-shell nuclei,
such as $^{28,29}$Si and $^{32}$S, as well as to  
a well-deformed rotor, $^{24}$Mg, to demonstrate quadrupole properties of the interactions.
We also perform some minimal modifications of the mainly $T=1$ centroids of the effective interaction DJ16 
in order to improve the description of O isotopes and of other nuclei.
We complete this study by proposing a spin-tensor analysis of the centroids in order to further compare 
the potentials.

\section{Microscopic two-body interactions}

For the derivation of the effective valence interactions, we begin with a NCSM calculation for the core system, 
$^{16}$O in the present work. One also performs NCSM calculations for core-plus-one and core-plus-two nucleon systems.  
%
The starting point is a translationally-invariant Hamiltonian for $A$ point-like nucleons
interacting via a realistic $NN$ interaction
\begin{equation}
\label{H_intrinsic}
H =\sum_{i<j=1}^{A}\frac{\vec{p_i}^2}{2m}- \frac{\vec{P}^2}{2mA}
         +\sum_{i<j=1}^{A}V_{ij}^{NN},
\end{equation}
where $m$ is the nucleon mass  (appoximated here as the average of the neutron and proton mass), 
$\vec{p}_i$ are nucleonic momenta, $\vec{P}=\sum_{i=1}^A \vec{p}_i$,
and $V_{ij}^{NN}$ denotes the bare $NN$ interaction. The pairwise Coulomb interaction is included between the protons.

Within the NCSM, the eigenproblem for $H$ is solved by the diagonalization of the
Hamiltonian matrix in a many-body spherical harmonic-oscillator basis (Slater determinants
built from single-particle oscillator functions), 
characterized by a given  energy quantum, $\hbar \Omega $.
%
The model space is restricted by the parameter $N_{\rm max}$, 
which means that the retained many-body configurations 
should satisfy the condition
that $\sum_{i=1}^A (2n_i+l_i) \le N_{\rm min} + N_{\rm max}$,
where $n_i$ is the single-particle radial harmonic oscillator quantum number, 
$l_i$ is the single-particle orbital angular momentum quantum number and 
$N_{\rm min}$ is the minimum of the summation that satisfies the Pauli principle for the chosen $A$-nucleon system.

One of the important advantages of the harmonic-oscillator potential is that it allows one to remove the spurious
center-of-mass motion, described by the Hamiltonian $H_{CM}=\vec{P}^2/(2mA) + 
\frac12 Am \Omega^2 \vec{R}^2$,
with ${\vec{R}=\frac1A \sum_{i=1}^A \vec{r_i}}$.
In practice, when the Slater determinant basis is used, the addition of a center-of-mass term 
$\beta (H_{CM}-\frac32 \hbar \Omega )$ with a large positive value of $\beta $ shifts the 
states with spurious center-of-mass excitations to higher energies~\cite{Lawson}.

Due to computational limits, the eigenproblem for (\ref{H_intrinsic})
can be successfully solved to high precision only for relatively ``soft'' $NN$ potentials.
In particular, several realistic $NN$ potentials, including Argonne V18 \cite{AV18}, 
CD-Bonn~\cite{CD-Bonn} and modern EFT interactions,
generate strong short-range correlations which require inaccessible basis dimensions to achieve useful results. 
Thus, a renormalization of the bare $NN$ interaction 
is required to accelerate the convergence~\cite{Barrett_2013}. 
In the present study, we perform the OLS transformation~\mbox{\cite{OLS1,OLS2,OLS3,OLS4}}
of the NCSM Hamiltonian. For comparison we also
perform calculations with the bare Daejeon16 $NN$ potential, as will be discussed later.

Adding (and subtracting later) the center-of-mass harmonic-oscillator Hamiltonian, 
as well as replacing $A$ by $a\le A$ in the summations, 
the Hamiltonian of Eq.~(\ref{H_intrinsic}) can be re-written as
\begin{equation}
\label{H_A_body}
 H_{a}+H_{CM} = \sum_{i=1}^{a}\left[\frac{\vec{p_i}^2}{2m} 
                        + \frac{1}{2}m\Omega^2\vec{r_i}^2\right]  
                        + \sum_{i<j=1}^{a}V_{ij}(\Omega,A),
\end{equation}
where $V_{ij}(\Omega,A)$ is the modified bare $NN$ interaction,
including dependence on $\Omega $ and $A$:
\begin{equation}
\label{V_Omega_A}
 V_{ij}(\Omega,A) = V_{ij}^{NN}-
                    \frac{m\Omega^2}{2A}(\vec{r}_i-\vec{r}_j)^2.
\end{equation}
If $a=A$, then $H_a$ coincides with the full Hamiltonian $H$ of Eq.~(\ref{H_intrinsic}).

\begin{table}[!t]
\vspace{-1ex}
\caption{\label{tab:F18_Ex_ncsm_Nmax4}The NCSM energies (in MeV) of the lowest
28 states $J^\pi_i$ of $^{18}$F calculated in the $4\hbar \Omega$ model space by
using OLS transformed and bare Daejeon16 $NN$ interaction with $\hbar \Omega=14$ MeV.}
\begin{ruledtabular}
\begin{tabular}{rrr|rrr}
$J^\pi_i$ & T & OLS & $J^\pi_i$ & T & bare \\ \hline
$3^+_1$ & 0  & $-130.407$ & $3^+_1$ & 0 & $-126.069$  \\
$1^+_1$ & 0  & $-130.400$ & $1^+_1$ & 0 & $-126.032$  \\
$5^+_1$ & 1  & $-129.314$ & $5^+_1$ & 1 & $-125.087$  \\
$0^+_1$ & 0  & $-129.122$ & $0^+_1$ & 0 & $-124.817$  \\
$2^+_1$ & 1  & $-127.433$ & $2^+_1$ & 1 & $-123.081$  \\
$2^+_2$ & 0  & $-127.265$ & $2^+_2$ & 0 & $-122.965$  \\
$1^+_2$ & 0  & $-126.180$ & $1^+_2$ & 0 & $-121.884$  \\
$0^+_2$ & 1  & $-126.092$ & $0^+_2$ & 1 & $-121.778$  \\
$2^+_3$ & 1  & $-125.649$ & $2^+_3$ & 1 & $-121.402$  \\
$4^+_1$ & 1  & $-125.281$ & $4^+_1$ & 1 & $-121.071$  \\
$3^+_2$ & 1  & $-124.812$ & $3^+_2$ & 1 & $-120.591$  \\
$3^+_3$ & 0  & $-124.737$ & $3^+_3$ & 0 & $-120.421$  \\
$1^+_3$ & 0  & $-120.704$ & $1^+_3$ & 0 & $-116.545$  \\
$4^+_2$ & 0  & $-119.269$ & $4^+_2$ & 0 & $-115.164$  \\
$2^+_4$ & 0  & $-118.668$ & $2^+_4$ & 0 & $-114.521$  \\
$1^+_4$ & 0  & $-117.351$ & $1^+_4$ & 0 & $-113.214$  \\
$4^+_3$ & 1  & $-116.416$ & $4^+_3$ & 1 & $-112.337$  \\
$2^+_5$ & 1  & $-115.744$ & $2^+_5$ & 1 & $-111.594$  \\
$3^+_4$ & 1  & $-115.650$ & $3^+_4$ & 0 & $-111.579$  \\
$1^+_5$ & 0  & $-115.283$ & $1^+_5$ & 1 & $-111.112$  \\
$2^+_6$ & 0  & $-115.231$ & $2^+_6$ & 0 & $-111.092$  \\
$2^+_7$ & 1  & $-114.917$ & $2^+_7$ & 1 & $-110.803$  \\
$1^+_6$ & 1  & $-114.885$ & $1^+_6$ & 1 & $-110.779$  \\
$3^+_5$ & 1  & $-114.820$ & $3^+_5$ & 1 & $-110.748$  \\
$3^+_6$ & 0  & $-107.854$ & $3^+_6$ & 0 & $-103.869$  \\
$0^+_3$ & 0  & $-106.258$ & $0^+_3$ & 0 & $-102.246$  \\
$1^+_7$ & 1  & $-105.969$ & $1^+_7$ & 1 & $-101.928$  \\
$2^+_8$ & 1  & $-105.262$ & $2^+_8$ & 1 & $-101.291$  \\
\end{tabular}
\end{ruledtabular}
\end{table}

\begin{table*}[ht] 
\vspace{-1ex}
\caption{\label{tab:spe_DJ16} Neutron ("$\nu$") and proton ("$\pi$") single-particle energies (in MeV) 
obtained from the OLS transformed Daejeon16 potential (for $A=18$ and $A=19$) and from bare Daejeon16.}
\begin{ruledtabular}
\begin{tabular}{c|ccc|ccc|ccc|}
 & \multicolumn{6}{c|}{OLS} & \multicolumn{3}{c|}{Bare} \\
\hline
 & \multicolumn{3}{c|}{$A=18$} & \multicolumn{3}{c|}{$A=19$} & \multicolumn{3}{c|}{$A=18$}  \\ 
 & \multicolumn{3}{c|}{$E_{\rm core}=-121.817$} & \multicolumn{3}{c|}{$E_{\rm core}=-121.783$} 
& \multicolumn{3}{c|}{$E_{\rm core}=-118.307$} \\[1mm] 
\hline
$(nlj)$ & $1s_{1/2}$ &$0d_{5/2}$ & $0d_{3/2}$ &  $1s_{1/2}$ & $0d_{5/2}$ & $0d_{3/2}$ 
& $1s_{1/2}$ &$0d_{5/2}$ & $0d_{3/2}$  \\[1mm] 
\hline
$\epsilon_{\nu }(nlj)$ & $-3.576$ & $-3.302$ & $6.675$ & $-3.572$ & $-3.299$ & $6.677$ & $-3.115$ & $-2.953$ & $6.889$ \\
$\epsilon_{\pi }(nlj)$ & $-0.077$ & $ 0.291$ & $9.974$ & $-0.073$ & $ 0.294$ & $9.976$ & $ 0.362$ & $ 0.621$ & $10.174$ \\
\end{tabular}
\end{ruledtabular}
\end{table*} 

The effective $NN$ interaction for the NCSM calculations
in an $a$-cluster approximation is constructed from
eigenstates of Eq.~(\ref{H_A_body}) in a sufficiently large basis space 
involving up to several hundred radial excitations for each orbital angular momentum. 
In the $a=2$ cluster approximation which we adopt here, 
a first OLS transformation of $H_2$ is performed~\cite{Dikmen2015} 
to produce a primary effective Hamiltonian, $H_2^P$, for the $A=18$
 systems within a certain model space defined by the $N_{\rm max}$ parameter.
The model space considered here for $^{18}$F has been fixed at $N_{\rm max}=4$  
with $\hbar \Omega =14$~MeV, as in the preceding work~\cite{Dikmen2015}.
The NCSM calculations for $^{18}$F have been performed
using the MFDn code~\cite{MFDn1,MFDn2,MFDn3,MFDn4,MFDn5}.
The lowest 28 eigenstates of $^{18}$F dominated by $N=0$ components 
(see Table~\ref{tab:F18_Ex_ncsm_Nmax4}, left {column) 
have been used to set up the secondary OLS transformation to the $sd$-shell valence space.
The resulting effective valence Hamiltonian, $H_{18}^{P'P}$, an effective one- and two-body operator for $N'_{\rm max}=0$ is derived.
By construction, the energies of $H_{18}^{P'P}$ for 2 valence nucleons in the $sd$ shell exactly coincide
with the eigenvalues of NCSM Hamiltonian for $^{18}$F in the full $N_{\rm max}=4$ oscillator space. 
See Ref.~\cite{Dikmen2015} for more details.
Then a NCSM calculation with the same $N_{\rm max}=4$ and  $\hbar \Omega =14$~MeV
 is performed for $^{16}$O to get the core energy and for $^{17}$O and $^{17}$F.
Subtracting the core energy from the latter calculation, one can get effective neutron
and proton one-body terms.
Subtraction of the core energy plus the one-body terms from the effective Hamiltonian for $^{18}$F, 
allows one to obtain the residual TBMEs to be used in the valence-space shell-model calculations.

The details of the approach and numerical values of the core, the single-particle energies and
the TBMEs, given by the chiral N3LO and JISP16 interactions, can be found in Ref.~\cite{Dikmen2015}.
The core energy and the single-particle energies obtained from the Daejeon16 $NN$ potential
are given in Tables~\ref{tab:spe_DJ16} (columns labeled as $A=18$), 
while the  TBMEs  are summarized in Table~\ref{tab:TBMEs_DJ16} in Appendix~\ref{app-TBMEs}. 
The files with TBMEs are also available online~\cite{git}.

\vspace{2mm}

{\it Bare versus OLS-renormalized potential}.
The Daejeon16 $NN$ interaction~\cite{DJ16}, used in the present study, is rooted in the SRG-evolved $\chi $EFT $NN$ 
interaction, supplemented by a specific adjustment of the off-shell properties via phase-equivalent transformations
to fit the spectroscopic data on light nuclei.
Being a ``soft'' potential, Daejeon16 provides good convergence of NCSM calculations 
as seen through the rapid approach to independence of the basis space parameters~\cite{DJ16}.  
To provide a useful first approximation to the converged NCSM basis spaces we elect to   
solve the eigenproblem for $^{18}$F at $N_{\rm max}=4$ using directly the Hamiltonian of Eq.~(\ref{H_intrinsic}) 
with the bare Daejeon16 interaction, avoiding the construction of 
a primary effective Hamiltonian.

As mentioned above, from the full NCSM calculations for $A = 18$ 
(with either the bare or the OLS-renormalized interactions), 
we identify and employ the 28 lowest eigenstates, dominated by $N=0$ components
(see Table~\ref{tab:F18_Ex_ncsm_Nmax4}), to generate the OLS transformation to the $sd$ shell
needed to construct the effective valence Hamiltonian.
For the valence Hamiltonian, we refer to results obtained with the bare Daejeon16 interaction 
used in the NCSM calculations for  $A = 16{-}18$ as the ``bare'' results 
even though the OLS transformation to the valence space has been implemented.  
When the OLS transformation has been implemented for both the NCSM calculations for $A = 16{-}18$, 
as well as for calculation of the effective valence space interaction, 
we refer to the results as ``OLS'' results.  
The corresponding core energy and single-particle energies 
are summarized in Table~\ref{tab:spe_DJ16}, while the TBMEs are presented in Appendix~\ref{app-TBMEs}
(column labeled as ``TBMEs of DJ16'' or ``TBMEs of DJ16$_{bare}$'' for TBMEs from OLS and bare results, respectively).

Although the core energies from OLS results and bare results are somewhat different, the TBMEs 
appear very similar.  
In order to employ a uniform set of procedures for all $NN$ interactions, 
including those which have strong short-range correlations, 
we will adopt the DJ16
TBMEs obtained from Daejeon16 via the double OLS transformation for further study below.

\vspace{2mm}

{\it Single-particle energies}. 
In phenomenological studies, single-particle energies for valence-space shell-model calculations
are conventionally taken from experimental spectra of closed-shell nuclei plus one proton or one neutron. 
The phenomenological USDB Hamiltonian uses the following optimized values:
$\varepsilon (0d_{5/2})=-3.9257$~MeV, $\varepsilon (1s_{1/2})=-3.2079$~MeV and 
$\varepsilon (0d_{3/2})=2.1117$~MeV. 

It is clear from Table~\ref{tab:spe_DJ16}, that theoretical single-particle
energies obtained in this section differ markedly from the phenomenological values.
First, we observe that the $d_{5/2}$ and $s_{1/2}$ orbitals are inverted. Second,
the spin-orbit splitting between $d_{3/2}$ and $d_{5/2}$ is about 10~MeV,
which is almost two times larger than the empirical value.
The reasons for these deficiencies stem from the fact that the Daejeon16 $NN$ potential was not fitted for
nuclei with~$A>16$; for example, the $d$-wave $NN$ interaction, which could be important for
generating the single-particle levels in the $sd$ shell, was nearly not involved in the fit of Daejeon16
to nuclear data. We shall address this issue in future studies.

For the major part of the present study we will adopt the USDB single-particle energies, 
the same for neutrons and protons, in order to focus our attention on the TBMEs. 
Fig.~\ref{fig:F18} shows the low-energy spectrum of $^{18}$F obtained from USDB and from various
microscopic interactions. The results labeled as BonnC, N3LO, JISP16 and DJ16 are obtained
from the respective TBMEs and the USDB single-particle energies. 
The spectrum labeled as DJ16A are obtained from a modified
DJ16 interaction, which will be explained in Section IV.D.

Results labeled as DJ16th are obtained from the DJ16 TBMEs with
theoretical single-particle energies from Table~\ref{tab:spe_DJ16}.
This is the spectrum which, by construction, coincides with the full NCSM results 
from the OLS-transformed Daejeon16 potential at $N_{\rm max}=4$ (Table~\ref{tab:F18_Ex_ncsm_Nmax4}).
One observes that DJ16th predicts a $3^+$ ground state, instead of $1^+$.
There are additional deficiencies in the spectra of other nuclei. 
We shall discuss some of the results obtained with DJ16th in the next Section.

\begin{figure}[tb]
 \centering
  \includegraphics[width=\columnwidth]{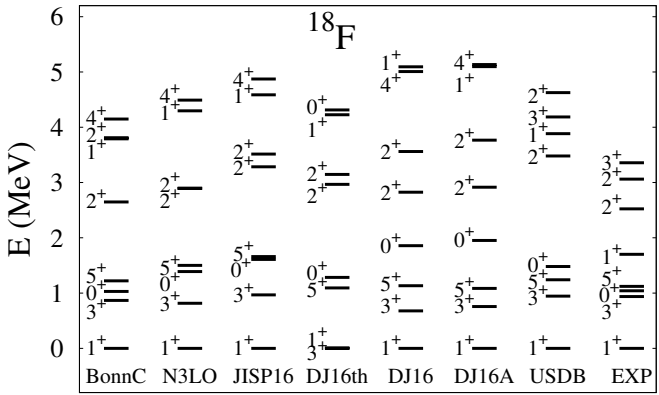} 
  \caption{\label{fig:F18} Low-energy spectra (8 lowest states) of $^{18}$F
obtained from USDB and from the microscopic effective interactions in comparison with experiment
(8 lowest positive-parity states).  
The USDB single-particle energies have been used in all calculations, except for DJ16th, where theoretical
values from Table~\ref{tab:spe_DJ16} are used with DJ16 TBMEs. 
DJ16A is a monopole-modified version of DJ16 (see Section~\ref{monopole-mod-DJ16A}).} 
\end{figure}

\vspace{2mm}

{\it $\hbar \Omega $ dependence.}
The microscopic effective interactions obtained from NCSM wave functions depend on the $\hbar \Omega $ value
of the harmonic oscillator potential. 
The NCSM calculations with bare potentials exhibit clear monotonic convergence patterns as a function of 
$\hbar \Omega $ and $N_{\rm max}$~\cite{Barrett_2013} 
and the results obtained with~$\hbar\Omega$ values corresponding to the minimum of 
the~$\hbar\Omega$ dependence at a given~$N_{\max}$ provide the best approximation for the converged energy.
At the same time, the NCSM calculations with OLS-transformed effective Hamiltonians 
usually converge even faster in model spaces with small and moderate~$N_{\max}$ values, 
however, the convergence patterns in this case are more complicated and may be non-monotonic~\cite{Maris2010}; 
the definition of the optimal~$\hbar\Omega$ value providing the best approximation to the converged result
in this case is not obvious.
Following the previous study~\cite{Dikmen2015},
all NCSM calculations discussed in this article are performed for $\hbar \Omega =14$~MeV. 
This value is close to the empirical value for $^{16}$O~\cite{Kirson07}.

\vspace{2mm}

{\it $A$ dependence.} In phenomenology, the matrix elements of an effective two-body potential calculated in a harmonic
oscillator basis should depend on the oscillator frequency $\Omega $. 
The latter is related to the mean square radius of the nucleus.
To approximately match the empirical root mean square radius with nucleons in a harmonic oscillator basis, 
the $A$ dependence of the $\hbar \Omega $ value can be expressed as
$$
\hbar \Omega \approx 40 \, A^{-1/3} \;\; {\rm MeV}.
$$
It can be shown that TBMEs of a $\delta $-force scale as $(\hbar \Omega )^{3/2}$,
while those of the two-body Coulomb interaction as $(\hbar \Omega )^{1/2}$.
TBMEs of a short-range potential are known to be mainly linear as a function
of $\hbar \Omega$. This is why phenomenological interactions are typically scaled as
$(A/A_0)^{-1/3}$, where $A_0$ is 
the mass number of a core plus two-nucleons system, 
which is essential for getting high quality results~\cite{CaurierRMP}.
For example, the phenomenological USDB interaction supposes $(A/A_0)^{-0.3}$ scaling ($A_0=18$)~\cite{USDab},
which is close to the estimate above.
At the same time, single-particle energies are most often assumed to be constant
throughout a given shell.

As was discussed in the beginning of this section,
the primary NCSM effective Hamiltonian depends on $A$ for which the OLS
transformation has been performed.
Therefore, the single-particle energies
and TBMEs obtained from a given $NN$ potential via two subsequent OLS transformations are 
mass dependent. 
The difference in the core energy and single-particle energies deduced for $A=18$ and $A=19$
from Daejeon16 is seen from Table~\ref{tab:spe_DJ16} (OLS columns) 
to be smaller than the phenomenologically successful $A^{-1/3}$ dependence.
A similar situation is observed for the single-particle energies and 
TBMEs deduced from JISP16 and N3LO potentials, as is reported in Ref.~\cite{Dikmen2015}.
 
The OLS-induced mass dependence of TBMEs obtained from the Daejeon16 potential is also very weak,
compared to the generally accepted empirical $A$ dependence.
Even the mass dependence of our theoretical single-particle energies produces negligible effect on
nuclear spectra and weakly influences the binding energies.
We expect that a more complete treatment, including bare and induced three-nucleon interactions 
as well as NCSM calculations in larger model spaces, will be needed to gain the stronger $A$ dependence 
that is observed in phenomenologically successful valence Hamiltonians.  
In the meantime, we will adopt the empirical $A$-dependent scaling for our derived Hamiltonian.

In the present study, while using the USDB single-particle energies, 
we impose the empirical scaling $(A/A_0)^{-0.3}$ for TBMEs
of all  microscopic effective interactions except for IMSRG. 
The IMSRG effective Hamiltonians have a much more sophisticated $A,Z$ dependence of their 
single-particle energies and TBMEs,
which cannot be given by an analytical expression for the use in the whole $sd$ shell,
since the Hamiltonian should be independently calculated for each particular nuclide. 
%
We use the original Hamiltonians for each $(A,Z)$
as derived in Ref.~\cite{StrPRL118,git-Stroberg} for comparison with the present results.

\vspace{2mm}

{\it Charge dependence}. In principle, one can derive the effective interaction from NCSM calculations 
not only for $^{18}$F, but also for $^{18}$O and $^{18}$Ne and, thus,  provide a full charge-dependent 
valence-space shell-model interaction.
The study of these charge-dependent aspects is in progress and will be published elsewhere. 
In the present work, we restrict ourselves
to the $T=0$ and $T=1$ TBMEs, as obtained from NCSM calculation for $^{18}$F, assuming charge symmetry and
charge independence of the nuclear effective Hamiltonian.
The USDB single-particle energies allow one to study nuclear spectra in the isospin-symmetry limit.
Charge-symmetry violation will be introduced in the next Section, 
when we use different proton and neutron theoretical single-particle energies
from Table~\ref{tab:spe_DJ16}.



\section{Theoretical valence-space effective Hamiltonians}
We start exploring the features of the effective interaction from NCSM by a straightforward
use of the theoretical valence-space effective Hamiltonians derived from the JISP16 and Daejeon16
$NN$ interactions, i.\:e., using the  theoretical single-particle 
energies and adopting no mass dependence of the TBMEs.
We will review here only O isotopes, so  only the neutron single-particle energies are of interest.
For JISP16, these are
$\varepsilon (\nu 0d_{5/2})=-2.270$~MeV, $\varepsilon (\nu 1s_{1/2})=-3.068$~MeV and 
$\varepsilon (\nu 0d_{3/2})=6.262$~MeV \cite{Dikmen2015}, while for Daejeon16 potential
the single-particle energies are given in Table~\ref{tab:spe_DJ16}.
We again note the inversion of the $0d_{5/2}$ and $1s_{1/2}$ single-particle states, 
as well as a large spin-orbit splitting of about 8~MeV (JISP16) and 10~MeV (Daejeon16).

The calculated ground state energies are shown in Fig.~\ref{fig:BE_O_th}. 
The results obtained with TBMEs from JISP16 and Daejeon16 using theoretical single-particle energies
are labeled as JISP16th and DJ16th, respectively.
We observe that DJ16th shows more bound ground states of $^{17-25}$O isotopes 
than JISP16th. 
The reason is that the theoretical single-particle energies for $0d_{5/2}$ and $1s_{1/2}$ from JISP16 
are about 1~MeV and 0.5~MeV, respectively, less negative than those from Daejeon16.
Even more, we notice that JISP16th significantly underbinds the lighter O isotopes. 

We also note that starting from $^{25}$O, the binding energies increase, 
thus JISP16th and DJ16th place correctly the drip line at $^{24}$O. 
This increase is due to the fact that the $d_{3/2}$ orbital starts to be filled in $^{25}$O, 
which brings a large gain in energy in very neutron-rich O isotopes.
The increase is especially significant for DJ16th, since the theoretical $\varepsilon (\nu 0d_{3/2})$ 
from Daejeon16 is about 0.4~MeV higher than that from JISP16.
The corresponding rms deviations in these relative binding energies are 1.9~MeV and 2.9~MeV for DJ16th and JISP16th, 
respectively.

For comparison, we present also the results labeled JISP16 and DJ16 obtained with empirical 
single-particle energies and mass dependence of the TBMEs. These results look better for light
O isotopes but significantly worse for heavier isotopes where they
start to deviate essentially from experiment.  In particular, these 
valence-space effective Hamiltonians do not reproduce the drip line in the O isotopes. We note that without 
the phenomenological mass dependence of the TBMEs we get even worse results for binding energies since the
valence space interaction in this case is stronger for heavier isotopes and hence the binding energies
are larger. We shall address the problem of binding energies by a slight modification of monopole
properties of the DJ16 interaction in the next Section.


The theoretical single-particle energies give rise to the results for binding energies which look
qualitatively correct though demonstrate clear quantitative   deviations from experiment. 
The deficiencies of our theoretical single-particle energies manifest in the nuclear spectra. As an example,
we show in Fig.~\ref{fig:O-spectra-theo} the spectra of $^{21,23}$O calculated with JISP16 and Daejeon16 using
the  theoretical single-particle energies (JISP16th and DJ16th results). 
The spectrum of $^{21}$O obtained with theoretical single-particle energies is far too compressed.
The dominant component of the $3/2^+_1$ state (by all interactions, including USDB) is due to the coupling of 
$d^4_{5/2}\times s_{1/2}$ and not due to the filling of the $d_{3/2}$ single-particle orbital.
In the spectrum of $^{23}$O, the $1/2^+_1$ and $5/2^+_1$ states are too close to each other 
(and even inverted in the case of JISP16th), while the $3/2^+_1$ state is too high. 
The latter is mainly due to the filling of the $d_{3/2}$ orbital.

The spectra  of $^{21,23}$O isotopes obtained with the phenomenological single-particle
energies are seen from Fig.~\ref{fig:O-spectra-theo} (JISP16 and DJ16 results) to reproduce the
experiment much better though $^{21}$O is still too compressed relative to experiment.  
Given the deficiencies of our current theoretical single-particle energies and with the motivation 
to further investigate the properties of our derived TBMEs, we will employ these USDB single-particle
energies and $A$-scaled TBMEs in all our calculations presented below except for the
results obtained with the IMSRG effective valence space interaction which
have been obtained using the original respective Hamiltonians from Ref.~\protect\cite{StrPRL118}.

\begin{figure}[!t]
  \includegraphics[width=\columnwidth]{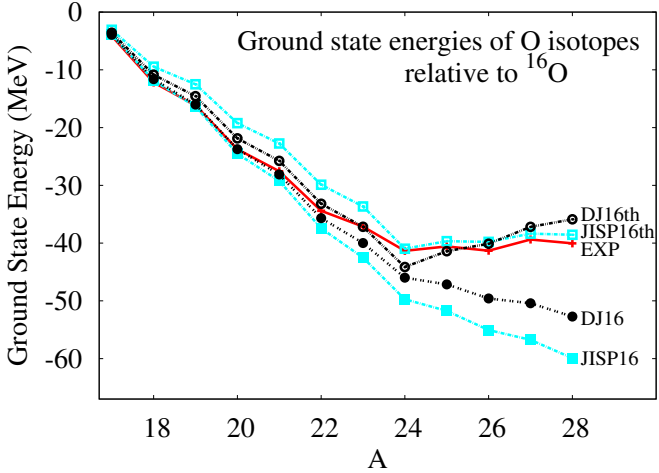} 
  \caption{\label{fig:BE_O_th} (Color online) Experimental ground state energies of O isotopes relative 
to the ground state energy of $^{16}$O in comparison with theoretical results, calculated 
using microscopic effective interactions.
Calculations using JISP16 and DJ16 are performed with the USDB single-particle energies and 
with the  $(A/A_0)^{-0.3}$ mass dependence of TBMEs.
Results denoted as JISP16th (empty cyan squares) and DJ16th (empty black circles) are obtained 
from the same valence-space TBMEs (without any mass dependence) 
and with the
theoretical single-particle energies.
The experimental data (extrapolations included) are from AME2012~\protect\cite{AME2012}.}
\end{figure}

\begin{figure*}[!]
  \includegraphics[width=0.48\textwidth]{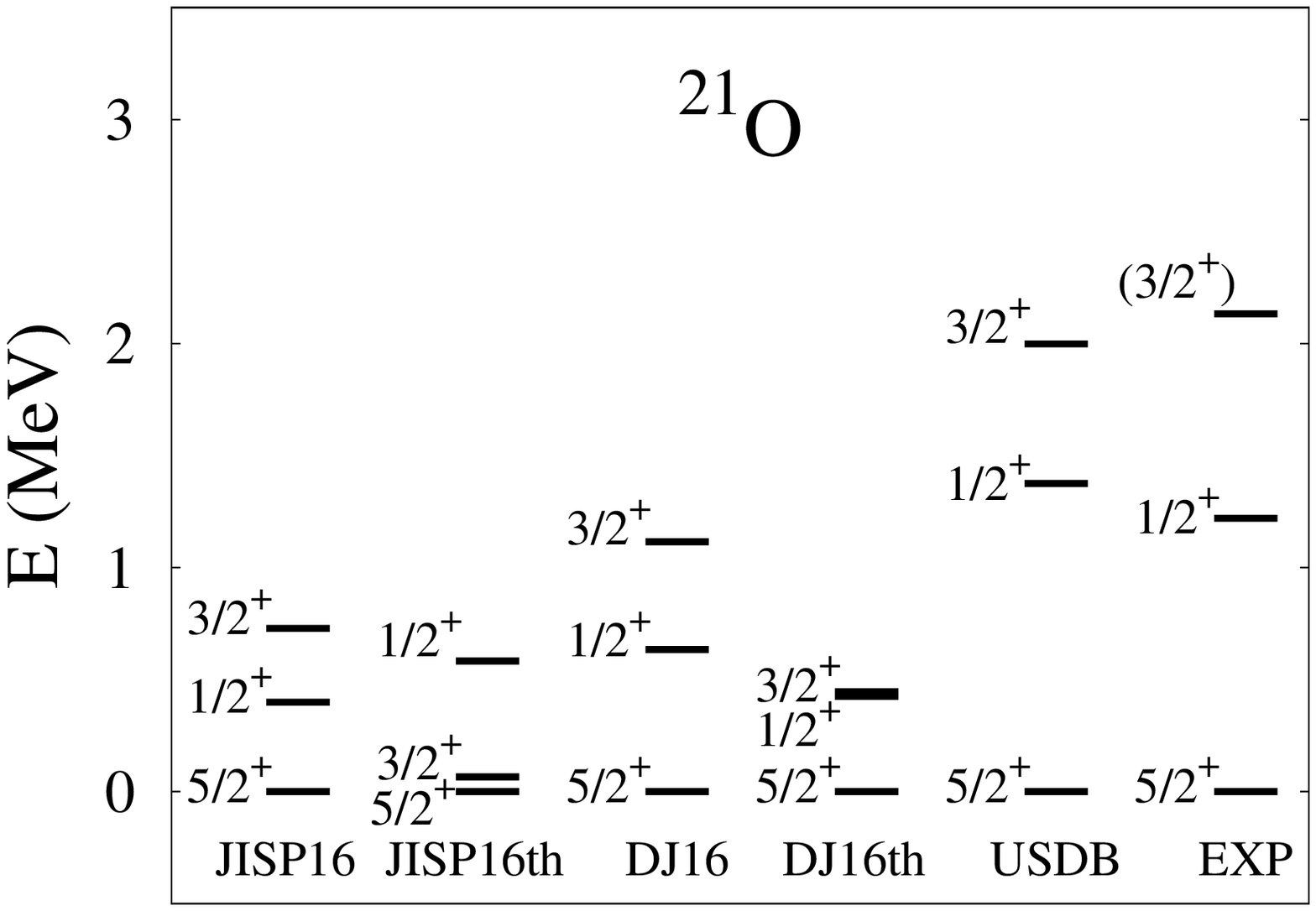} 
  \includegraphics[width=0.48\textwidth]{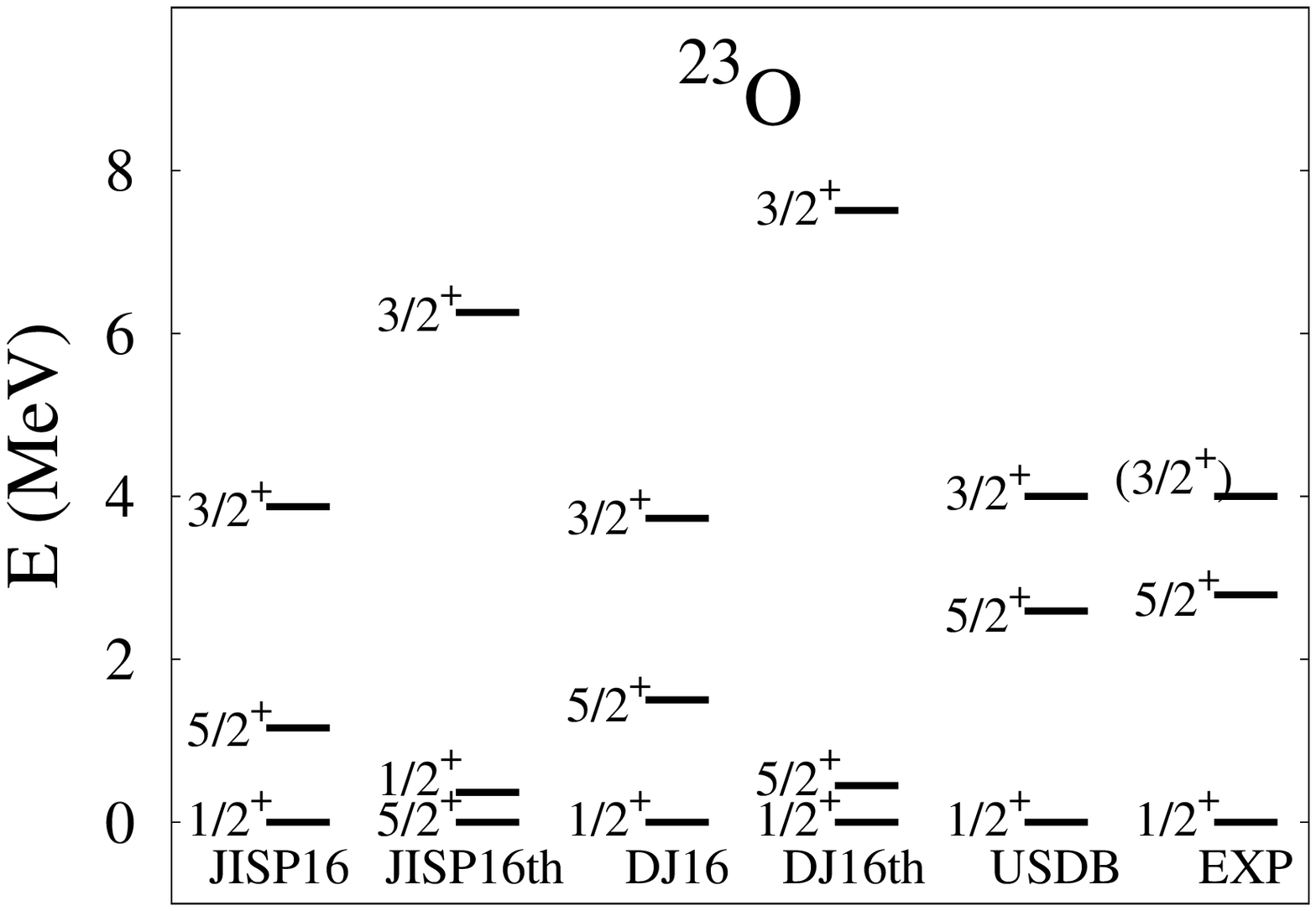}
  \caption{\label{fig:O-spectra-theo} Experimental low-energy spectrum of $^{21,23}$O in comparison with theoretical results, 
calculated using the phenomenological USDB interaction and 
the microscopic effective interactions. Calculations with JISP16 and DJ16 employ 
the USDB single-particle energies and the $(A/A_0)^{-0.3}$ mass dependence of TBMEs, while
spectra denoted as JISP16th and DJ16th are obtained from the same valence-space TBMEs 
(without any mass dependence) and with the theoretical single-particle energies.}
\end{figure*}
%
%

\section{Monopole properties}

\subsection{Monopole Hamiltonian}

The monopole part~\cite{BaFr64} of the valence-space shell-model Hamiltonian plays an important role for \mbox{spectroscopic} properties
since it needs to encapsulate the robust evolution of the spherical nuclear mean field 
as a function of valence nucleons~\cite{PoZu81,ZuDu95}.
In the model space of one or two major oscillator shells, the monopole Hamiltonian
{contains} only terms involving proton and neutron number operators  and can be expressed as
\begin{multline}
\label{hamiltonian}
\displaystyle \hat H_{mon} = \sum\limits_k\epsilon^{\nu }_k \, \hat n^{\nu }_k +
\sum\limits_k\epsilon^{\pi }_k \, \hat n^{\pi }_k +
\sum\limits_{kk'} V^{\nu \pi }_{kk'} \,  \hat n^{\nu }_k \hat n^{\pi }_l  \\ +
\displaystyle \sum\limits_{k\le k'} 
\frac{\hat n^{\nu }_k (\hat n^{\nu }_{k'} - \delta_{kk'})}{1+\delta_{kk'}} \,  V^{\nu \nu } _{kk'}+
\sum\limits_{k\le k'} \frac{\hat n^{\pi }_k (\hat n^{\pi }_{k'}-\delta_{kk'})}{1+\delta_{kk'}} \, 
V^{\pi \pi }_{kk'}  ,
\end{multline}
where $\hat n^{\pi }_k $ and $\hat n^{\nu }_k $ are proton and neutron number operators, $k$ $(k')$ refer to
a complete set of quantum numbers of a harmonic oscillator orbital, e.\:g., $k\equiv (n_k l_k j_k)$,
and $V^{\rho \rho' }_{kk'}$ are centroids of the two-body interaction,
\begin{equation}
\label{centroid}
V^{\rho \rho' }_{kk'} = \frac{ \sum\limits_J 
\langle k^{\phantom{'}}_{\rho } k'_{\rho'} |V| k^{\phantom{'}}_{\rho } k'_{\rho'}\rangle^{\phantom{'}}_J (2J+1)}
{ \sum\limits_J (2J+1)}  ,
\end{equation}
where $\rho $, $\rho' $ denote protons ($\pi $) or neutrons ($\nu $) and  the total angular momentum of a two-body state $J$ 
runs over all values allowed by the Pauli principle.

The nucleon separation energies, calculated using the monopole Hamiltonian, are called
{\it effective single-particle energies (ESPEs)}~\cite{OtHo01}. 
They provide an easy and quite useful way to analyze centroids of an interaction.
For example, if we assume a traditional shell-model (or normal) filling of the orbitals and evaluate ESPEs for closed sub-shell nuclei,
keeping the mass-dependence of TBMEs approximately the same for nuclei with $A$ and $A\pm 1$,
the evolution of ESPEs with respect to a reference nucleus ($A_r$), can be expressed as
\begin{equation}
\label{ESPE}
\tilde \varepsilon^{\rho }_k (A) = \varepsilon^{\rho }_k (A_r) + 
\sum\limits_{k', \rho' }V^{\rho \rho '}_{kk'} \, n^{\rho '}_{k'},
\end{equation}
where $k'$ runs over filled valence space orbitals and $n^{\rho '}_{k'}$ is the occupation
number of the orbital $k'$ for nucleons of the type $\rho '$. 

Subtracting the monopole part of the Hamiltonian from the full valence space Hamiltonian $\hat{H}$, 
we obtain the "higher-multipole" Hamiltonian, 
{\bf  ${\hat H_{mult}  = \hat H- \hat H_{mon}}$}, which
contains particle-particle correlations --- pairing, quadrupole-quadrupole correlations and so on
(see Ref.~\cite{DuZu96} for details).


\begin{figure*}[!t]
 \centering
  \includegraphics[height=.22\textwidth]{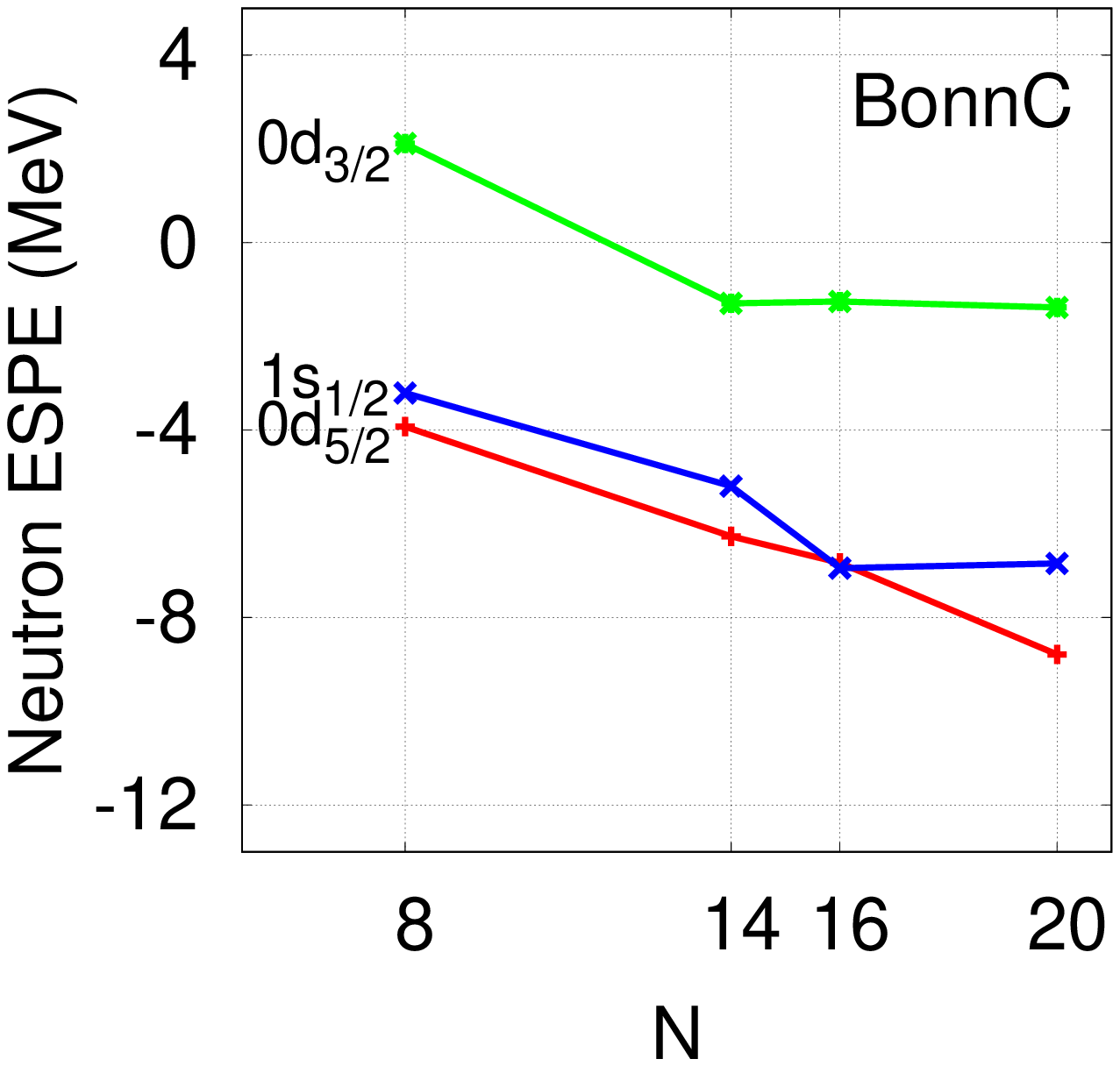}   
  \includegraphics[height=.22\textwidth]{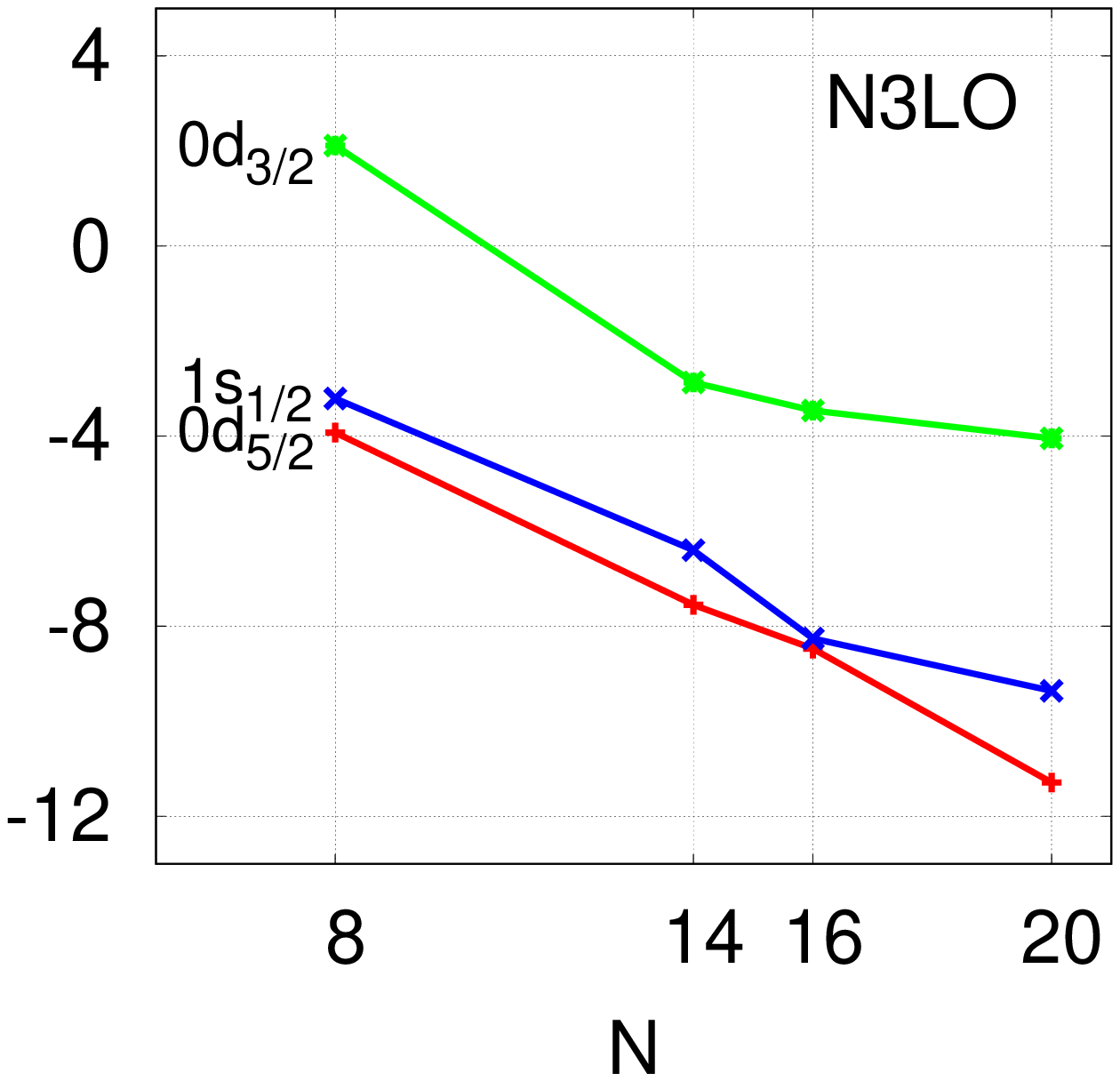}
  \includegraphics[height=.22\textwidth]{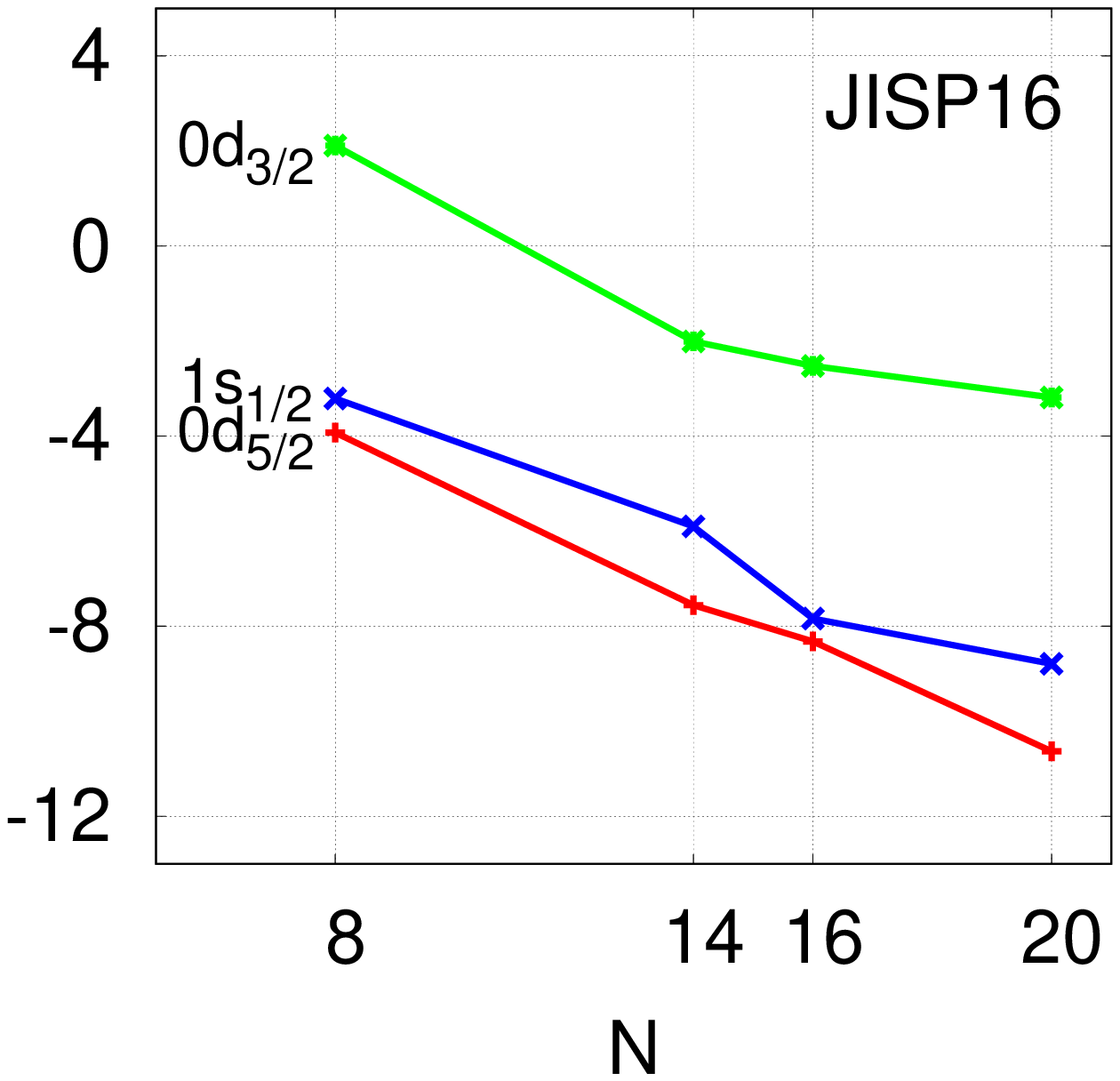} \\[3mm]    
  \includegraphics[height=.22\textwidth]{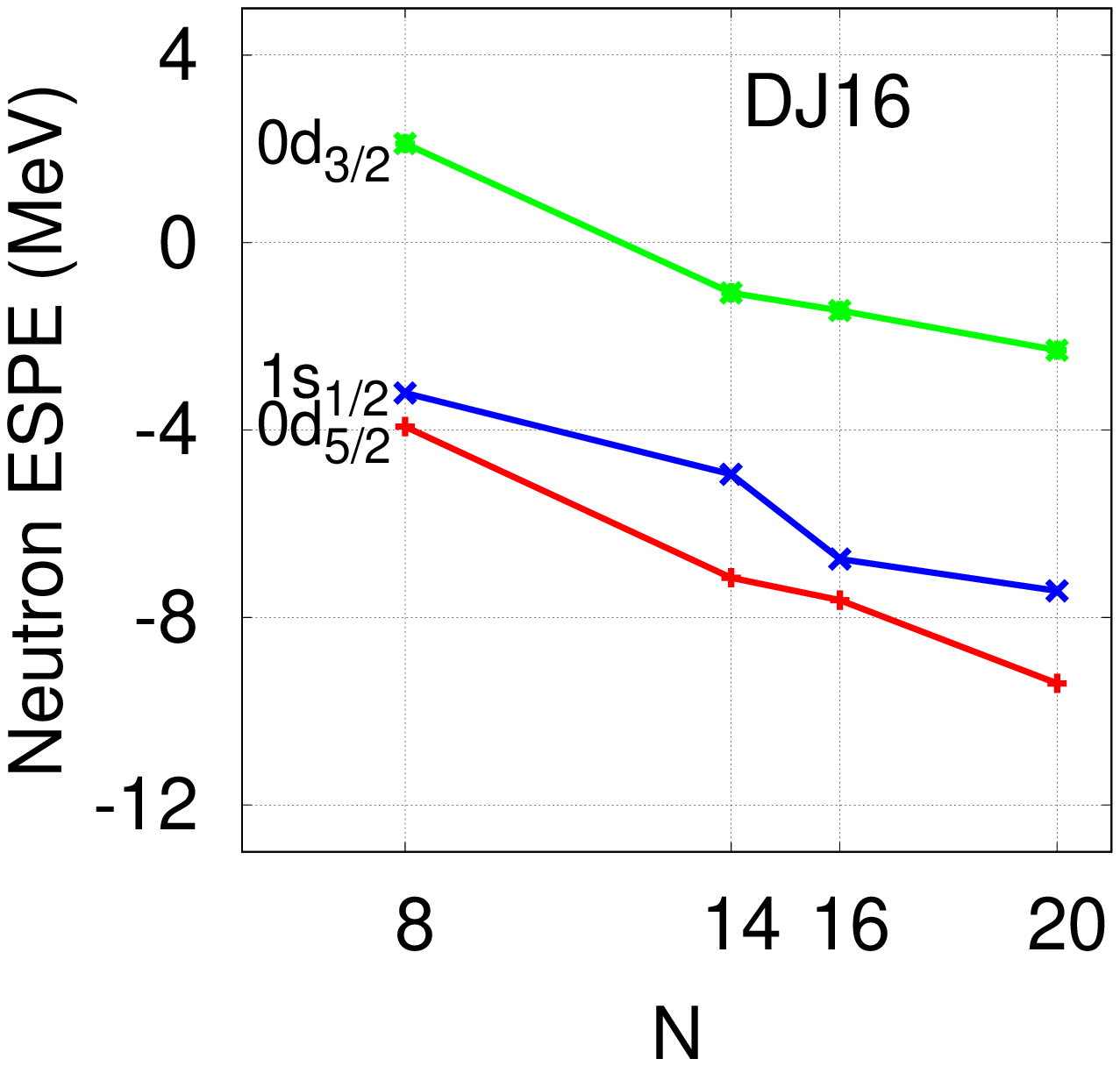}  
  \includegraphics[height=.22\textwidth]{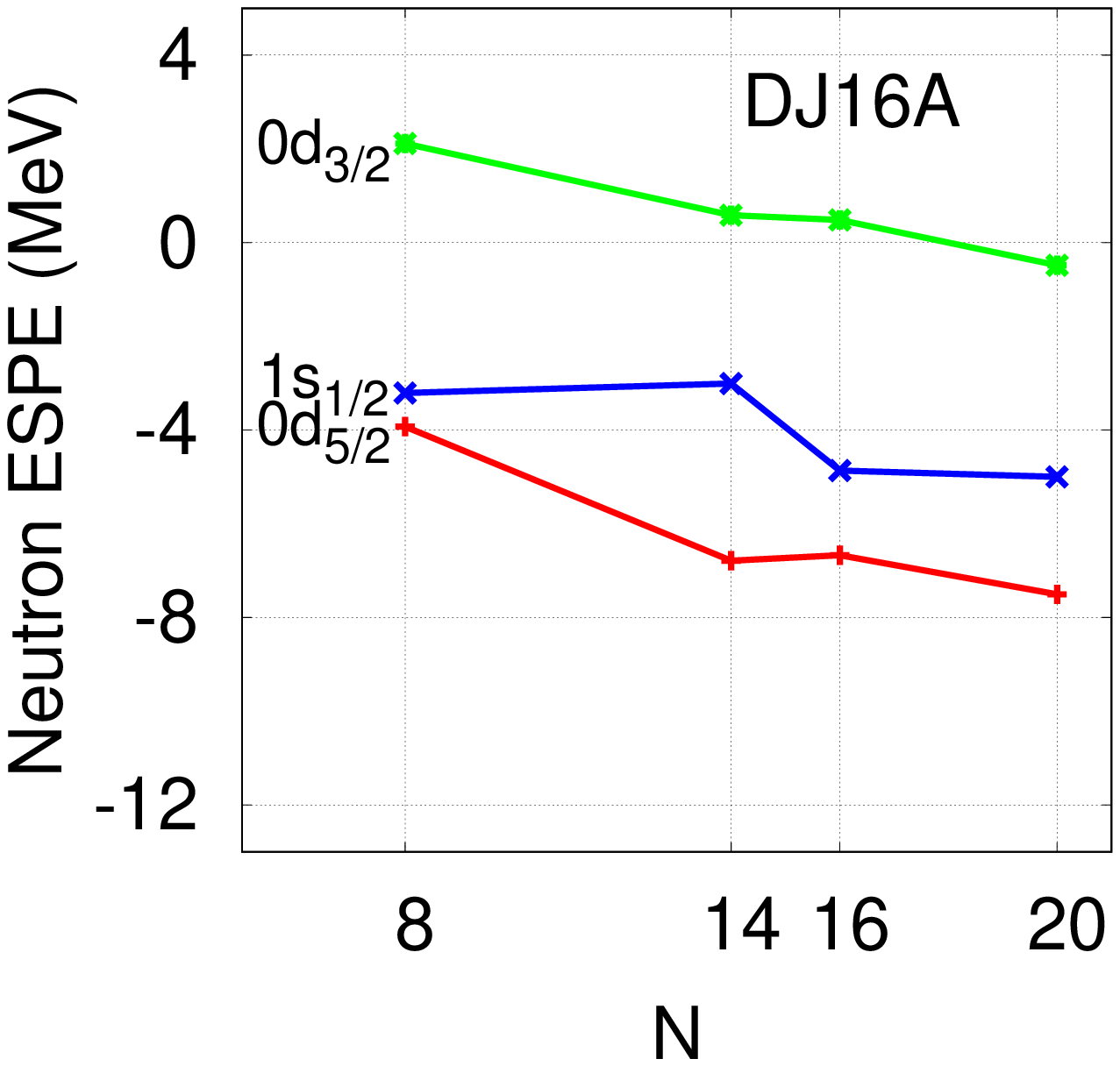}  
  \includegraphics[height=.22\textwidth]{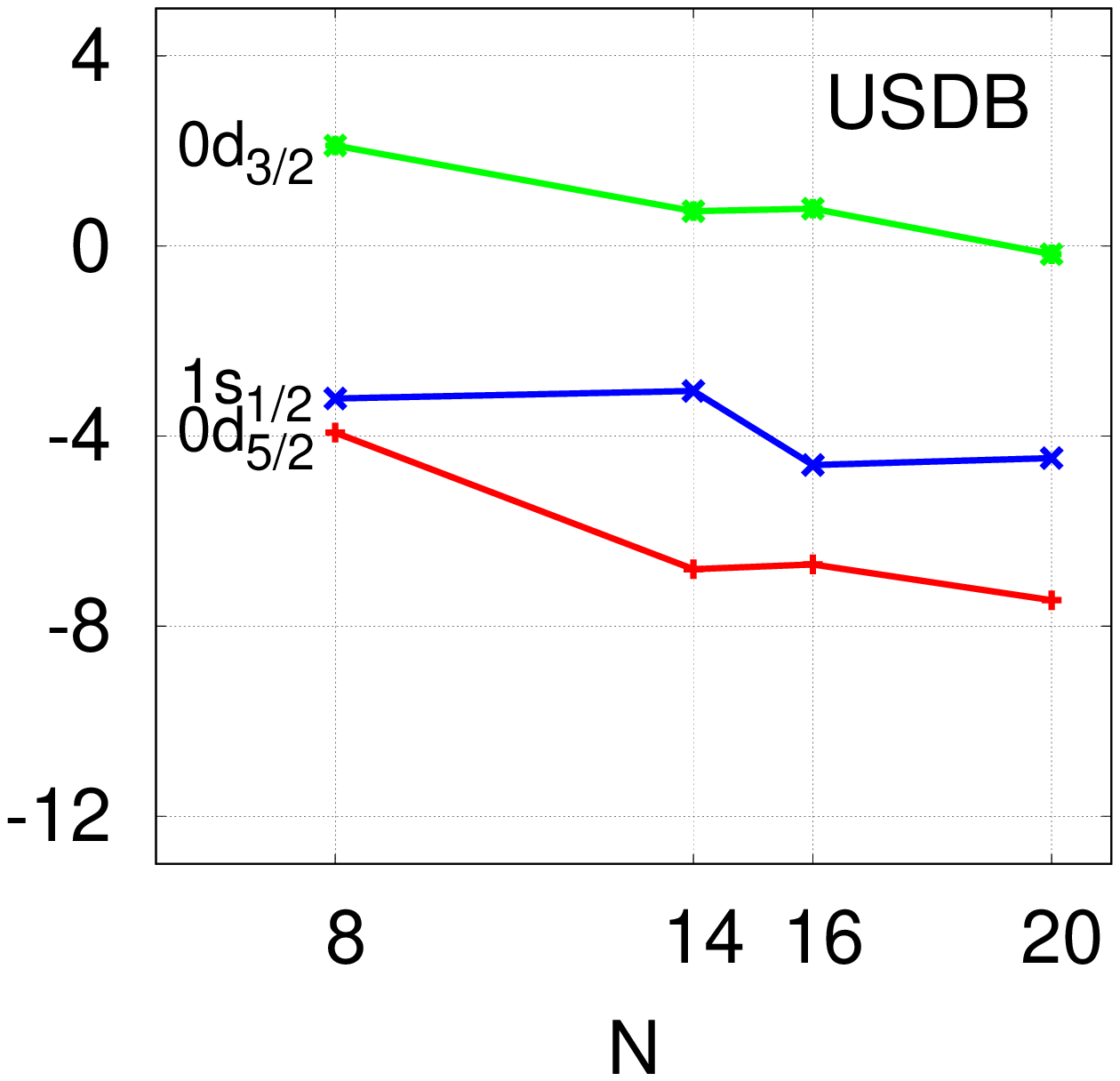}  
  \includegraphics[height=.22\textwidth]{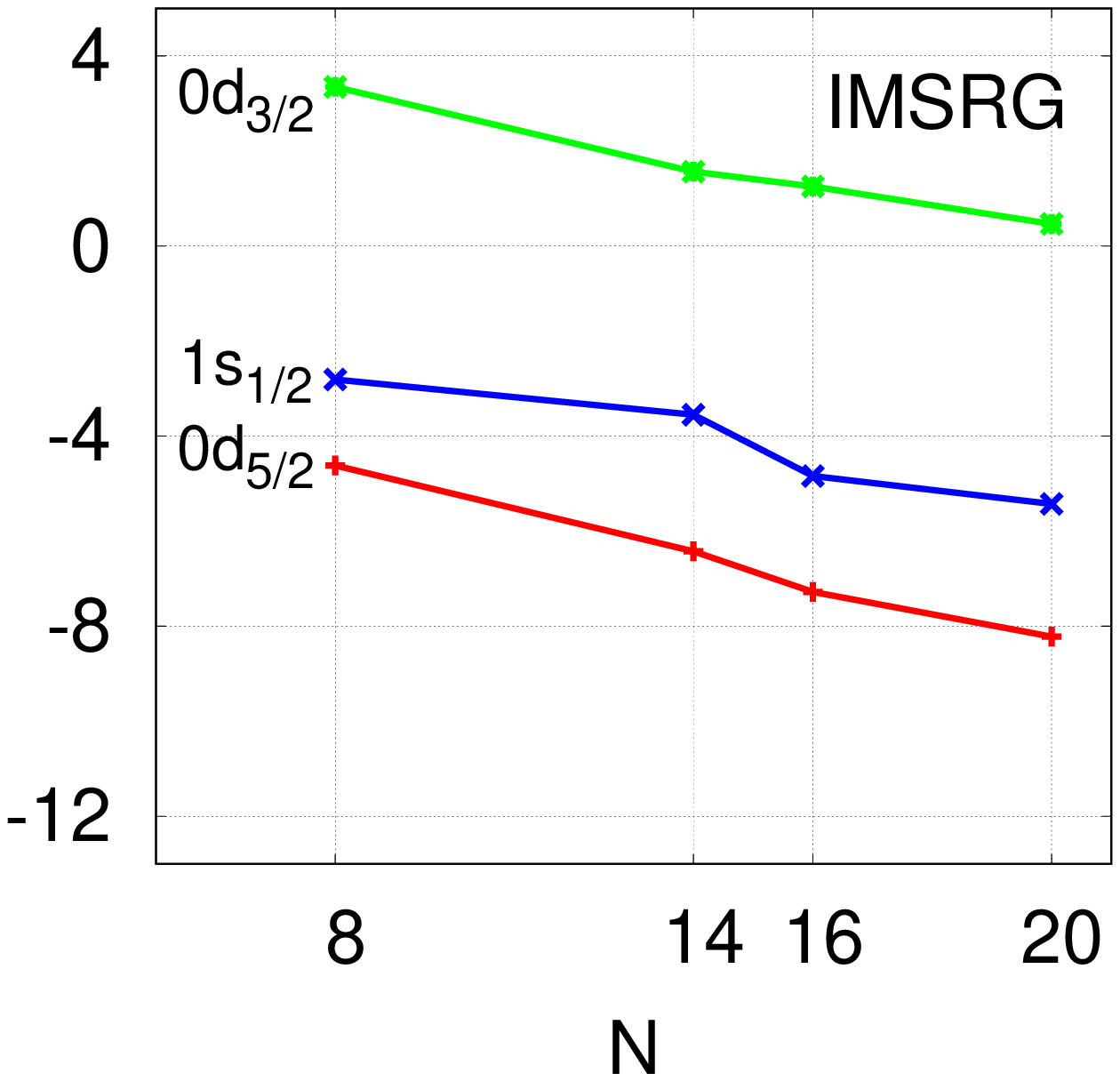} 
  \caption{\label{fig:O-ESPEs} (Color online) Variation of ESPEs in O isotopes with neutron number $N$ calculated 
using the empirical (USDB) and the microscopic effective interactions 
(which employ the scaling with $A$ described in the text).
DJ16A is a monopole-modified version of DJ16 (see text for details).  
The results within IMSRG have been obtained using the original Hamiltonians from Ref.~\protect\cite{StrPRL118}
for O isotopes.
}
\end{figure*}

The monopole part of the Hamiltonian describes the {\it spherical nuclear mean field}, 
which plays a lead role in the filling of orbitals and (sub)shell gaps.
Its single-particle states, or ESPEs, provide an important ingredient for the arrangement
of shells and the interplay between spherical and deformed configurations in nuclei.
The higher multipole part of the interaction provides the so-called {\it correlation energy} 
for particle-hole excitations across the shell gap.
Large shell gaps 
are a prerequisite in order to obtain rigid magic numbers. 
A reduction of the spherical shell gaps may lead to the formation of
a deformed ground state, if the correlation energy of a given excited (intruder) configuration 
is large enough to overcome the naive cost in energy for producing the excited configuration.

It has been recognized long ago~\cite{PoZu81} that the main defect of the traditional microscopic
effective interactions derived from two-body $NN$ potentials is
an unsatisfactory monopole term, resulting in the absence of sufficiently large sub-shell gaps and
providing too much binding. 
This in turn leads to the lack of sphericity in closed sub-shell nuclei and failures in the description
of open-shell nuclei.
Given the importance of the monopole component of an effective interaction, 
we now provide a detailed analysis of the ESPEs of the valence-space interactions under consideration here.

\subsection{\boldmath $T=1$ centroids}

We begin by investigating the properties of the $T=1$ centroids of the microscopic effective interactions
in comparison with those of the phenomenological USDB interaction.
To this end, we show in Fig.~\ref{fig:O-ESPEs} the neutron ESPEs in O isotopes 
as a function of the neutron number.
Since we use only the monopole Hamiltonian, the ESPEs have been evaluated in closed (sub)-shell nuclei
($^{16}$O, $^{22}$O, $^{24}$O and $^{28}$O) assuming a normal filling of the orbitals with the order
determined by single-particle energies with respect to the core nucleus
(a Hartree--Fock approximation).
The ESPEs are thus represented by jagged lines. 
The slope of each segment is given by the corresponding centroids
of the two-body interaction, as seen from Eq.~(\ref{ESPE}).
 
In all cases, except for IMSRG, the starting point is the $A$-independent
single-particle energies from the USDB Hamiltonian, quoted in the previous section.
The TBMEs of USDB and of the microscopic effective interactions are scaled  
as $(A/A_0)^{-0.3}$ with $A_0=18$.  This mass dependence is the one
exploited for USDB (as well as USD and USDA) and we keep it for the microscopic effective interactions
recognizing the empirical need for rescaling and anticipating that future corrections to the theory, 
such as the inclusion of 3-nucleon interactions, should replace this rescaling as in the case of IMSRG.

\begin{figure*}[t!]
  
  \includegraphics[height=.22\textwidth]{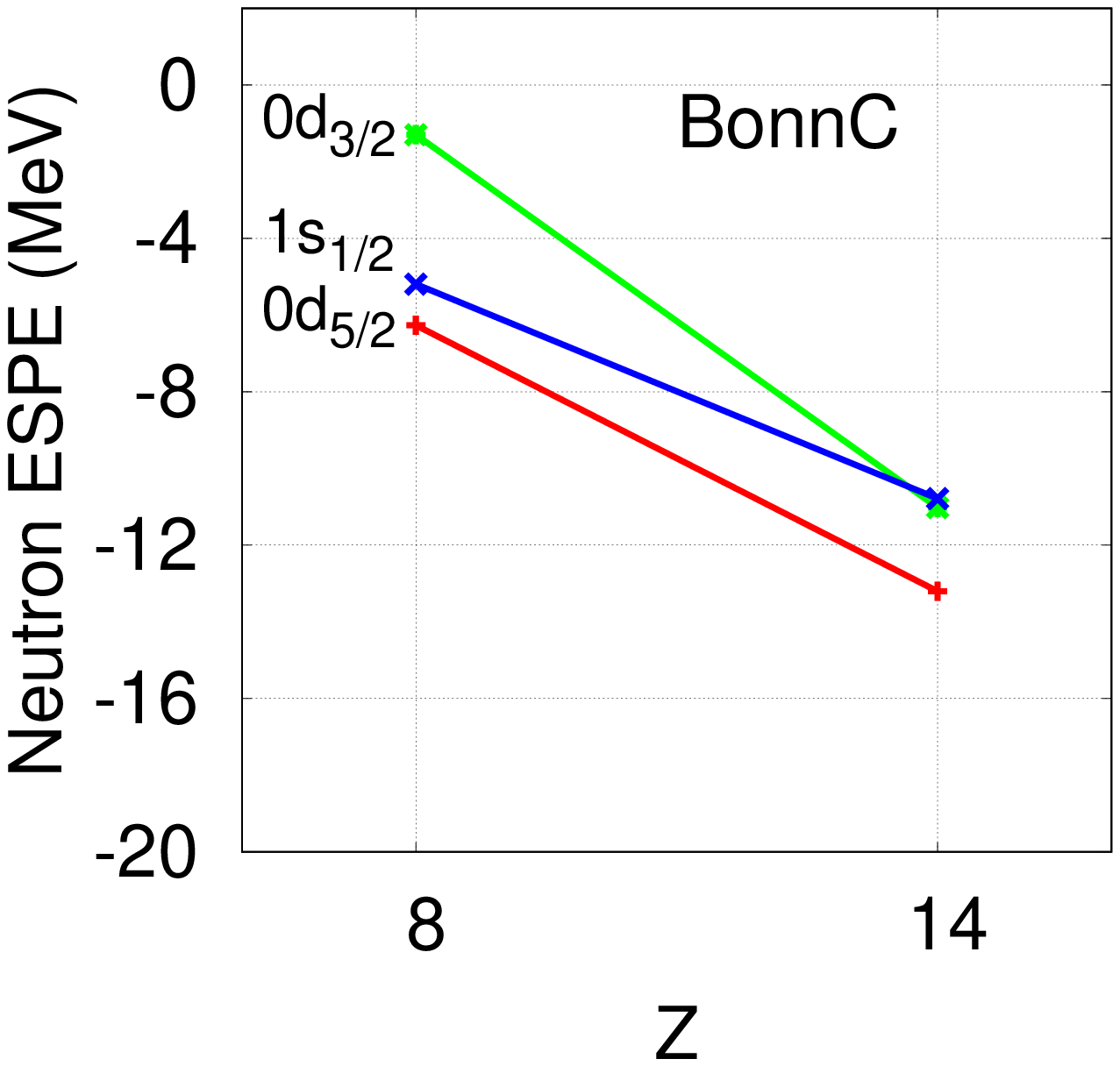}  
  \includegraphics[height=.22\textwidth]{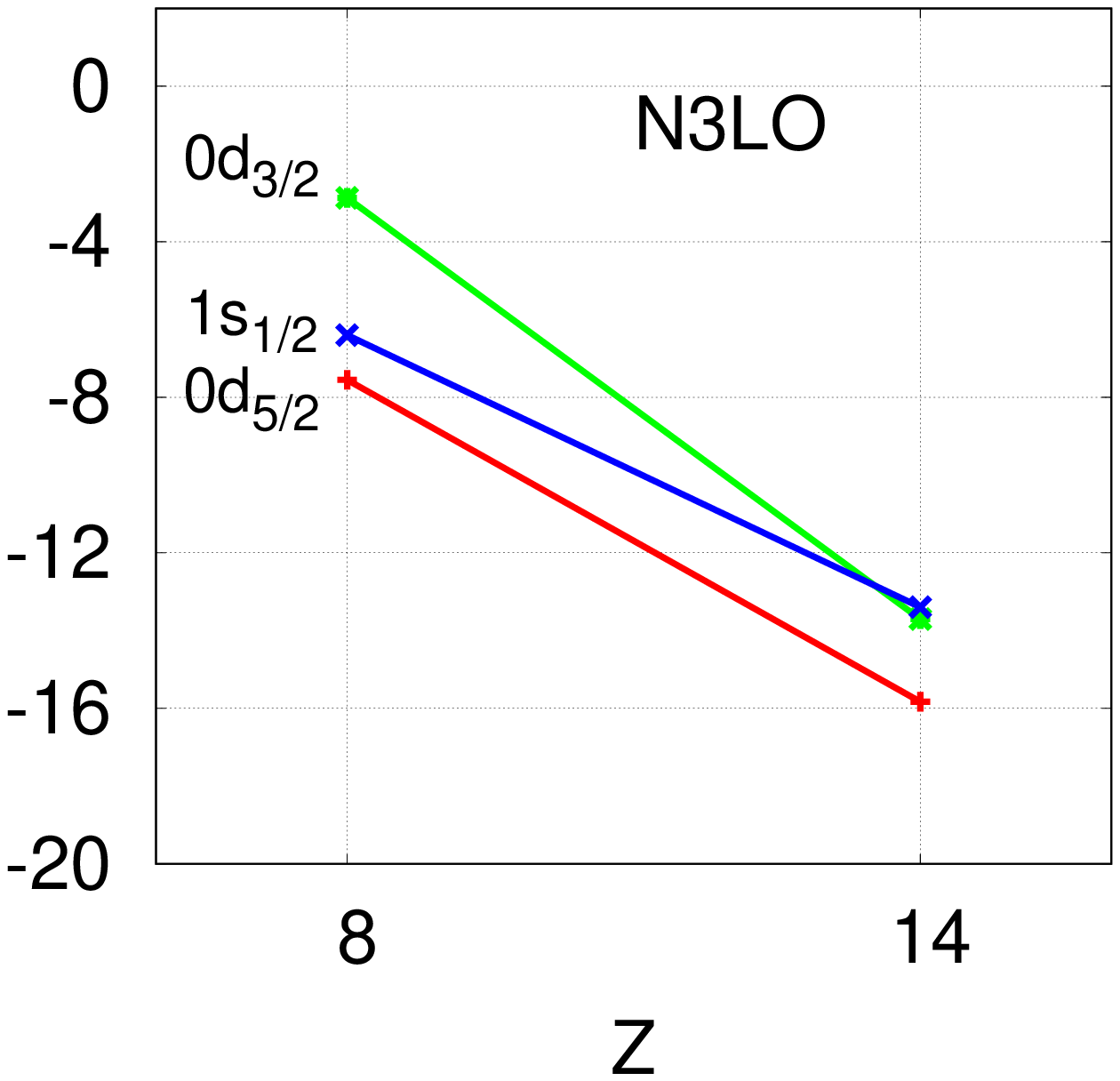} 
  \includegraphics[height=.22\textwidth]{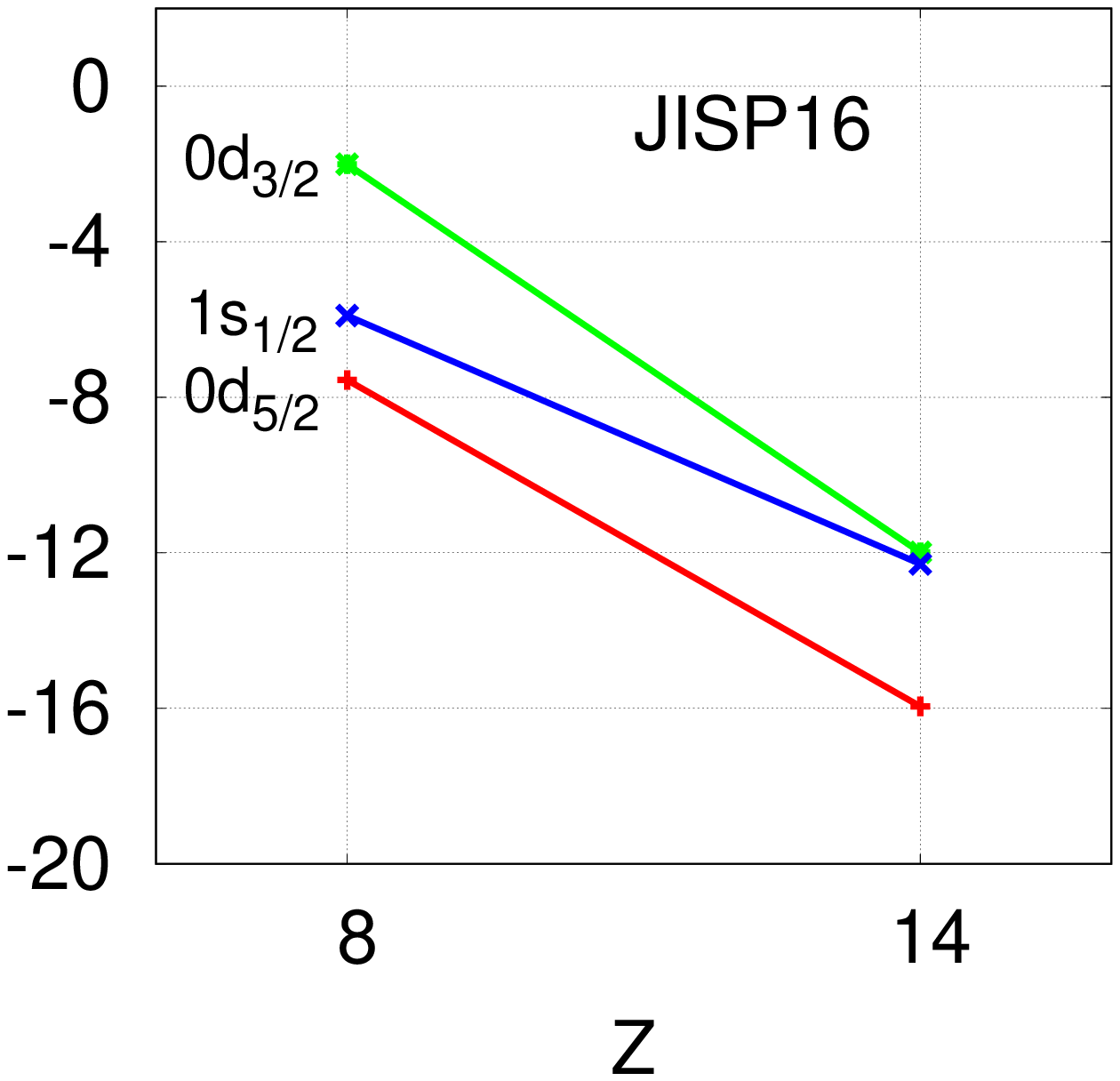} \\[3mm]
  \includegraphics[height=.22\textwidth]{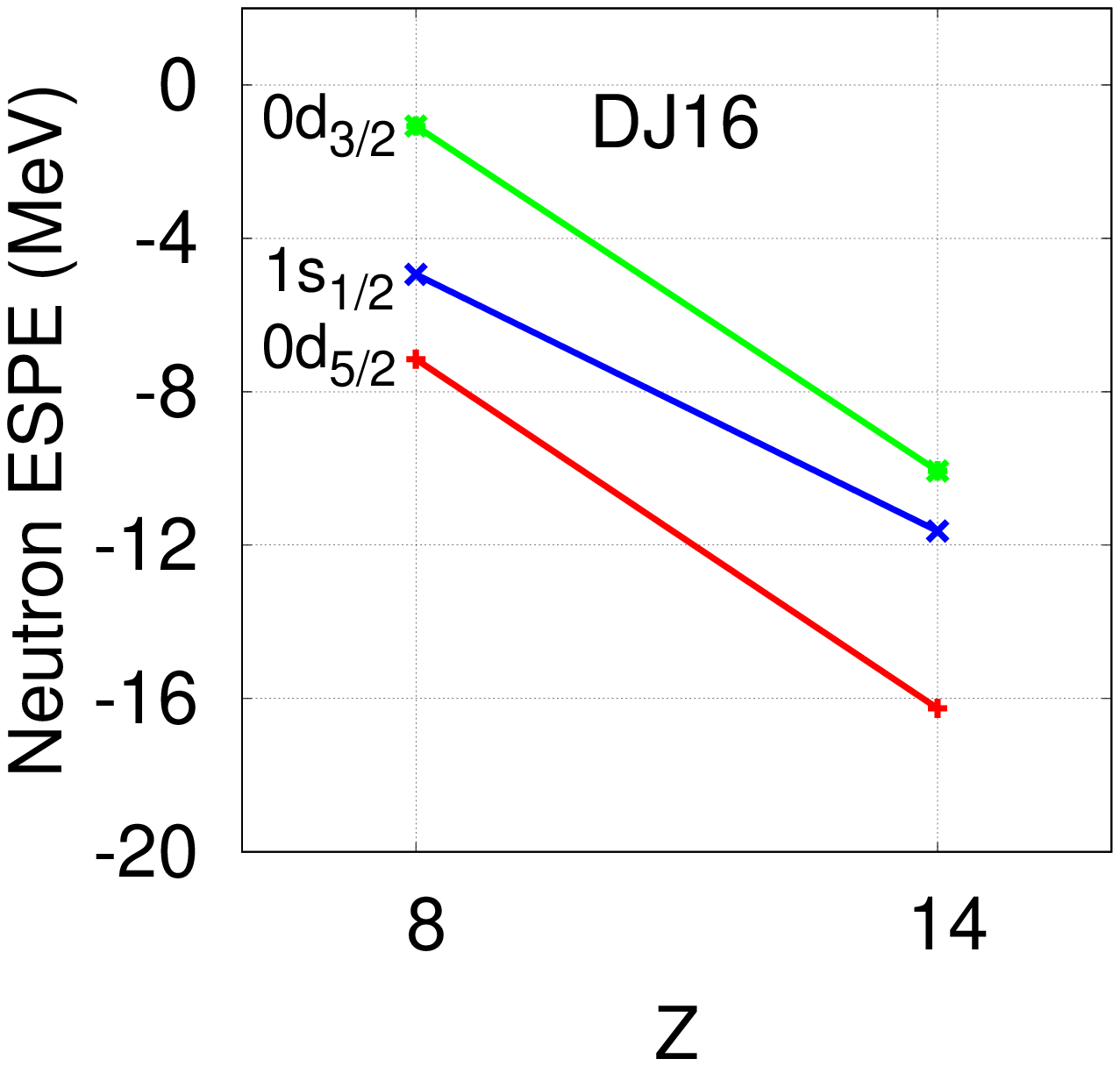} 
 \includegraphics[height=.22\textwidth]{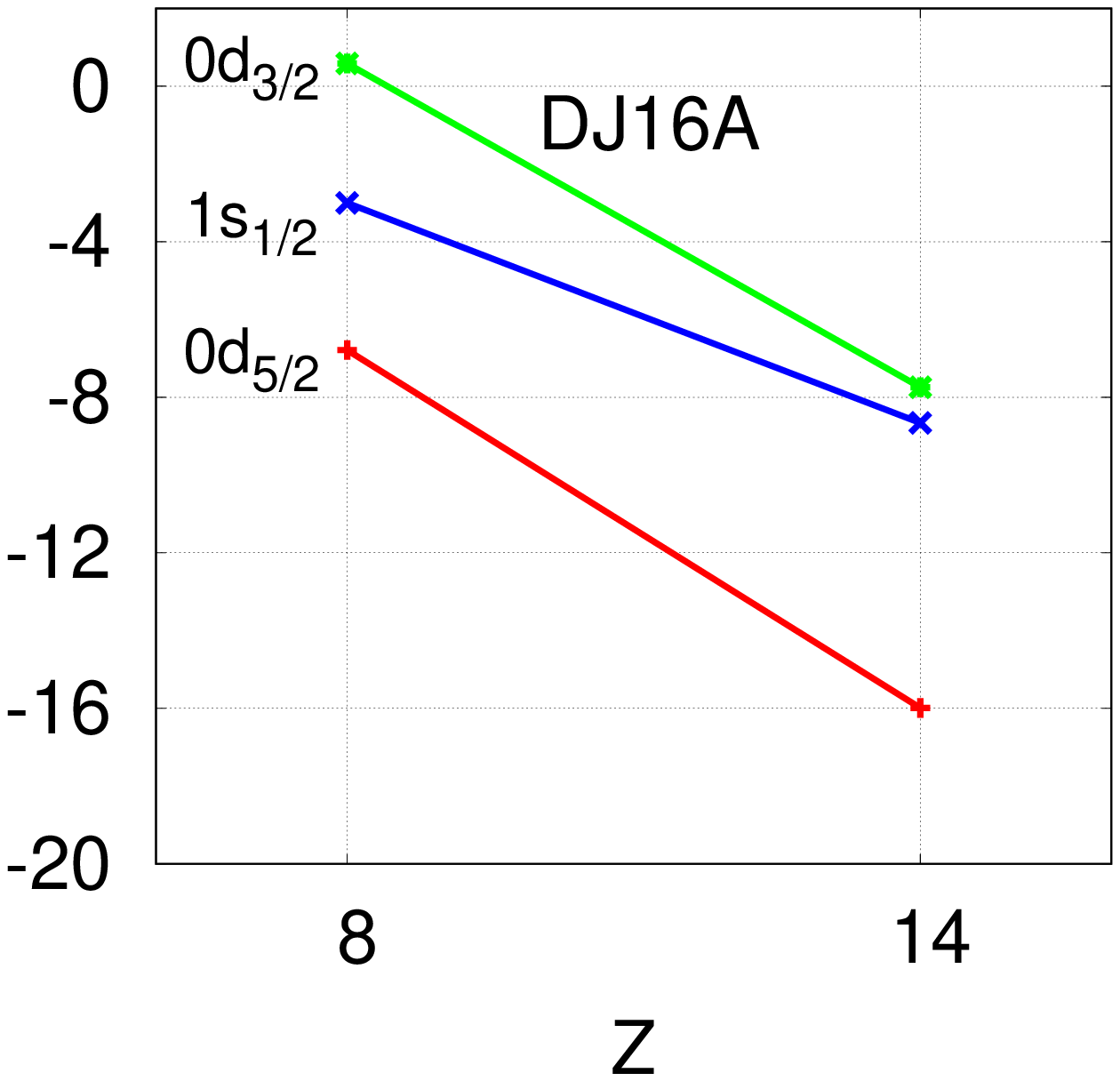}
 \includegraphics[height=.22\textwidth]{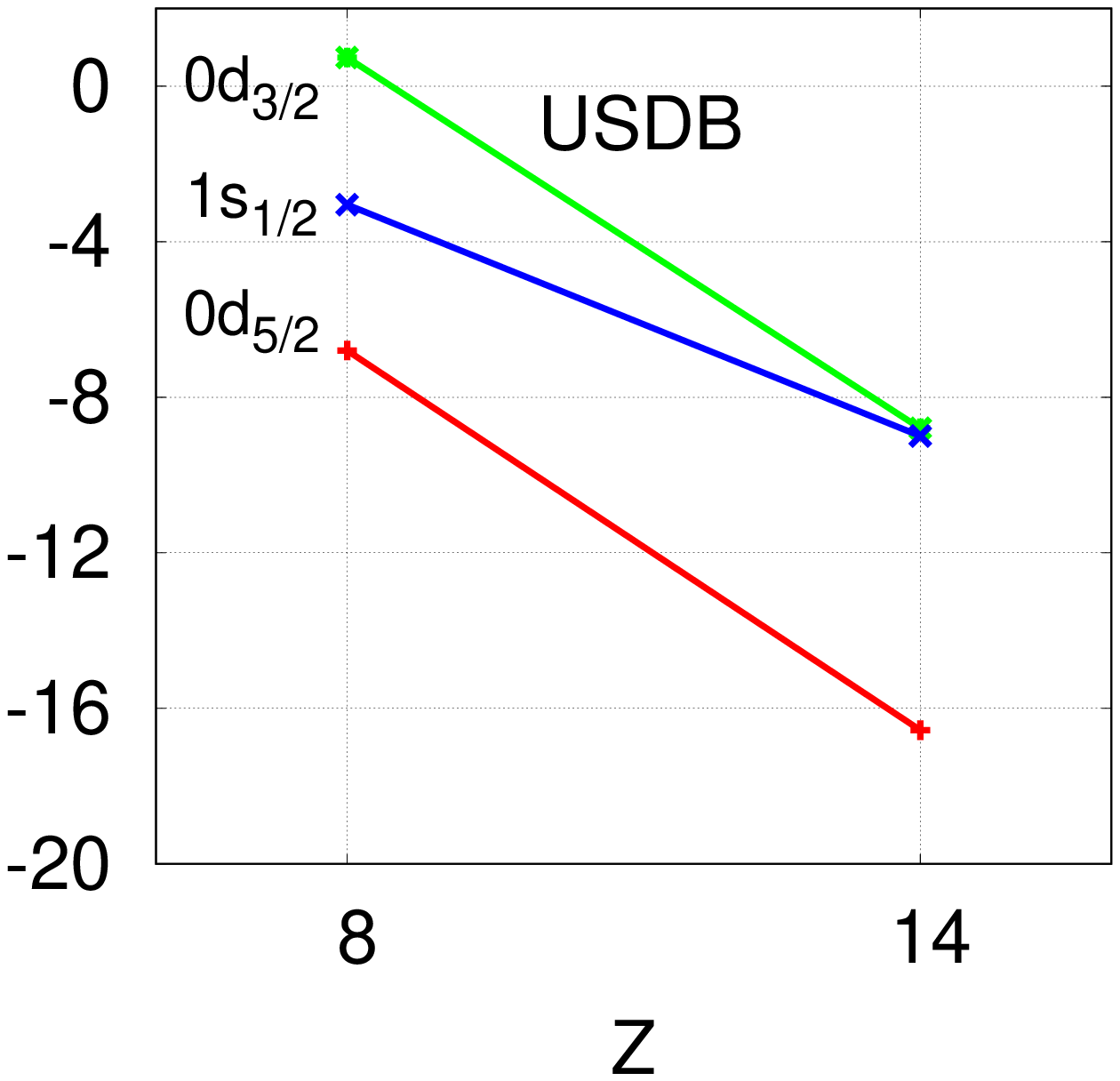}
  \includegraphics[height=.22\textwidth]{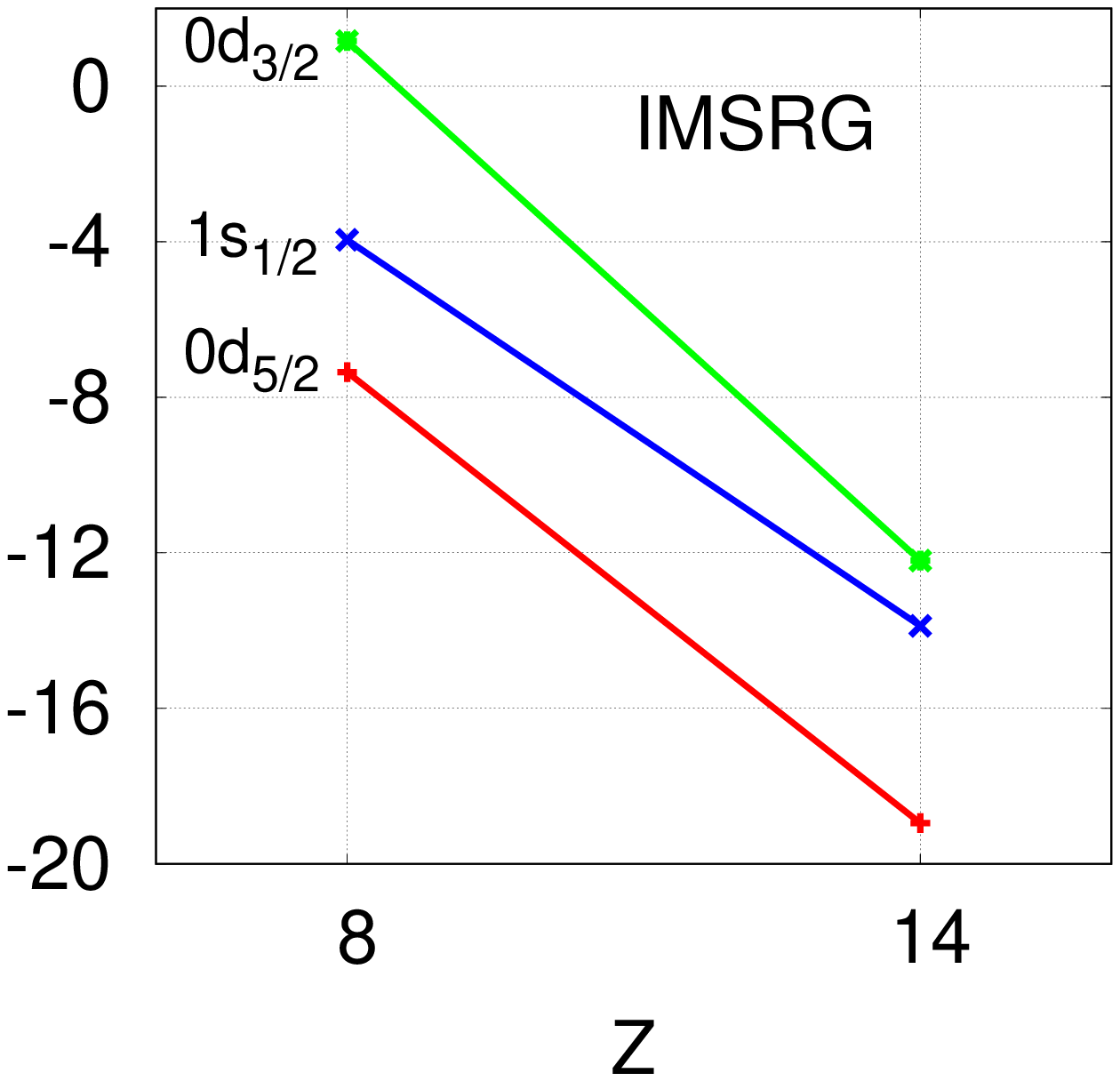}
  \caption{\label{fig:N14-ESPEs} (Color online) Variation of the neutron ESPEs in $N=14$ isotones calculated 
using the phenomenological (USDB) and the microscopic effective interactions.
}
\end{figure*}

While the neutrons fill $0d_{5/2}$, $1s_{1/2}$ and $0d_{3/2}$ orbitals 
(from $^{16}$O to $^{22}$O, then to $^{24}$O and on to $^{28}$O), the ESPEs acquire shifts 
due to additional increments provided by the 
$V^{T=1}_{d_{5/2} d_{5/2}}$, $V^{T=1}_{s_{1/2} d_{5/2}}$ and $V^{T=1}_{d_{5/2} d_{3/2}}$ centroids, respectively.

Two important features of the phenomenological USDB interaction
are easily seen. First, there is the appearance of the $N=14$ sub-shell closure in $^{22}$O. 
Similarly, there is a clear 
$N=16$ shell gap in $^{24}$O, resulting in the corresponding magic structure of that nucleus.

Observing the ESPEs obtained from four different microscopic effective interactions
(BonnC, N3LO, JISP16 and DJ16),
we can easily see that while all of them reproduce the $N=16$ shell gap in $^{24}$O,
the $N=14$ gap between $1s_{1/2}$ and $0d_{5/2}$ in $^{22}$O is rather small.
The worst situation is for the $G$-matrix based effective interaction (from the BonnC potential), 
a bit larger shell gap is produced from N3LO, still bigger with JISP16 and 
the best result is obtained with DJ16.
The numerical values of the $N=14$ shell gap in O isotopes are summarized in Table~\ref{tab:O_shell_gaps}.

\begin{table}[!b]
\centering
\caption{Evolution of the $N=14$ sub-shell gap and neutron $d_{5/2}{-}d_{3/2}$ spin-orbit  splitting in the O isotopes 
from $^{16}$O to $^{28}$O as obtained from the ESPEs of different Hamiltonians.}\vspace{1ex}
\label{tab:O_shell_gaps}
\begin{ruledtabular}
\begin{tabular}{l|r|r|r|r|r|r|r|r} 
& \multicolumn{4}{c|}{Gap $\nu (s_{1/2}{-}d_{5/2})$} & \multicolumn{4}{c}{Gap $\nu (d_{3/2}{-}d_{5/2})$} \\[1mm]
\hline
& \multicolumn{4}{c|}{MeV} & \multicolumn{4}{c}{MeV} \\[1mm]
\hline
 & $^{16}$O & $^{22}$O & $^{24}$O & $^{28}$O &  $^{16}$O & $^{22}$O & $^{24}$O & $^{28}$O \\[1mm]  
\hline
BonnC       & 0.72 & 1.07 & $-0.11$ & 1.94 & 6.04 & 4.97 & 5.58  & 7.40 \\
N3LO        & 0.72 & 1.15 &  0.21 & 1.93 & 6.04 & 4.68 & 5.01  & 7.24 \\
JISP16      & 0.72 & 1.66 &  0.48 & 1.84 & 6.04 & 5.55 & 5.80  & 7.44 \\
DJ16        & 0.72 & 2.21 &  0.88 & 1.98 & 6.04 & 6.08 & 6.18  & 7.11 \\
DJ16A       & 0.72 & 3.78 &  1.80 & 2.51 & 6.04 & 7.37 & 7.15  & 7.02 \\
USDB        & 0.72 & 3.75 &  2.09 & 2.99 & 6.04 & 7.53 & 7.49  & 7.28 \\
IMSRG       & 1.81 & 3.40 &  2.44 & 2.79 & 7.97 & 8.52 & 8.04  & 8.68 \\
\end{tabular}
\end{ruledtabular}
\end{table}

\begin{table*}[t!]
\centering
\caption{Evolution of the neutron sub-shell shell gaps in $N=14$ isotones from $^{22}$O to $^{28}$Si as obtained from
the ESPEs of different Hamiltonians.}
\label{tab:N14_shell_gaps}\vspace{1ex}
\begin{ruledtabular}
\begin{tabular}{l|r|r|r|r|r|r|r|r|r} 
& \multicolumn{3}{c|}{Gap $\nu (s_{1/2}{-}d_{5/2})$} & \multicolumn{3}{c|}{Gap $\nu (d_{3/2}{-}d_{5/2})$} 
& \multicolumn{3}{c}{Gap $\nu (d_{3/2}{-}s_{1/2})$}  \\
\hline
& \multicolumn{3}{c|}{MeV} & \multicolumn{3}{c|}{MeV} & \multicolumn{3}{c}{MeV} \\[1mm]
\hline
& $^{22}$O & $^{28}$Si & Diff & $^{22}$O &  $^{28}$Si & Diff & $^{22}$O & $^{28}$Si & Diff \\[1mm]  
\hline
BonnC       & 1.07 & 2.42 & {\bf 1.35} & 4.97 & 2.18 & $\mathbf{-2.79}$ & 3.90 & $-0.23$ & $\mathbf{-4.13}$ \\
N3LO        & 1.15 & 2.43 & {\bf 1.28} & 4.68 & 2.13 & $\mathbf{-2.55}$ & 3.53 & $-0.31$ & $\mathbf{-3.84}$ \\
JISP16      & 1.66 & 3.66 & {\bf 2.00} & 5.55 & 3.96 & $\mathbf{-1.59}$ & 3.89 & 0.30    & $\mathbf{-3.59}$ \\
DJ16        & 2.21 & 4.62 & {\bf 2.41} & 6.08 & 6.19 & {\bf 0.11}       & 3.87 & 1.57    & $\mathbf{-2.30}$ \\
DJ16A       & 3.78 & 7.33 & {\bf 3.55} & 7.37 & 8.25 & {\bf 0.88}       & 3.59 & 0.92    & $\mathbf{-2.67}$ \\
USDB        & 3.75 & 7.57 & {\bf 3.82} & 7.53 & 7.77 & {\bf 0.23}       & 3.78 & 0.20    & $\mathbf{-3.58}$  \\
IMSRG       & 3.40 & 5.07 & {\bf 1.67} & 8.52 & 6.75 & $\mathbf{-1.76}$ & 5.11 & 1.68    & $\mathbf{-3.43}$ \\
\end{tabular}
\end{ruledtabular}
\end{table*}

The evolution of the $N=14$ shell gap is governed by the difference between
$V^{T=1}_{d_{5/2} d_{5/2}}$ and $V^{T=1}_{d_{5/2} s_{1/2}}$ centroids of the TBMEs.
The detailed spin-tensor structure of those centroids will be discussed below.

We also note that the spin-orbit splitting between $0d_{3/2}$ and $0d_{5/2}$ ESPEs
is relatively well reproduced by all microscopic interactions only in $^{28}$O (see also Table~\ref{tab:O_shell_gaps}).
At $N=14$ and $N=16$, the spin-orbit splitting is about $1.5{-}2$~MeV smaller than what is provided by USDB.
The spin-orbit splitting closest to USDB is that from DJ16. 

In Ref.~\cite{Otsuka3N}, the behavior of the ESPEs obtained from microscopic effective interactions,
based on a $NN$ potential, was ascribed to the missing $3N$ forces (see Fig.~2 of that reference).
Indeed, the ESPEs from the IMSRG Hamiltonian~\cite{StrPRL118}, 
derived from the chiral $NN$ plus $3N$ interaction, better reproduce 
those from USDB than our  microscopic effective interactions which all omit 3N interactions.
The variations in the $N=14$ shell gap are clearly seen for $^{22}$O in Fig.~\ref{fig:O-ESPEs}
which lead to significant differences in spectroscopic properties of this nucleus and its neighbors 
as will be discussed below.

It can also be noticed from Fig.~\ref{fig:O-ESPEs} that the slopes of the ESPEs
obtained with microscopic interactions are on average steeper 
than those obtained from USDB. These very attractive centroids 
will manifest themselves in the overbinding of O isotopes, as we will also see below.

\subsection{Proton-neutron centroids}

To analyze the proton-neutron centroids, we consider evolution of the neutron ESPEs in
a series of isotones (as protons fill the $d_{5/2}$ orbital).
The neutron ESPEs in $N=14$ isotones from $^{22}$O to $^{28}$Si are shown in Fig.~\ref{fig:N14-ESPEs}.
The starting point of these calculations are ESPEs in $^{22}$O as obtained by different effective interactions
and shown in Fig.~\ref{fig:O-ESPEs}.
The numerical values of the $N=14$ shell gaps from the monopole part of the interactions
are summarized in Table~\ref{tab:N14_shell_gaps}.
Already in $^{22}$O the $N=14$ gaps given by the microscopic interactions are not large enough.
In addition, there is also a significant increase of the $N=14$ shell gap from $^{22}$O towards $^{28}$Si, 
produced by the USDB interaction (by 3.82~MeV).
This feature is not reproduced by the proton-neutron centroids of the microscopic effective interactions, 
which show a much weaker increase of the $N=14$ shell gap. 
We note, however, a nice trend from BonnC and N3LO results (1.35~MeV and 1.28~MeV, respectively), 
to JISP16 (2 MeV) and DJ16 results (2.41~MeV).
This means that the difference between the corresponding centroids, 
$ V^{pn}_{d_{5/2} d_{5/2}}$ and $V^{pn}_{s_{1/2} d_{5/2}}$, becomes closer to the USDB 
value when using interactions tuned to light nuclei with the intention of minimizing effects of 
neglected $3N$ interactions.

From Table~\ref{tab:N14_shell_gaps} it is seen that the spin-orbit splitting between neutron 
$0d_{3/2}$ and $0d_{5/2}$ states in $^{28}$Si stays about the same as in $^{22}$O only for USDB and DJ16.
Other interactions produce a reduction of the spin-orbit splitting.

We remark that the IMSRG ESPEs exhibit trends different from those of the USDB ESPEs.
As in the case of the other microscopic effective interactions considered here,
the $N=14$ sub-shell gap between O and Si is not as large as the guidance from USDB,
which may manifest itself in certain deficiencies in the description of spectra in the Na--Mg region.
The overly attractive proton-neutron centroids of IMSRG will result in the overbinding of those nuclei as well.
This suggests that the inclusion of $3N$ forces by IMSRG improves its $T=1$ monopoles (i.\:e., makes them more repulsive,
creating necessary shell gaps). This comes out not to be sufficient yet to improve the proton-neutron monopoles.

\subsection{Monopole modifications to DJ16 TBMEs\label{monopole-mod-DJ16A}}

With further guidance from USDB,
we observe that modifications of the microscopic interactions are needed to improve agreement with experimental data.
We will consider modifications to DJ16, since there are indications 
from the results presented above that those modifications are less severe than for other interactions.
We have already noticed that the $T=1$ centroids of DJ16 are significantly different from the USDB centroids, 
while the proton-neutron centroids of the two interactions are comparatively similar. 
In the present study 
we therefore propose modifying mainly the $T=1$ monopole term of DJ16 in order to
see the effect on the spectroscopy.
Guided by the USDB ESPEs  from Fig.~\ref{fig:O-ESPEs}, we have added to the original DJ16 centroids:
80~keV to $V^{T=1}_{d_{5/2} d_{5/2}}$, 
350~keV to  $V^{T=1}_{d_{5/2} s_{1/2}}$, 300~keV to $V^{T=1}_{d_{5/2} d_{3/2}}$ and 200~keV to 
$V^{T=1}_{d_{3/2} s_{1/2}}$.
We have also made the $T=0$ centroid $V^{T=0}_{d_{5/2} d_{5/2}}$ 80 keV more attractive to compensate for
the repulsion in the $T=1$ centroid (the corresponding proton-neutron centroid is thus unchanged),
and 100~keV to $V^{T=0}_{d_{5/2} s_{1/2}}$.
These additions are equally distributed among the TBMEs of different $J$. 
The resulting neutron ESPEs are shown in Figs.~\ref{fig:O-ESPEs}--\ref{fig:N14-ESPEs} and labeled as DJ16A. 
We can observe
that the spherical mean-field from DJ16A is rather close to that provided by USDB as intended with these modifications.
Therefore, the differences in the spectroscopy will be mainly related to differences in the other 
multipole terms of the effective interaction.

\section{O isotopes}

\begin{figure}[!t]
  \includegraphics[height=.23\textheight]{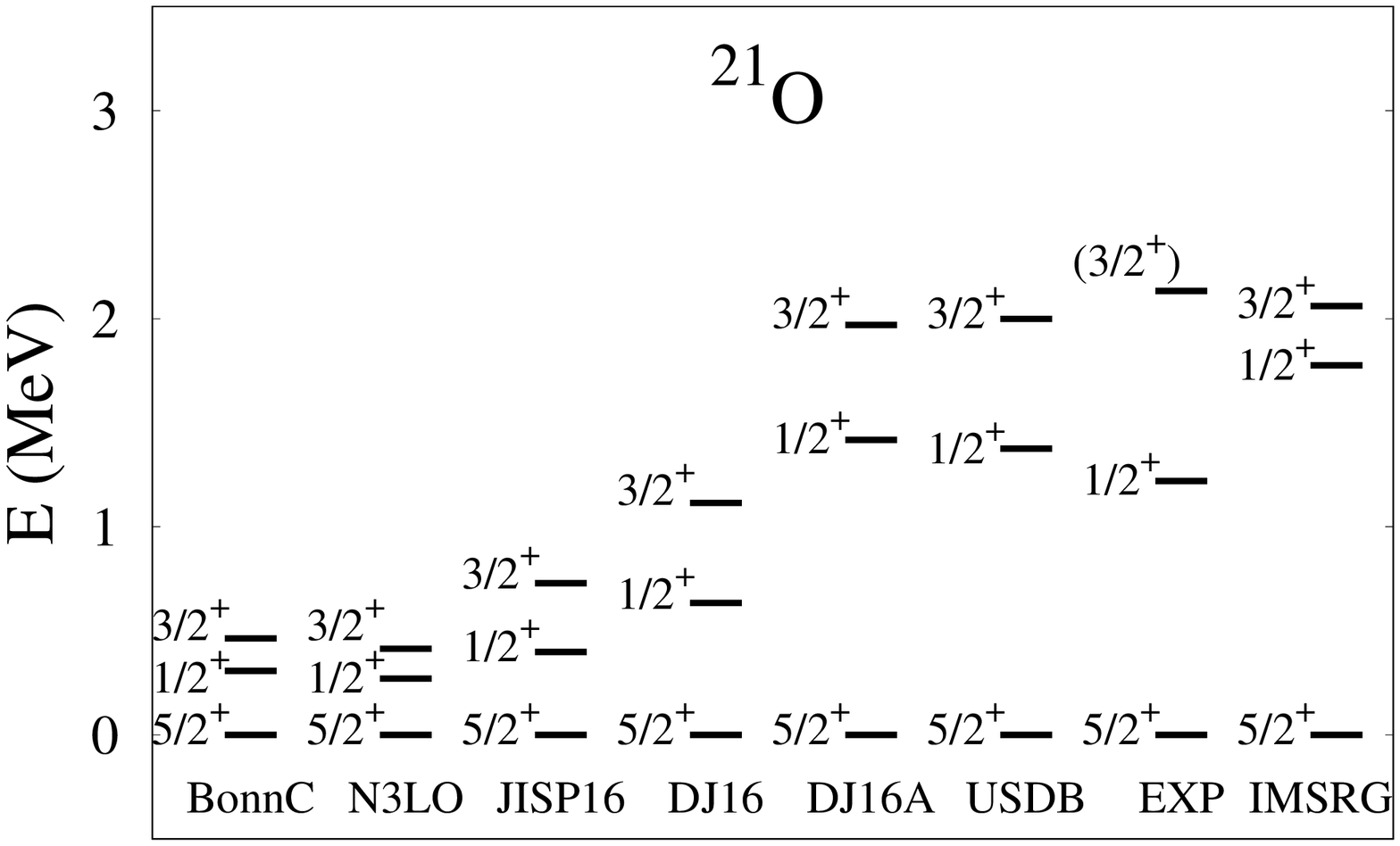} 
 \includegraphics[height=.23\textheight]{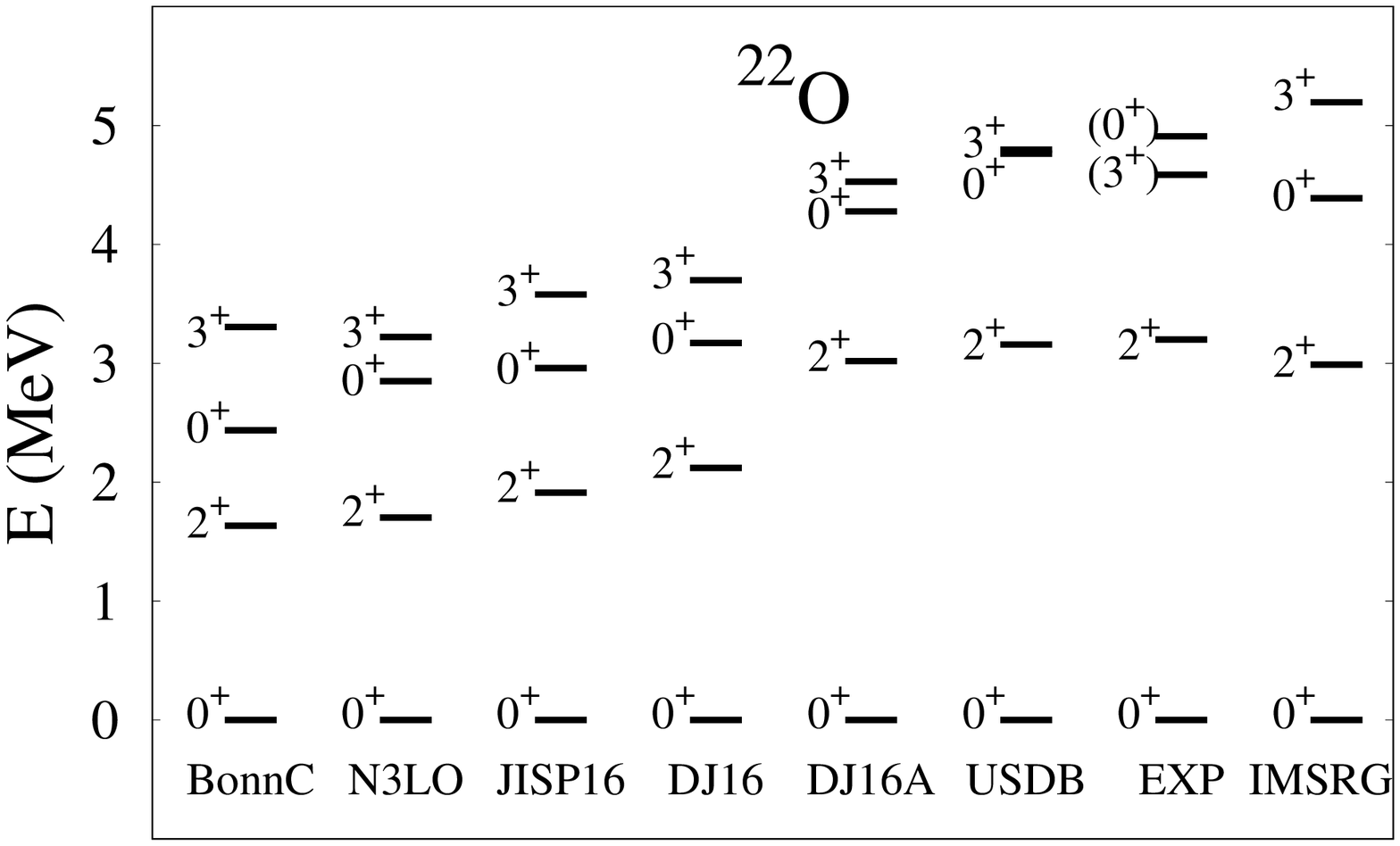} \\
  \includegraphics[height=.23\textheight]{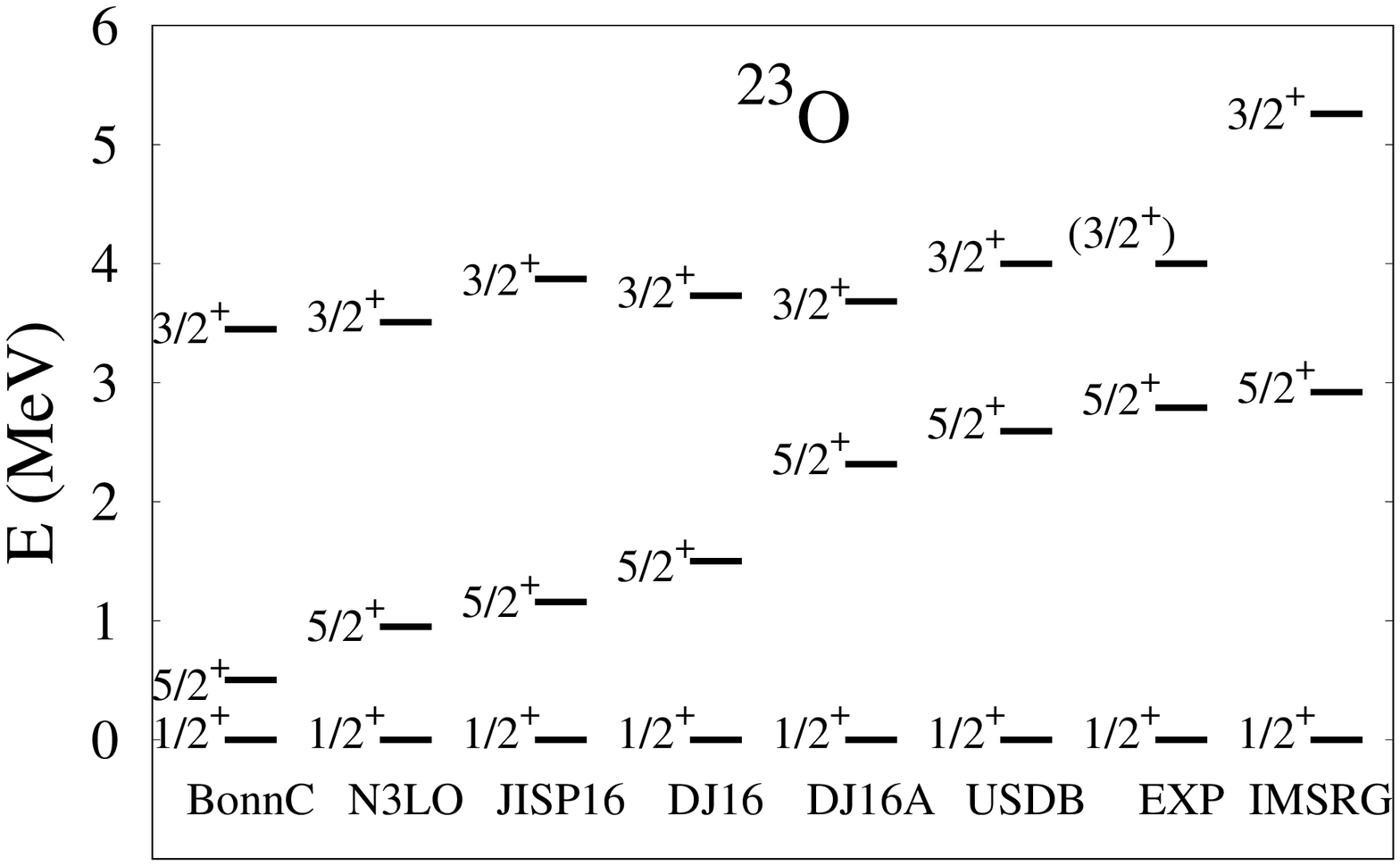} 
  \includegraphics[height=.23\textheight]{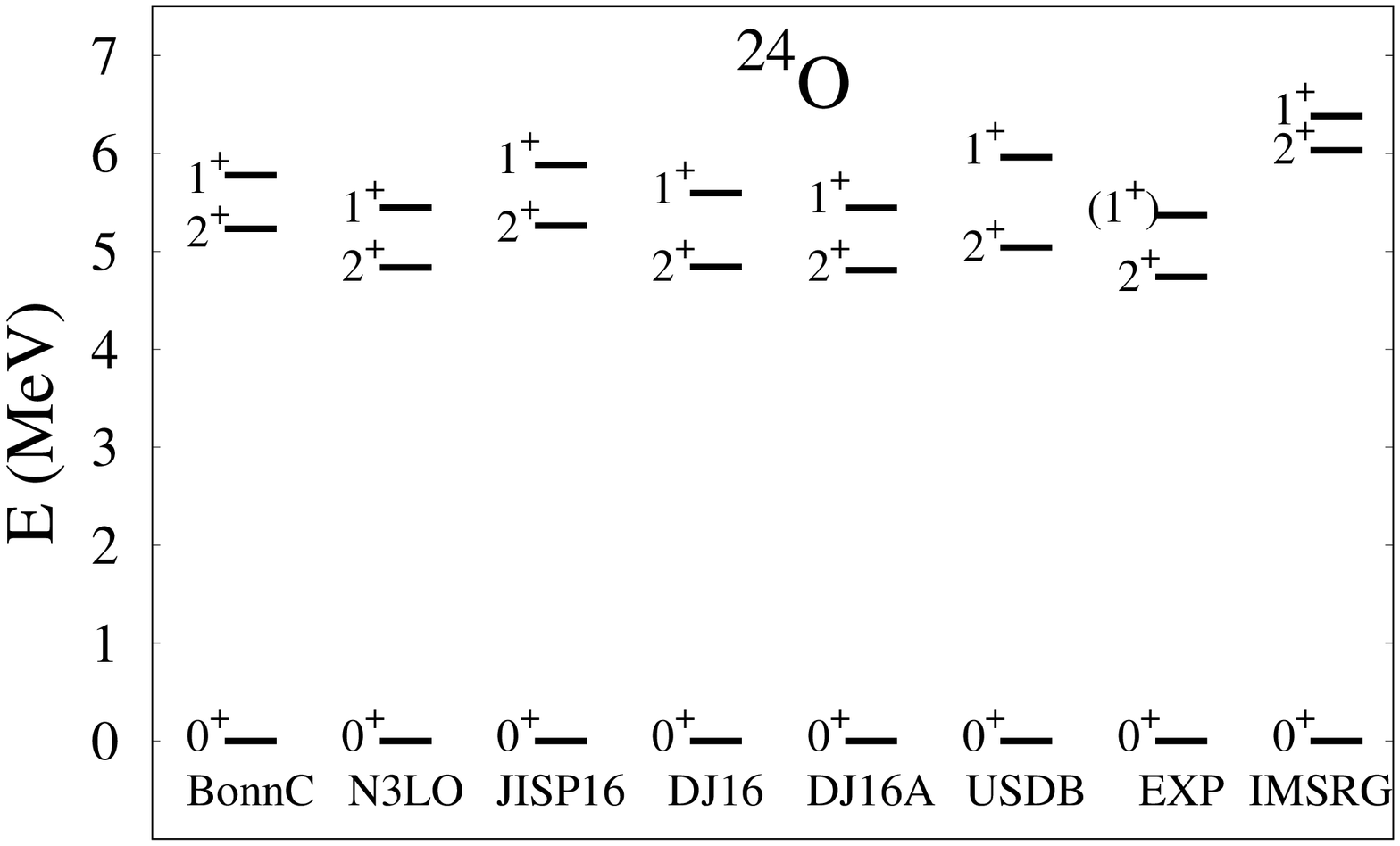} 
  \caption{\label{fig:O-spectra} Experimental low-energy spectrum of $^{21-24}$O in comparison with theoretical results, 
obtained from USDB and from the microscopic effective interactions.
The experimental data are from Ref.~\protect\cite{nndc}.}
\vspace{-8ex}
\end{figure}
\begin{figure}[!t]
  \includegraphics[width=\columnwidth]{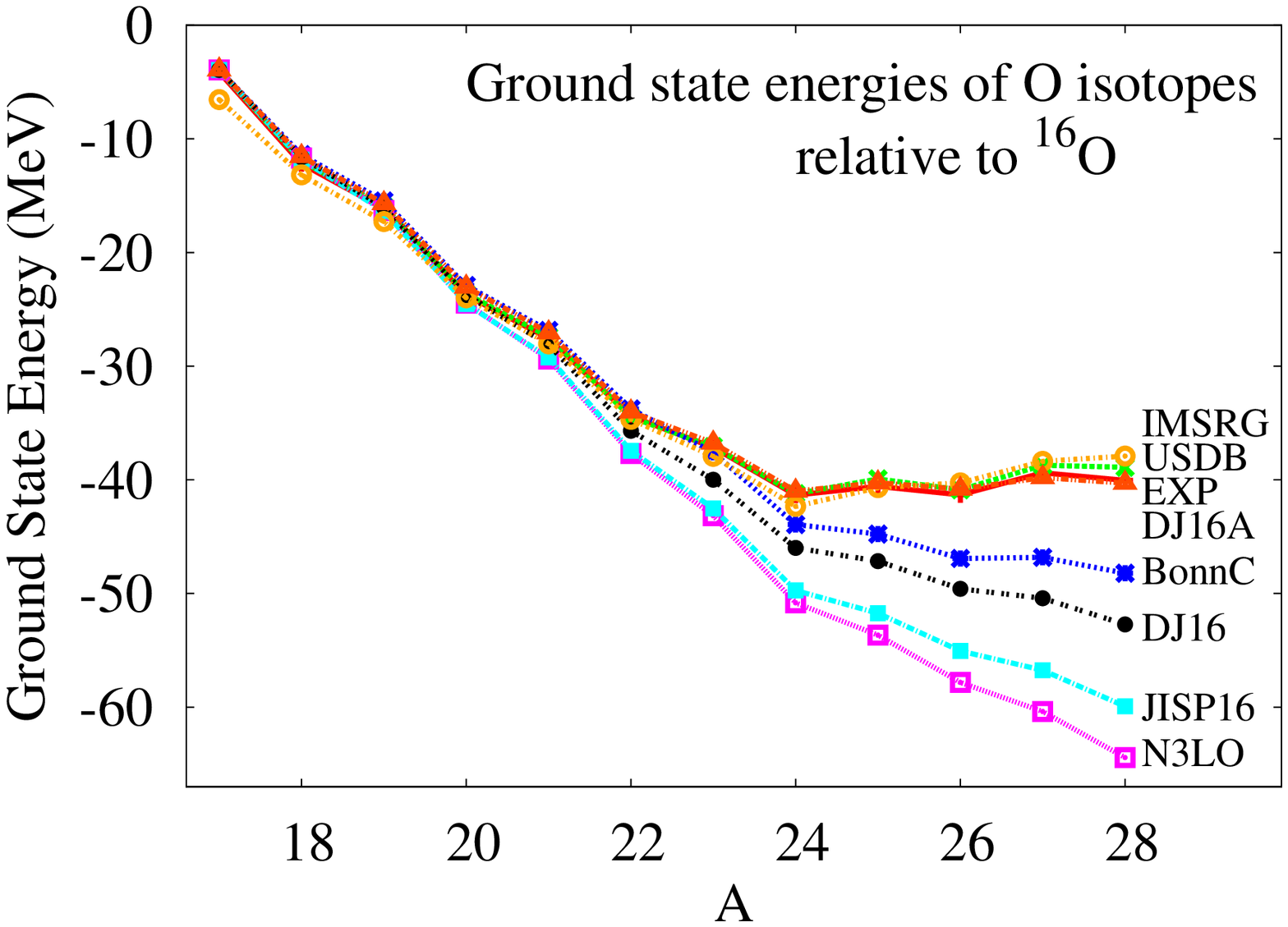} 
  \caption{\label{fig:BE_O} (Color online) Experimental ground state energies of O isotopes relative to the ground state energy of 
$^{16}$O, in comparison with theoretical results, obtained from USDB and from the microscopic effective interactions.
The experimental data (extrapolations included) are from AME2012~\protect\cite{AME2012}.}
\end{figure}

%

To demonstrate the importance of the neutron $N=14$ and $N=16$ shell gaps, we present the spectra of the $^{21-24}$O 
isotopes obtained from shell-model diagonalization  (Fig.~\ref{fig:O-spectra}).

Let us first discuss the spectrum of $^{22}$O. 
A large ${N=14}$ shell gap produced by USDB ensures the \mbox{properties} of a ``doubly-magic'' nucleus 
with a sufficiently high-lying $2^+_1$ state.
It is obvious that the microscopic interactions cannot produce a sufficiently high first $2^+$ state
because of the small $N=14$ shell gap. As the size of the $N=14$ shell gap progressively increases from BonnC to DJ16  
(Fig.~\ref{fig:O-ESPEs} and Table~\ref{tab:O_shell_gaps}),
we see that the excitation energy of the first $2^+$ state increases accordingly.
A similar correlation of the $N=14$ shell gap with the position of the lowest  $1/2^+$, $3/2^+$ states 
in the spectrum of $^{21}$O and of the lowest  $5/2^+$ state 
in the spectrum of $^{23}$O can be observed (Fig.~\ref{fig:O-spectra}). 
These nuclei have one neutron hole or one particle beyond a doubly-magic $^{22}$O.
Due to the small $N=14$ shell gap, those excited states appear low in the spectra, 
especially for the BonnC and N3LO interactions.  
They also move generally upwards in the spectra, 
as the interactions are changed from BonnC or N3LO to JISP16 and on to DJ16.

For comparison, we show also the IMSRG results obtained with the Hamiltonians from Refs.~\cite{Stroberg2016,StrPRL118}.
The corresponding spectra are in good agreement with experiment due, in large measure, 
to the satisfactory monopole component of the interaction. 
With IMSRG, a few low-lying states are positioned slightly higher in energy than their experimental counterparts.

At the same time, all interactions agree in the existence of the $N=16$ gap. This in turn
results in good agreement of the lowest states of $^{24}$O with experiment, 
as seen from Fig.~\ref{fig:O-spectra}. 

The ground state energies of O isotopes relative to the ground energy of $^{16}$O are shown in Fig.~\ref{fig:BE_O}.
Here, one observes that while the USDB Hamiltonian locates the neutron drip-line at $^{24}$O
in agreement with the experimental data, the microscopic interactions BonnC, N3LO, JISP16, and DJ16 
extend it at least to $^{28}$O, overbinding the neutron-rich O isotopes. 
This extra binding is related to the \mbox{absolute} values of the corresponding centroids 
(or slopes of the ESPEs presented in Fig.~\ref{fig:O-ESPEs}).
To further elucidate the deficiencies of the microscopic effective interactions, we will present 
a spin-tensor analysis below.

Let us remark that if we remove the empirical mass dependence of TBMEs and keep them constant for all $A$,
the spectra of the O isotopes are nearly unchanged, while the ground state energies become even more negative,
leading to even more overbinding of neutron-rich isotopes.

\begin{table*}[!t]
\centering
\caption{Root-mean-square deviations (in keV) between experimental and theoretical binding energies of O isotopes 
and between experimental and theoretical excitation energies of low-lying states of a few $sd$-shell nuclei
shown in Figs.~\ref{fig:O-spectra}--\ref{fig:Mg24} as obtained from different interactions.}
\label{tab:rms}\vspace{1ex}
\begin{ruledtabular}
\begin{tabular}{l|r|r|r|r|r|r} 
Interaction & BE(O) & $^{21-24}$O & $^{19,21,23,25,27}$F and $^{39}$K & $^{22}$Na & $^{28}$Si,$^{32}$S  & $^{24}$Mg \\
\hline
BonnC       &  3882 & 1460 & 1019 &  925  & 1186 & 1116 \\
N3LO        & 11621 & 1316 & 1069 & 1043  & 1331 & 1275 \\
JISP16      &  9673 & 1151 &  925 &  736  &  993 &  939 \\
DJ16        &  5960 &  931 &  700 &  540  & 1146 & 1096 \\
DJ16A       &   449 &  274 &  285 &  389  &  891 &  806 \\
USDB        &   467 &  251 &  437 &  169  &  234 &  313 \\
IMSRG       &  1177 &  738 &      &  413  & 1497 &  \\
\end{tabular}
\end{ruledtabular}
\end{table*}

\begin{figure*}[t!]
  \includegraphics[height=.23\textheight]{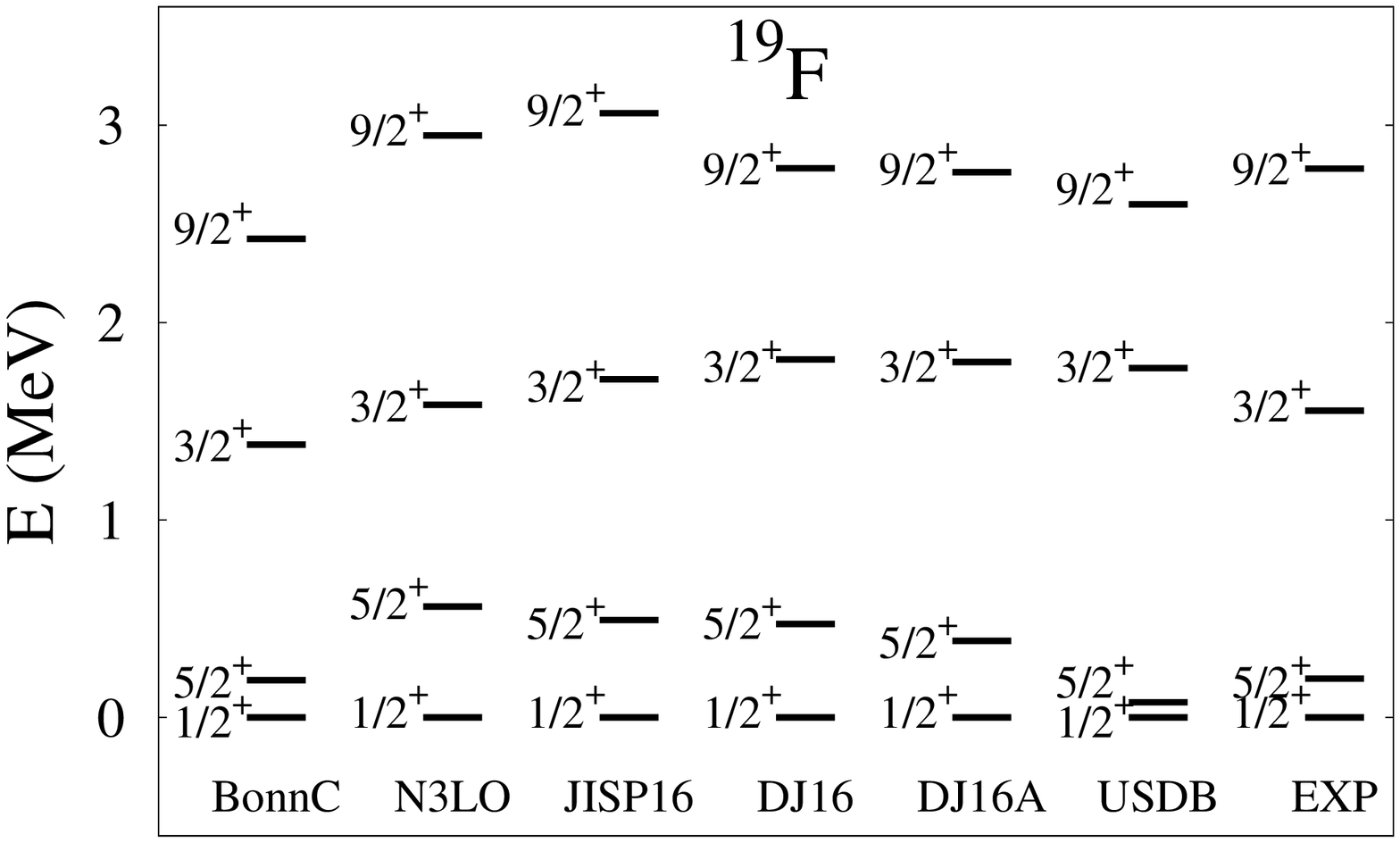}
  \includegraphics[height=.23\textheight]{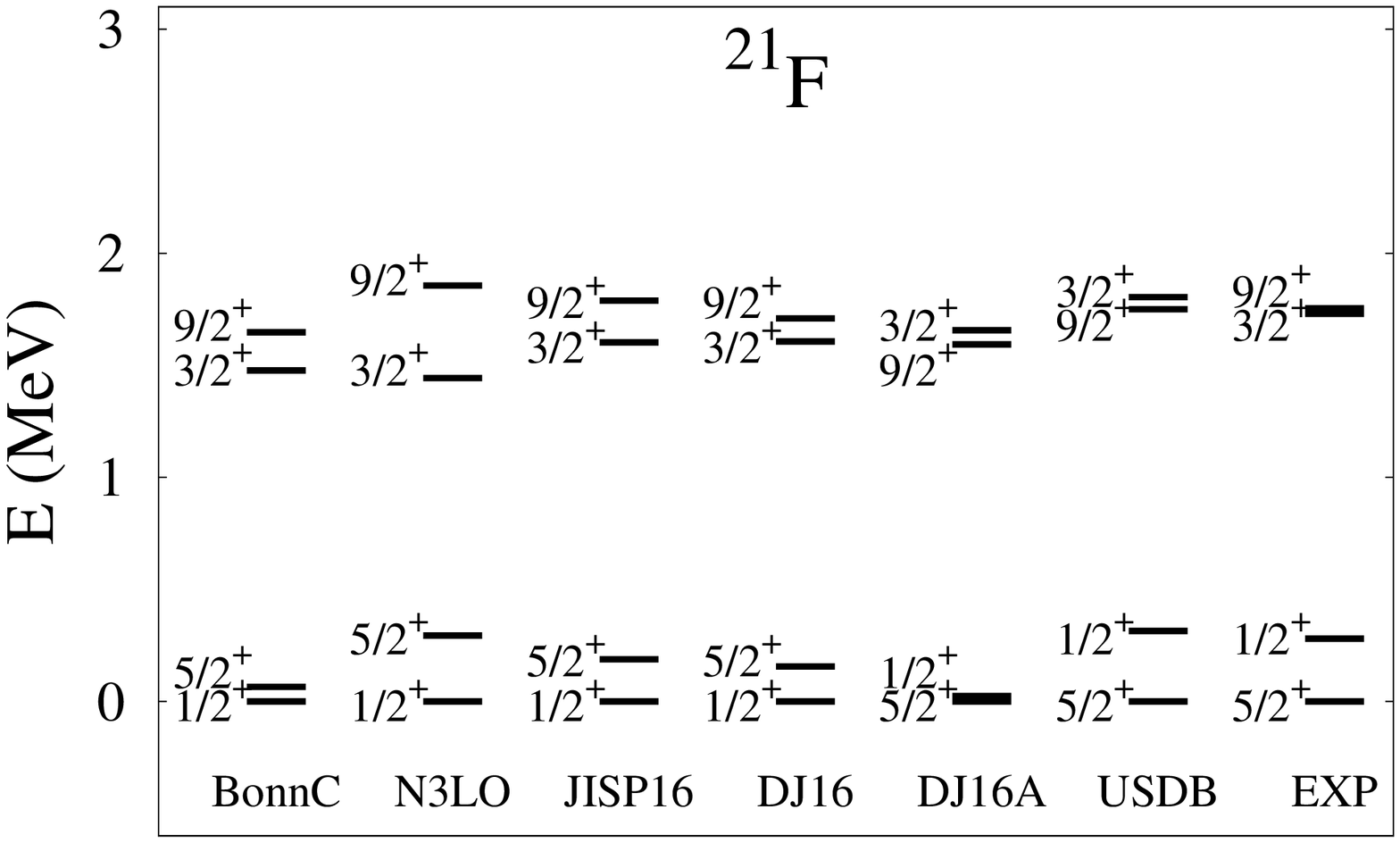} \\
  \includegraphics[height=.23\textheight]{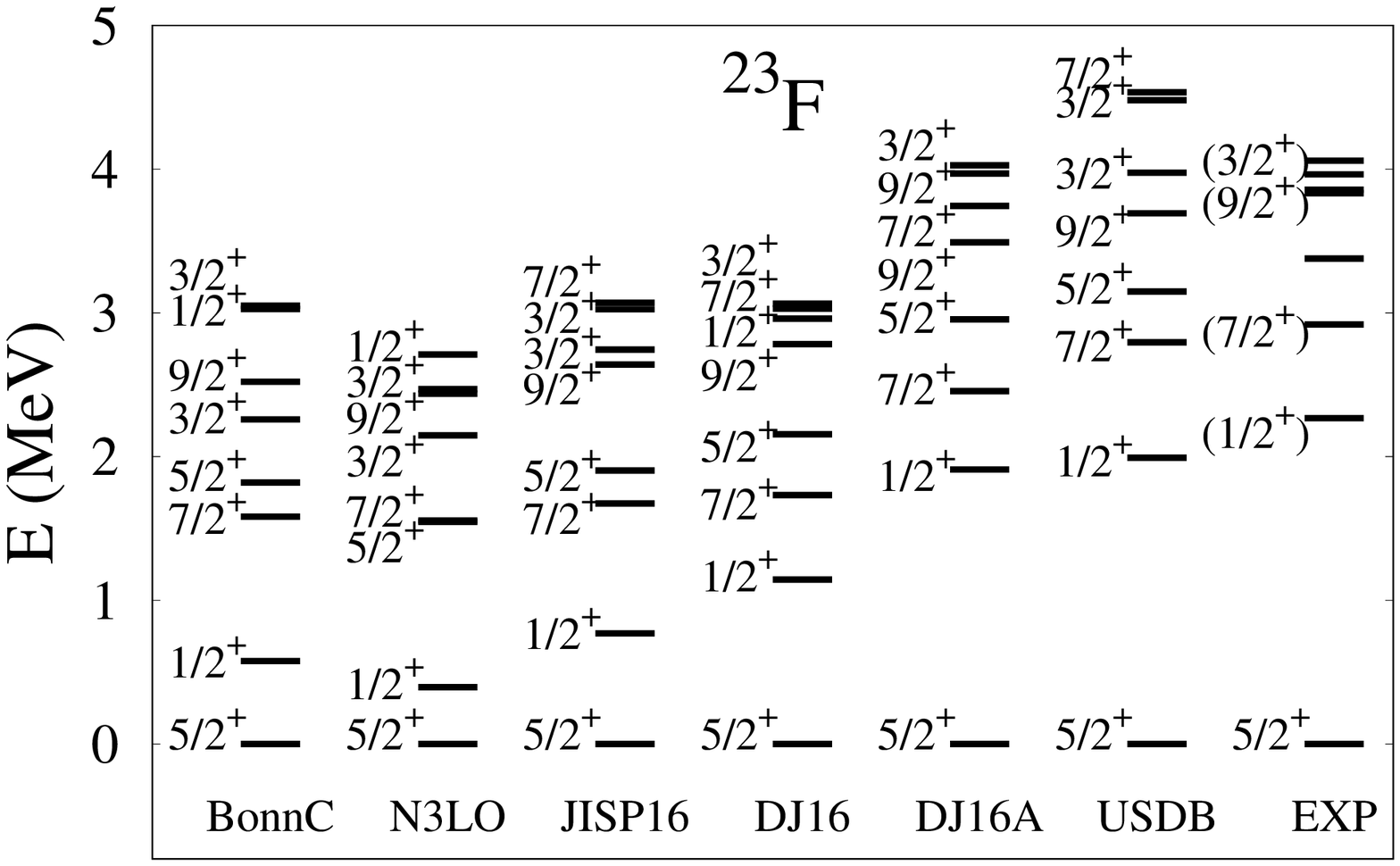}
  \includegraphics[height=.23\textheight]{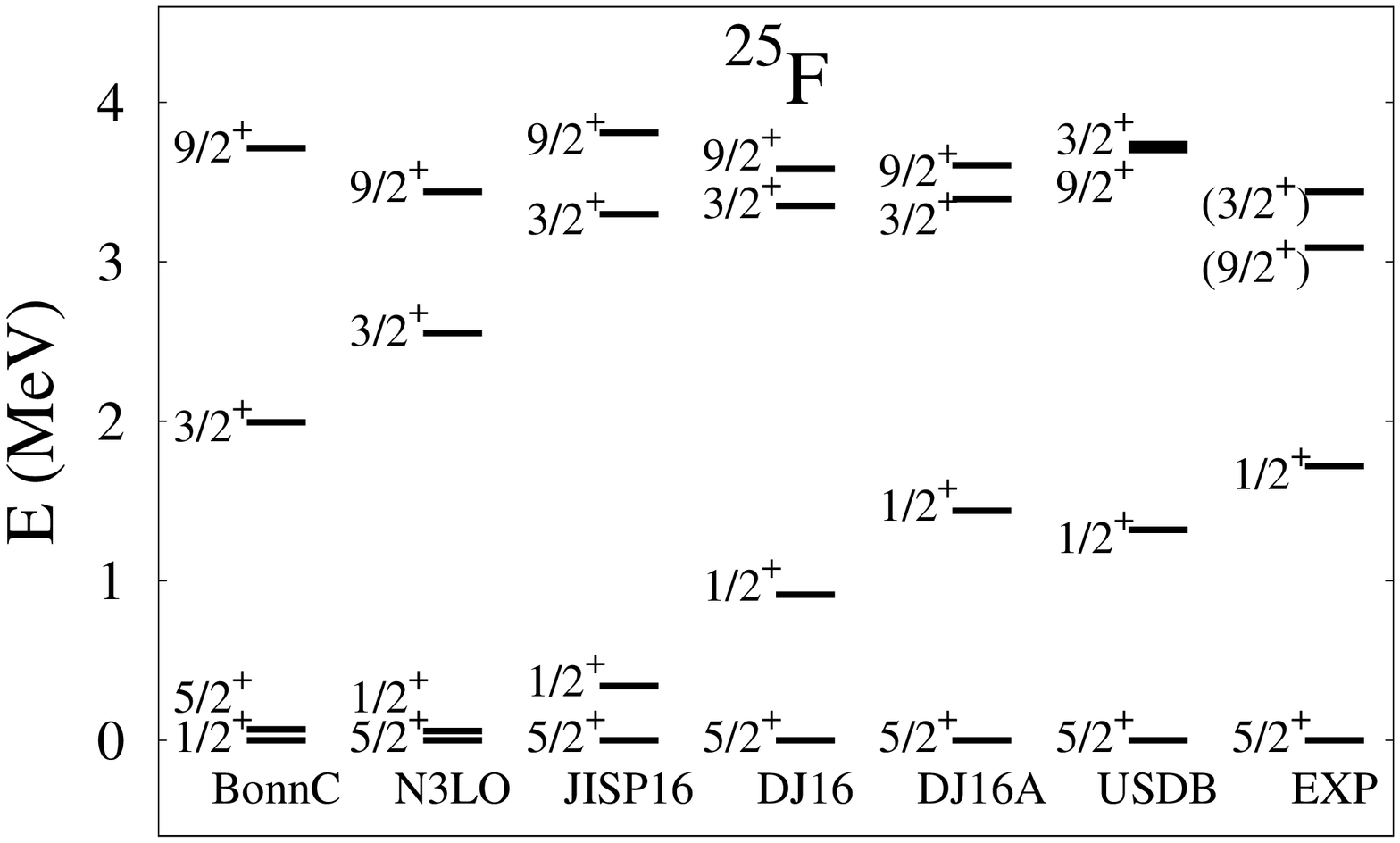} \\
  \includegraphics[height=.23\textheight]{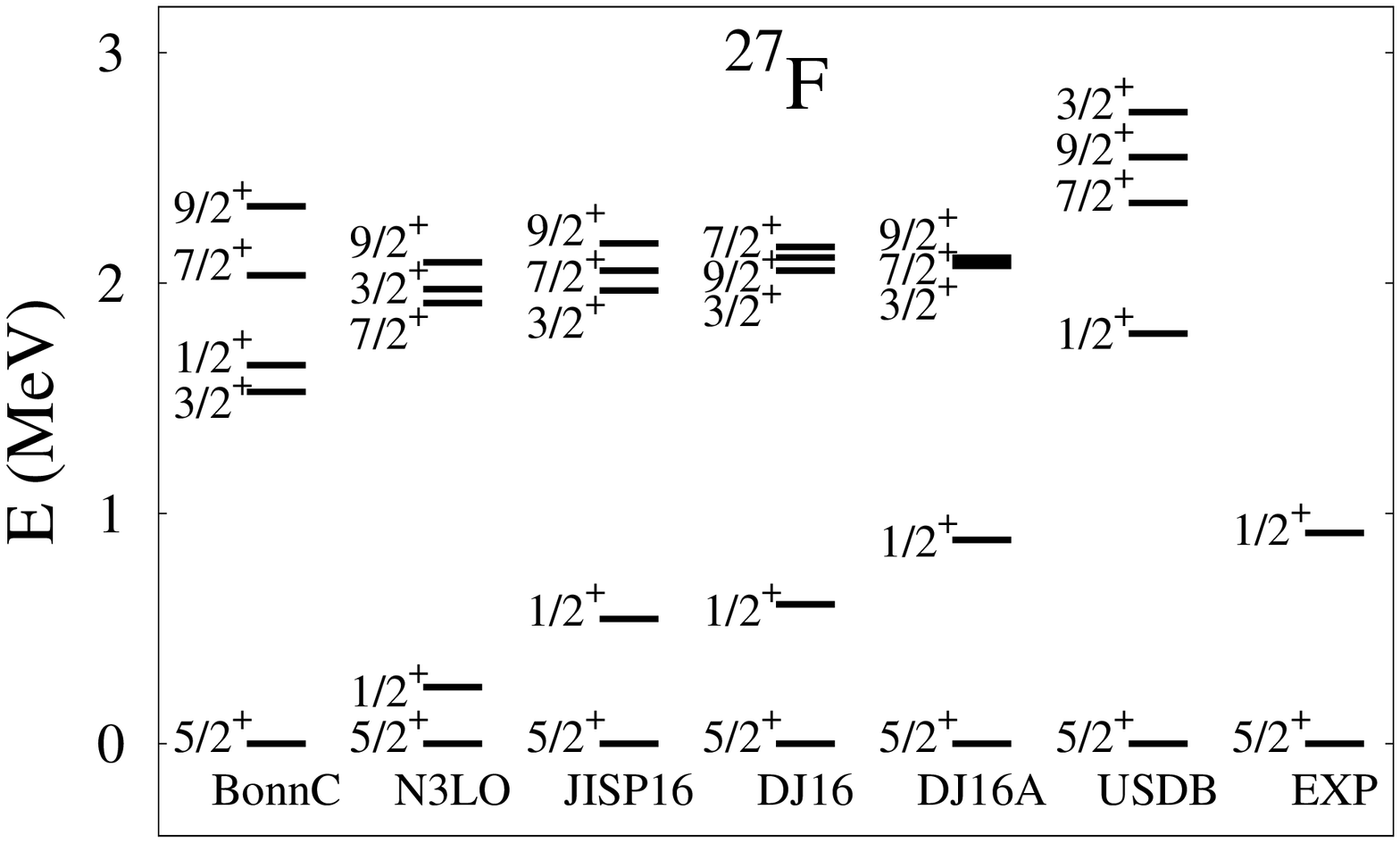}
  \includegraphics[height=.23\textheight]{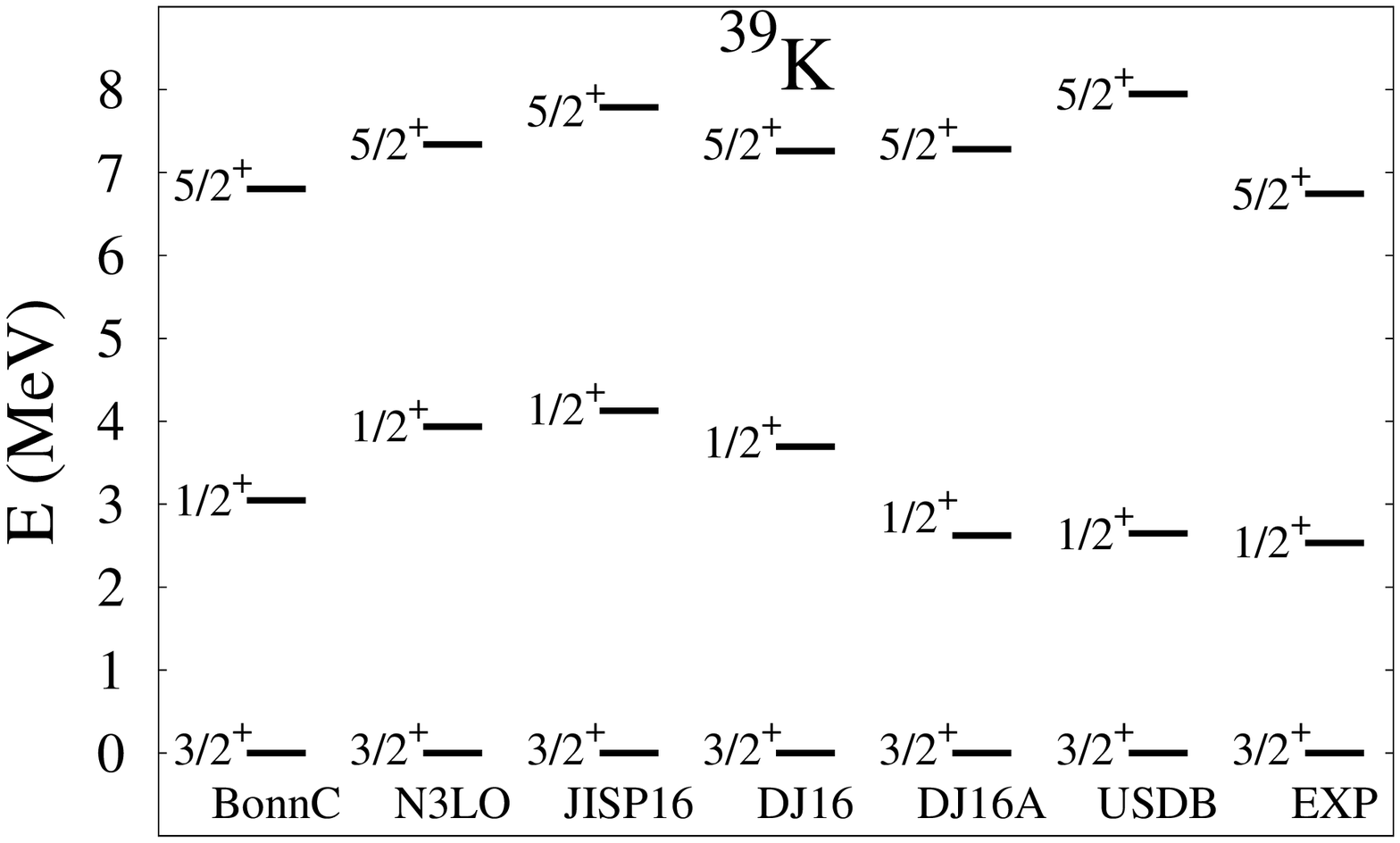}
  \caption{\label{fig:F-isotopes} Low-energy spectra of odd-$A$ $^{19-27}$F and $^{39}$K, 
obtained from USDB and microscopic effective interactions, in comparison with the experimental  data on positive-parity states 
from Refs.~\protect\cite{nndc,VanLe2017,Doorn2017}.
For $^{39}$K, we show experimentally deduced centroids from Ref.~\protect\cite{Doll76}.}
\end{figure*}

In Ref.~\cite{Otsuka3N}, the authors point out the need for
$3N$ forces to improve on the saturation properties (see Fig.~4 of their paper and discussion in the text).
Indeed, the IMSRG calculation using $NN$ and $3N$ forces shows a robust agreement with the experimental
binding energies. 
It slightly overbinds O isotopes in the beginning of the $sd$ shell and slightly underbinds the very neutron-rich 
isotopes as we review in Fig.~\ref{fig:BE_O} to compare with our results.

We also observe that empirical modification of the centroids, introduced to produce DJ16A, cures
most of the defects of the O isotopes' spectra and generates binding energies in remarkable agreement with experiment.
The rms deviations between experiment and theory (binding energies and excitation spectra)
are summarized in Table~\ref{tab:rms}. Only states shown in the figures of the present article are taken
into account in the evaluation of the rms deviations.

\section{\boldmath Odd-$A$ F isotopes and $^{39}$K}

The odd-$A$ F isotopes are important because, while neutrons are affected by the pairing force, 
the proton single-particle centroids can provide direct information on the proton-neutron monopoles. 
In practice, it is difficult to get the experimental centroids 
due to the sparsity and imprecision of available data on the spectroscopic factors.
The low-energy theoretical spectra of odd-$A$ F isotopes are shown in Fig.~\ref{fig:F-isotopes} in comparison
with experiment.
Only in $^{23,25}$F, the low-lying $5/2^+$, $1/2^+$ and $3/2^+$ states may contain 
appreciable proton $d_{5/2}$, $s_{1/2}$ and $d_{3/2}$ single-particle components, respectively.

\begin{table*}[!t]
\centering
\caption{Experimental and calculated interaction energies, Int($J$), 
as defined in Ref.~\protect\cite{VanLe2017} [see also Eqs.~\eqref{IntJ} and~\eqref{IntJfree}] in $^{26}$F
and their angular-momentum average matrix element $V^{\pi \nu}_{d_{5/2}d_{3/2}}$. All values are in MeV.}
\label{tab:F26}\vspace{1ex}
\begin{ruledtabular}
\begin{tabular}{r|r|r|r|r|r|r|r|r} 
Int($J$) 	    & EXP     	  & BonnC   & N3LO    & JISP16  & DJ16 & DJ16A & USDB & IMSRG \\[1mm]
\hline
Int(1) 	    & $-1.85(13)$ & $-2.43$ & $-2.26$ & $-2.20$ & $-2.20$ & $-2.13$ & $-1.99$ & $-2.24(7)$ \\
Int(2) 	    & $-1.19(14)$ & $-1.34$ & $-2.01$ & $-1.81$ & $-1.59$ & $-1.39$ & $-1.43$ & $-1.86(5)$ \\
Int(3) 	    & $-0.45(19)$ & $-0.50$ & $-0.55$ & $-0.36$ & $-0.26$ & $-0.18$ & $-0.46$ & $-0.53(4)$ \\
Int(4) 	    & $-1.21(13)$ & $-1.53$ & $-1.66$ & $-1.61$ & $-1.56$ & $-1.50$ & $-1.75$ & $-1.56(4)$ \\
\hline
$V^{\pi \nu}_{d_{5/2}d_{3/2}}$ & $-1.06(8)$  & $-1.30$ & $-1.48$ & $-1.36$ & $-1.27$ & $-1.17$ & $-1.34$ & $-1.41(2)$ \\
\end{tabular}
\end{ruledtabular}
\end{table*}

The low-lying states of $^{19}$F are relatively well-reproduced by all interactions,
which is usually the case for a nucleus with a small number of valence particles.
In the case of $^{21}$F, there is an inversion of the lowest $1/2^+$ and $5/2^+$ states
in the spectra obtained by all microscopic interactions
when compared to the experiment or USDB. 
One of the possible reasons is the insufficient $N=14$ shell gap seen in Fig.~\ref{fig:N14-ESPEs}.
Modifications made to the monopoles to produce
the DJ16A interaction, succeed in yielding the $5/2^+$ ground state, but produce inverted
higher lying $3/2^+$ and $9/2^+$ states.

The small $N=14$ shell gap in $^{22}$O, as obtained from the microscopic effective interactions, 
manifests itself in a compressed spectrum of $^{23}$F relative to experiment shown in Fig.~\ref{fig:F-isotopes}. 
In particular, the first $1/2^+$ and $3/2^+$ states, which contain large components of proton $s_{1/2}$ and $d_{3/2}$ single-particle states,
are too low with respect to the experimental data and to the USDB calculation. 
Again, we see a continuous improvement from BonnC potential to N3LO, JISP16 and DJ16.
There is a rather good agreement between DJ16A, USDB and experiment.

The spectrum of the neutron-rich $^{25}$F is better described
by DJ16 and DJ16A as compared to other microscopic interactions. 
It is interesting that the position of the $1/2^+$ first excited state in $^{27}$F, 
observed in Ref.~\cite{Doorn2017}, is well reproduced by DJ16A 
and even better described by JISP16 and DJ16 than by USDB.
No configurations outside the $sd$ shell-model space, which was suggested in Ref.~\cite{Doorn2017}, 
are required.

Finally, as the $sd$ shell becomes almost filled, it is the spectrum of $^{39}$K  which may 
shed light on the evolution of the nuclear mean field.
The experimental spectrum of $^{39}$K (Fig.~\ref{fig:F-isotopes}) 
shows the centroids of the single-particle states, as extracted from Ref.~\cite{Doll76}. 
They can be directly compared to the theory.
Although the microscopic interactions show $1/2^+$ state a bit high, there is
in general a robust agreement with the experiment.

The rms deviations of the excitation energies relative to experiment for the  odd-$A$ fluorine isotopes 
and $^{39}$K can be found in the 4th column of Table~\ref{tab:rms}.
Note that DJ16A provides the smallest rms deviation for this particular selection of nuclei.

\section{\boldmath $^{26}$F}

\begin{figure}[htpb]
\includegraphics[width=\columnwidth]{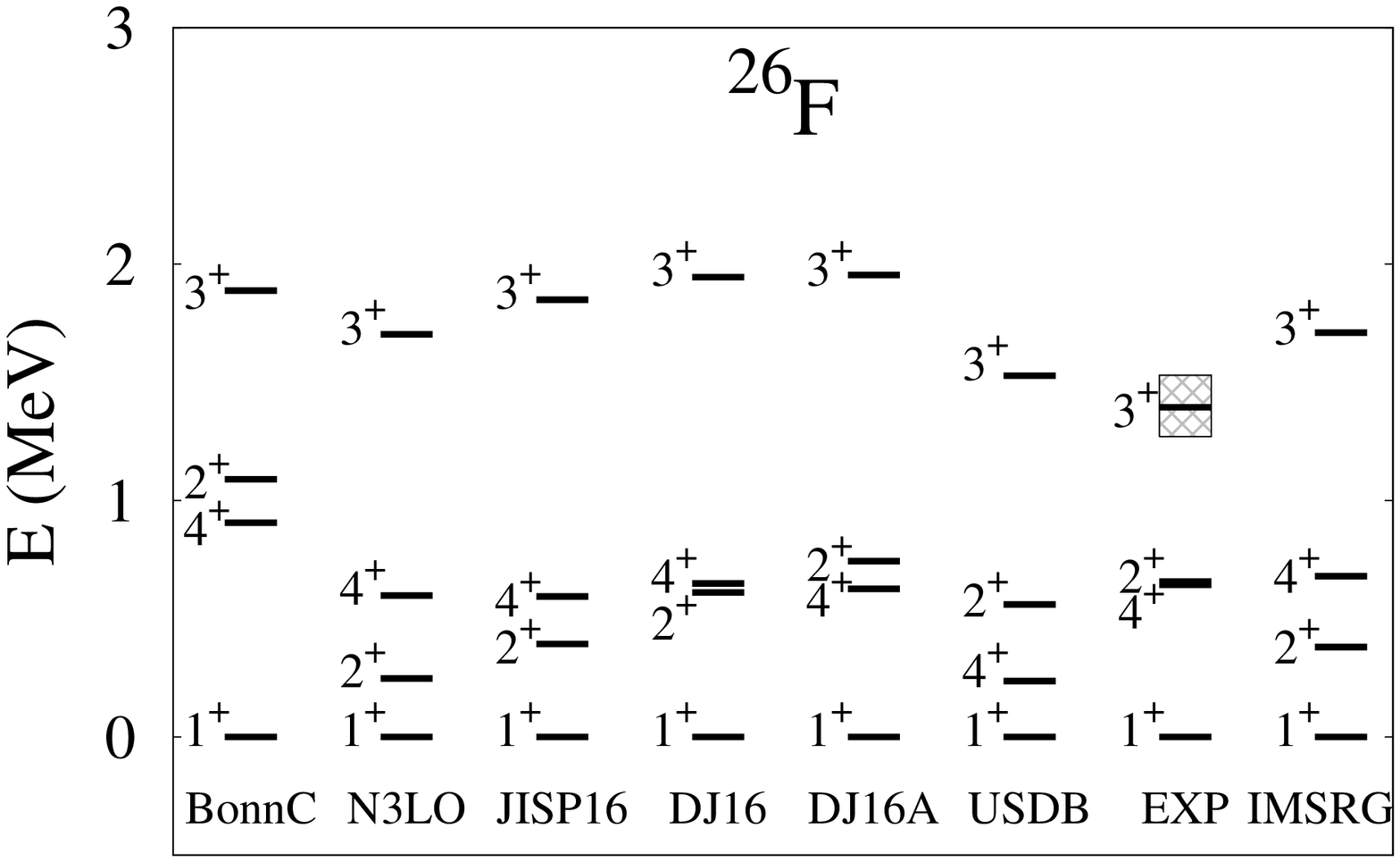} 
  \caption{\label{fig:F26-spectrum} 
Low-energy spectrum of $^{26}$F  obtained from USDB and microscopic effective interactions, 
in comparison with the experimental  data from Ref.~\protect\cite{VanLe2017}.
The spectrum entitled ``IMSRG'' is quoted also from Ref.~\protect\cite{VanLe2017}.}
\end{figure}


Recently, significant attention has been given to a drip-line nucleus $^{26}$F~\cite{LeSo2013,VanLe2017}.
The low-energy theoretical spectra of $^{26}$F are shown in Fig.~\ref{fig:F26-spectrum} in comparison with
the experiment from Ref.~\cite{LeSo2013,VanLe2017}.
If one assumes that, as in the independent particle picture, the low-energy multiplet of states
$(1^+ {-} 4^+)$ is mainly provided by  $(\pi d_{5/2} )(\nu d_{3/2})$ configuration, then one can estimate
empirical proton-neutron matrix elements as proposed in Refs.~\cite{LeSo2013,VanLe2017},
\begin{equation}
{\rm Int}(J)= {\rm BE}(^{26}{\rm F})_J-{\rm BE}(^{26}{\rm F}_{\rm free}),
\label{IntJ}
\end{equation}
Where 
\begin{equation}
{\rm BE}(^{26}{\rm F}_{\rm free})={\rm BE}(^{25}{\rm F})_{gs} + {\rm BE}(^{25}{\rm O})_{gs} - {\rm BE}(^{24}{\rm O})_{gs} .
\label{IntJfree}
\end{equation}
Refs.~\cite{LeSo2013,VanLe2017} observe
that both a phenomenological interaction (USDA) and the microscopic effective
interaction from IMSRG produce those matrix elements which are systematically more attractive than the experimentally
deduced ones. They ascribe this systematic difference  to the influence of the continuum on the proton-neutron interaction.
The interaction energies~Int($J$), estimated within a schematic independent-particle approach~\cite{VanLe2017}, 
are shown in Table~\ref{tab:F26}, together with their respective centroids ($V^{\pi \nu}_{d_{5/2}d_{3/2}}$).
We observe that the centroid from DJ16 agrees slightly better with the experimental value than the USDB result.
The closest to the experimental averaged matrix element is provided by DJ16A, although we note
that DJ16A proposes values of Int(1), Int(2), Int(4) too attractive and an insufficiently 
attractive value of Int(3).


\begin{figure*}[!ht]
\centerline{\includegraphics[width=.86\textwidth]{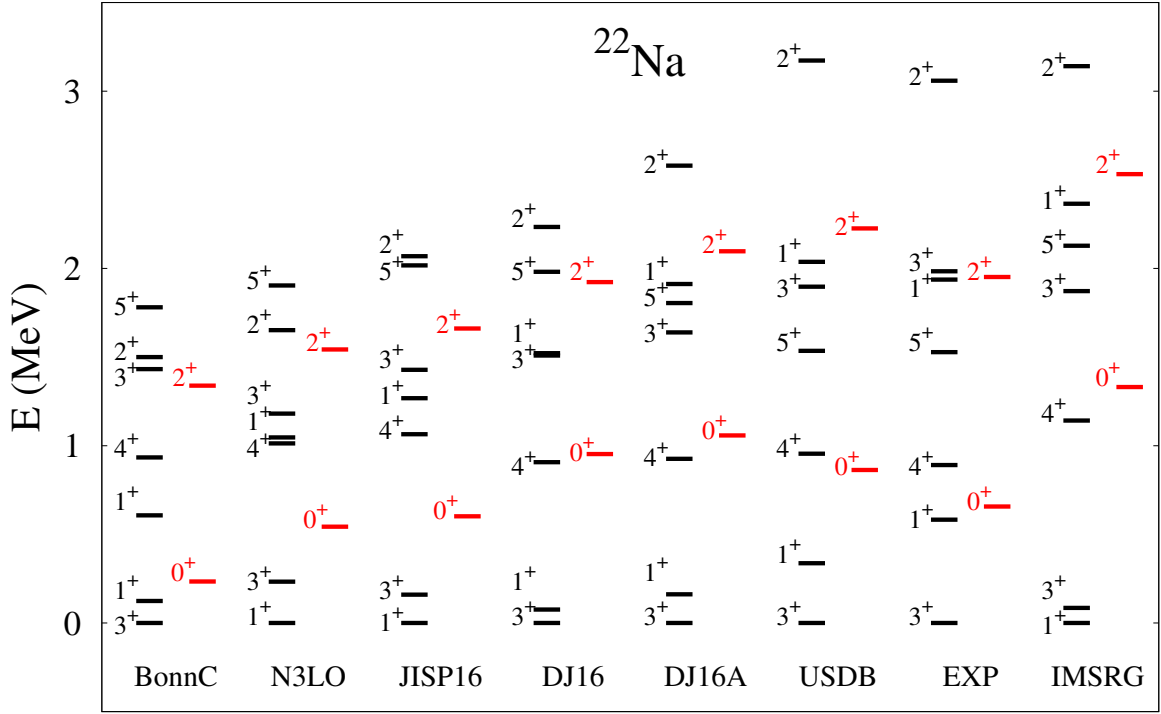} }
  \caption{\label{fig:Na22-spectrum}(Color online) 
Low-energy spectrum of $^{22}$Na  obtained from USDB and microscopic effective interactions, 
in comparison with the experimental  data on positive-parity states from Ref.~\protect\cite{nndc}.
$T=0$ states are shown in black, while $T=1$ states are plotted in red.}
\end{figure*}

\section{\boldmath $^{22}\rm Na$}

The case of $^{22}$Na with 3 protons and 3 neutrons in the valence space is considered to be 
an important benchmark of the $3N$ forces~\cite{Zuker3N}. We observe in Fig.~\ref{fig:Na22-spectrum} 
that the $T=0$ spectrum from DJ16
provides a better agreement with experiment in comparison with those from N3LO and JISP16:
the ground state is correctly found to be $3^+$ and the two lowest $2^+,T=0$ states shift higher in energy
towards closer agreement with experiment.
The rms deviations between theory and experiment for excitation energies are given in Table~\ref{tab:rms} (the fifth column). 
It is evident that in spite of a continuous reduction of the rms deviations from BonnC to DJ16 and DJ16A,
it is the USDB which provides the best TBMEs for the spectrum of $^{22}$Na.

\section{\boldmath$\rm Si$ and S isotopes}
\begin{figure}[htpb]
  \includegraphics[width=\columnwidth]{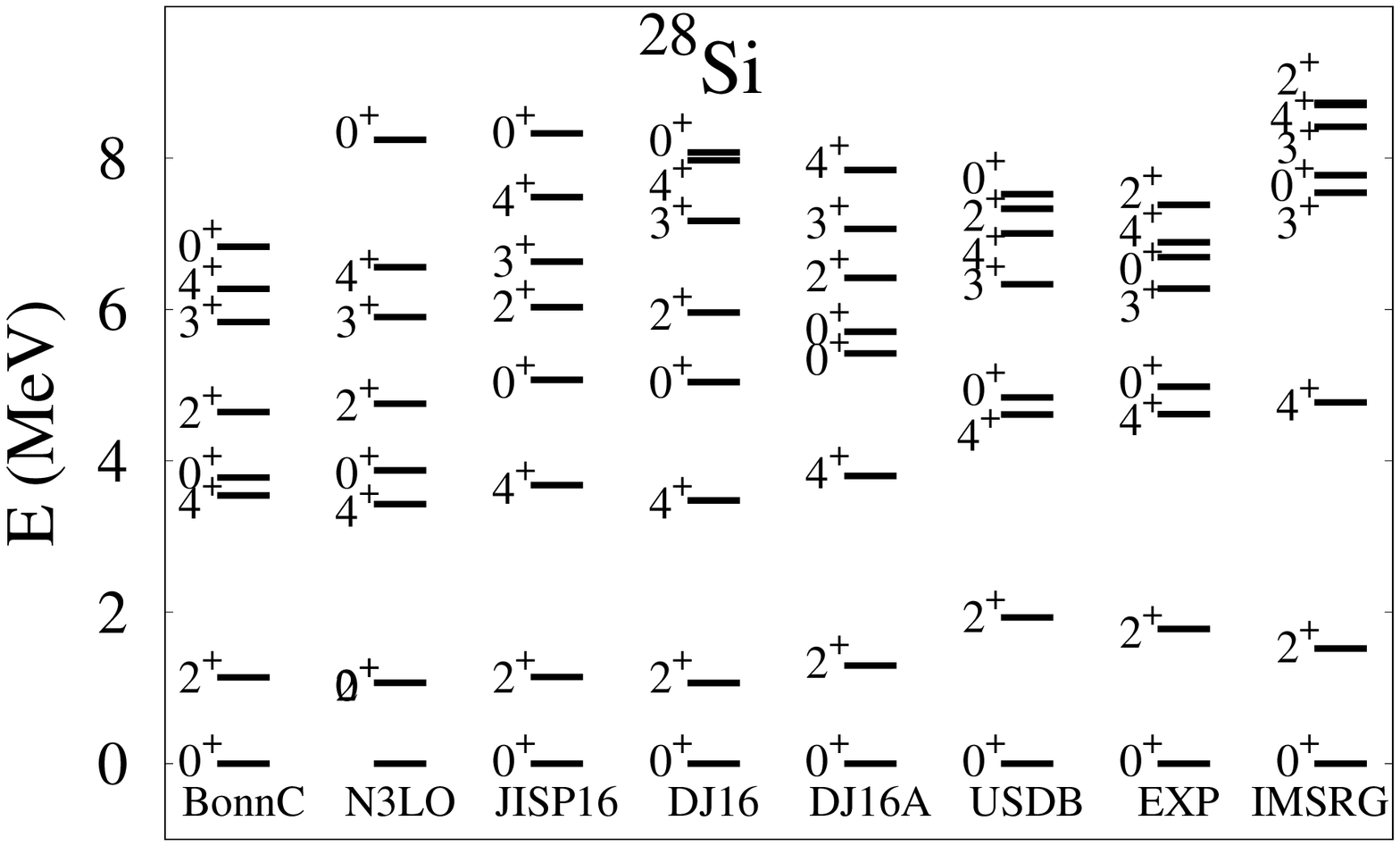} \\
  \includegraphics[width=\columnwidth]{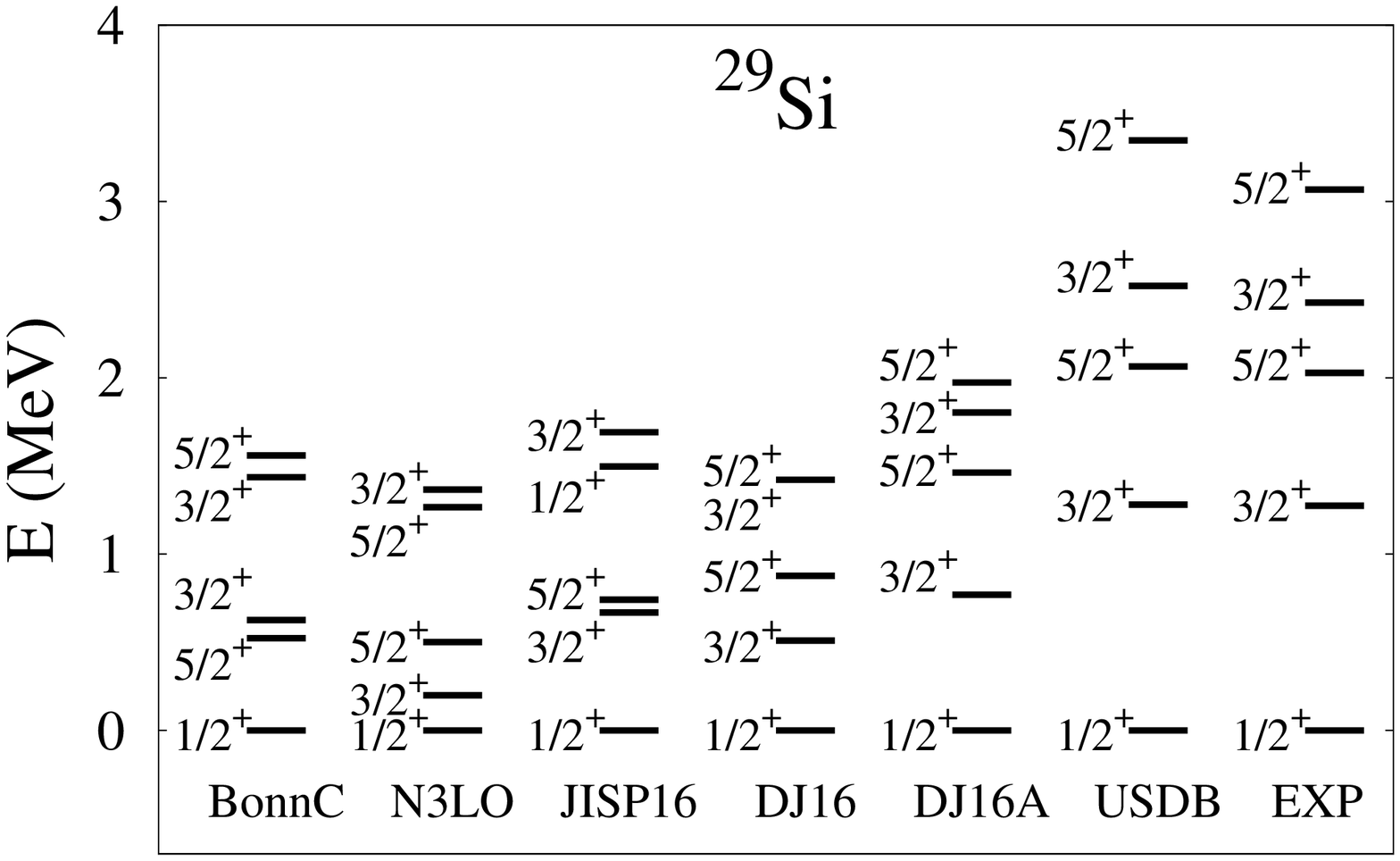} 
  \caption{\label{fig:Si28} 
Low-energy spectra of $^{28,29}$Si  obtained from USDB and microscopic effective interactions, 
in comparison with the experimental  data on positive-parity states from Ref.~\protect\cite{nndc}.}
\end{figure}

\begin{table*}[!]
	\renewcommand\footnoterule{} 
	\centering 
	\caption{Selected electromagnetic transition rates (in e$^2\cdot\rm fm^4$) and quadrupole 
	moments (in $\rm e\cdot fm^2$) in $^{24}$Mg, $^{28}$Si and $^{32}$S. 
		The experimental data are from Ref.~\protect\cite{nndc}. 
                No IMSRG Hamiltonian was available for $^{24}$Mg at~\protect\cite{git-Stroberg}.} 
	\label{tab:EM} \vspace{1ex}
        \begin{ruledtabular} 
	\begin{tabular}{l|r|r|r|r|r|r|r|r} 
 & Exp & USDB & BonnC & N3LO & JISP16 & DJ16 & DJ16A & IMSRG \\
\midrule[0.25pt]   
\hline
$\vphantom{\int^{t}}^{24}$Mg & & & & & & & & \\
$B(E2;2^+_1 \to 0^+_1)$ 
                                          & 88(4)   &  95 & 108 & 107 & 106 & 104 & 101 & \\  
$B(E2;4^+_1 \to 2^+_1)$ 
                                           & 160(16) & 124 & 143 & 140 & 138 & 138 & 137 & \\  
$B(E2;6^+_1 \to 2^+_1)$ 
                                           &         & 115 & 140 & 135 & 133 & 135 & 135 & \\
$Q(2^+_1)$ 
                                           & $-16.6(6)$ & $-19.3$ & $-18.3$ & $-18.8$ & $-19.1$ & $-19.7$ & $-19.5$ & \\
\midrule[0.25pt]        
\hline
\midrule[0.25pt]  
$\vphantom{\int^{t}}^{28}$Si & & & & & & & & \\
$B(E2;2^+_1 \to 0^+_1)$ 
                                           & 67(3) & 100 & 140 & 144 & 134 & 140 & 125 & 119\\  
$B(E2;4^+_1 \to 2^+_1)$ 
                                           & 83(9) & 140 & 187 & 194 & 176 & 188 & 174 & 161 \\  
$B(E2;0^+_2 \to 2^+_1)$ 
                                          & 48(3) &  82 &   7 &   4 &   1 &   1 & 7 & 18 \\
$Q(2^+_1)$ 
                                           & $+16(3)$ & $+20.9$ & $+24.0$ & $+24.4$ & $+23.4$ & $+24.0$ & $+23.0$ & $+22.2$  \\
\midrule[0.25pt]  
\hline
\midrule[0.25pt]  
$\vphantom{\int^{t}}^{32}$S & & & & & & & & \\
$B(E2;2^+_1 \to 0^+_1)$ 
                                           & 61(4)  &  60 & 107 & 117 & 107 & 108 & 84 & 38  \\  
$B(E2;4^+_1 \to 2^+_1)$ 
                                          & 85(18) &  85 &  80 & 121 & 109 & 136 & 121 & 67 \\  
$B(E2;0^+_2 \to 2^+_1)$ 
                                          & 71(7)  &  67 &  67 &  27 &  57 &  33 & 19 & 51 \\
$Q(2^+_1)$ 
                                          & $-15.4(20)$ & $-12.9$ & $-11.4$ & $+9.7$ & $+5.4$ & $-0.9$ & $-13.8$ & $-9.7$ \\   
	\end{tabular} 
\end{ruledtabular} 
\vspace{1ex}
\end{table*} 

The spectra of $^{28}$Si and $^{32}$S are widely regarded as more collective than those of our previously considered nuclei.
Fig.~\ref{fig:Si28} (upper panel) shows the theoretical spectra of $^{28}$Si in comparison with the experiment.
One observes that the microscopic effective interactions, obtained from BonnC, N3LO, JISP16 and Daejeon16,
produce the $2^+_1$ and $4^+_1$ states lower in energy than measured experimentally while
the USDB results are in good agreement with experiment. 
The source of these differences lies partly in the size of the $N=14$ shell gaps in $^{28}$Si, 
as we observed in Fig.~\ref{fig:N14-ESPEs}.

In $^{28}$Si, according to USDB the $2^+_1$ and $4^+_1$ states have large admixtures of available $sd$-shell configurations.
A similar  feature appears in the results obtained by the other interactions.
At the same time, the $0^+_1$ and $0^+_2$ states from USDB are dominated by the closed-subshell component 
(6 protons and 6 neutrons are in the $d_{5/2}$ orbital, $|\pi d^6_{5/2} \, \nu d^6_{5/2} \rangle $) 
with 21\% and 40\%, respectively. 
BonnC and N3LO do not populate this configuration to more than 1\% in all three lowest $0^+$ states
which is probably the consequence of their small $N=14$ shell gaps.
In the cases of JISP16 and DJ16, it is the third $0^+$ state which has the prominent 
$|\pi d^6_{5/2} \, \nu d^6_{5/2} \rangle $ configuration with probabilities of 13\% and 23\%, respectively.
DJ16A produces results similar to USDB, namely, the closed subshell configuration is present
at 10\% in the $0^+_1$ state and 46\% in the $0^+_3$ state, which is close in energy to the $0^+_2$. 
Due to this near-degeneracy of the two states,  $0^+_3$ and~$0^+_2$, 
one could compare the properties of this DJ16A $0^+_3$ state with the properties of 
the experimental $0^+_2$ state (or $0^+_2$ from USDB).

To gain insights into collectivity, we have also calculated $E2$-transition rates between the low-energy states 
and the quadrupole moment of the first $2^+$ state. Standard shell-model effective charges have been used,
$e_{\pi }=1.5\,e$ and $e_{\nu }=0.5\, e$. 
In a more complete treatment in the future, we will derive effective $E2$ operators in the same framework as 
the associated valence effective interactions~\cite{Vary:2018jxg}.  

As seen from Table~\ref{tab:EM}, in the case of $^{28}$Si, the
microscopic interactions predict enhanced \mbox{collectivity} in the $B(E2)$ transition rates from
$2^+_1$ and $4^+_1$ states over that measured experimentally.
The transitions from $2^+_1$ to $0^+_2$ are found to be quite weak. 
However, we remark that we obtain $B(E2;0^+_3 \to 2^+_1)=46$~e$^2\cdot\rm fm^4$ with DJ16A which is close
to the experimental value of $B(E2;0^+_2 \to 2^+_1)=48(3)$~e$^2\cdot\rm fm^4$.

IMSRG produces the lowest $2^+_1$ and $4^+_1$ states in $^{28}$Si in good agreement with experiment.
With this interaction, the closed subshell component, $|\pi d^6_{5/2} \, \nu d^6_{5/2} \rangle $, represents
6\% of the ground state and 24\% of the $0^+_2$ state.
The $B(E2)$ values for the stronger transitions are smaller than those obtained from other microscopic effective
interactions and are closer to the experiment and to the USDB results. 
However all the remaining levels in $^{28}$Si are generated by IMSRG too high in energy 
resulting in a larger rms energy deviation from experiment than that of all other interactions.

The small $N=14$ shell gaps partly manifest themselves again in the theoretical spectra of $^{29}$Si
obtained with various microscopic effective interactions: as may be expected, most interactions produce spectra 
that are more compressed than the experimental spectrum (Fig.~\ref{fig:Si28}, lower panel).
Consistent with this attention to the $N = 14$ shell gap,  some improvement can be noticed for DJ16A.

\begin{figure}[t!]
  \includegraphics[width=\columnwidth]{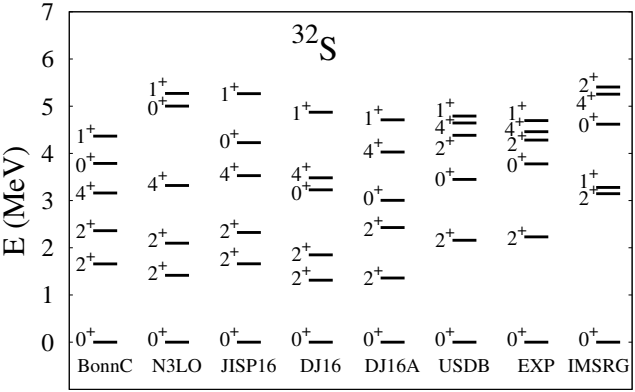} 
  \caption{\label{fig:S32} 
Low-energy spectrum of $^{32}$S  obtained from USDB and microscopic effective interactions, 
in comparison with the experimental  data on positive-parity states from Ref.~\protect\cite{nndc}.}
\end{figure}

The theoretical and experimental spectra of $^{32}$S are 
shown in Fig.~\ref{fig:S32}.
The microscopic effective interactions obtained from the BonnC, N3LO, JISP16 and 
\mbox{Daejeon16} potentials again show the $2^+_1$ and $2^+_2$ states too low in energy,
while the $0^+_2$ state generated by  N3LO and JISP16 lies quite high.
According to USDB, the two lowest $0^+_{1,2}$ states contain 30\% and 38\% of the spherical configuration,
$|\pi d^6_{5/2} \pi s^2_{1/2} \, \nu d^6_{5/2} \nu s^2_{1/2} \rangle $,
while the microscopic interactions, including DJ16A, 
predict those configurations to be occupied on the order of $4{-}10$\% and $33{-}60$\%,
respectively. 
All microscopic interactions considered, except IMSRG,
provide the $B(E2)$ values for the quadrupole transitions 
between the lowest states which overestimate the experimental value, see Table~\ref{tab:EM}. 
The calculations with N3LO and JISP16 do not reproduce the negative value of the quadrupole moment
of the $2^+_1$ state.

The IMSRG low-energy spectrum of $^{32}$S is somewhat more extended in energy compared to experiment, 
with the $1^+_1$ state appearing too low. The latter may be due to the interplay of a few TBMEs,
while the former may stem from the underlying structure of the ESPEs and multipole components of the interaction.
As seen from the $B(E2)$ values, the quadrupole collectivity is less pronounced than found experimentally.
The two lowest $0^+_{1,2}$ states contain 43\% and 17\%, respectively, of the spherical 
$|\pi d^6_{5/2} \pi s^2_{1/2} \, \nu d^6_{5/2} \nu s^2_{1/2} \rangle $ configuration.

As seen from Figs.~\ref{fig:Si28}--\ref{fig:S32}, DJ16A produces only marginal improvement of the spectra of 
$^{28}$Si and $^{32}$S. 
More significant improvement in the spectra with DJ16A is visible for $^{29}$Si.
It seems apparent that the minimal monopole modifications are not sufficient to fully describe 
the spectra in the middle of the $sd$ shell. 
For example, other components of the interaction may also need modification.

\section{Quadrupole properties}

\begin{figure*}[!t]
\includegraphics[width=\textwidth]{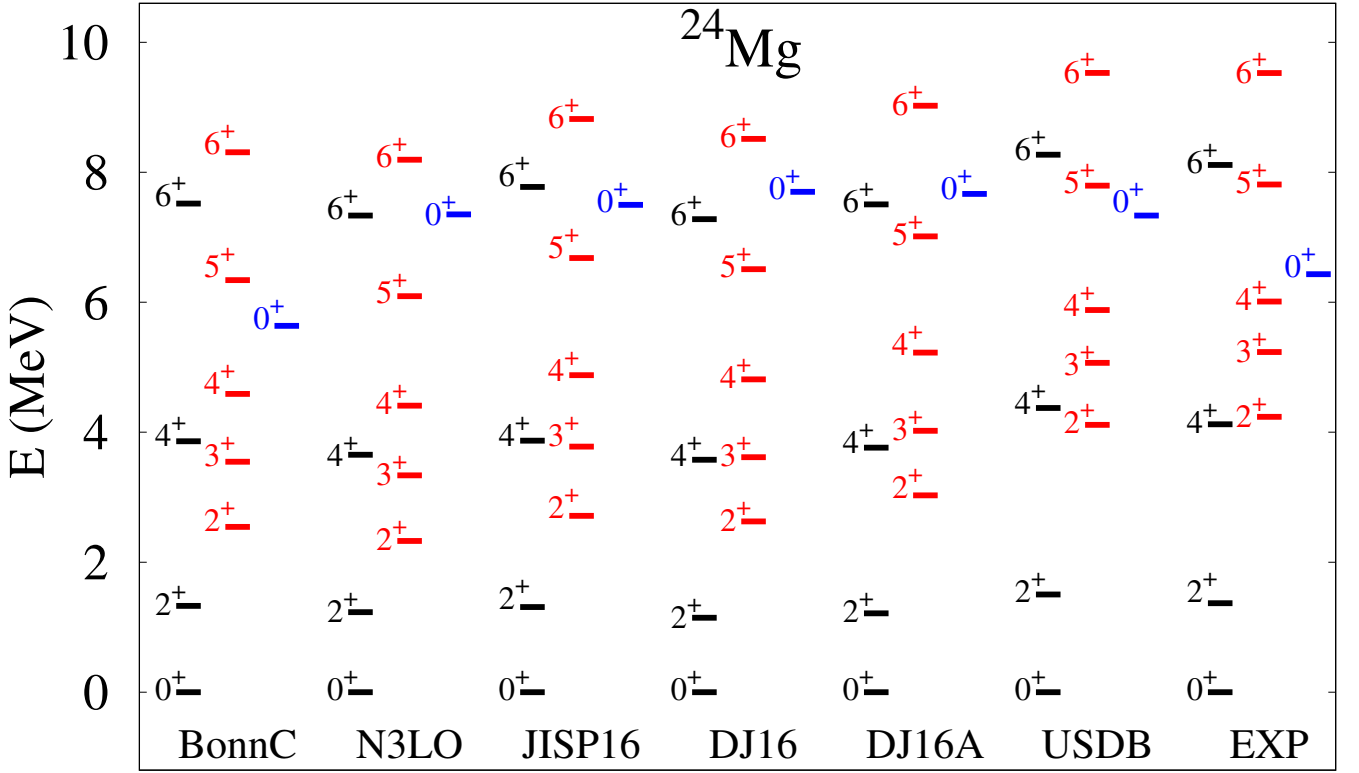}
\caption{\label{fig:Mg24}(Color online) 
Low-energy spectrum of $^{24}$Mg  obtained from USDB and microscopic effective interactions, 
in comparison with the experimental  data on positive-parity states from Ref.~\protect\cite{nndc}.
Different rotation-vibration bands are distinguished by color.
}
\end{figure*}

To characterize the proton-neutron quadrupole component of the microscopic interactions, 
we consider the spectrum of a well-known  $sd$-shell rotor, $^{24}$Mg, see Fig.~\ref{fig:Mg24}. 
We remark that the ground-state band is relatively well described by all interactions. 
The microscopic interactions produce a somewhat lower lying and slightly more stretched $\gamma $-band, as
compared to the experimental result and that from USDB.
No IMSRG Hamiltonian is available yet for $^{24}$Mg at~\cite{git-Stroberg}.

The electromagnetic properties of $^{24}$Mg are summarized in Table~\ref{tab:EM}. 
We see that all interactions predict strong in-band $E2$ transitions in good agreement with the experiment.
The calculated quadrupole moments of the $2^+_1$ state are consistent with the measured values.

In summary, we find that the quadrupole components of the microscopic interactions are satisfactory.\\

\section{Spin-tensor decomposition of the two-body interaction}

In order to better understand the deficiencies in the centroids of the microscopic interactions, 
we performed a spin-tensor decomposition of the effective interactions under consideration.
Any two-nucleon interaction can be expanded as a sum of scalar products of
scalar, vector and rank-2 spherical tensors,
\begin{equation}
V = \sum\limits_{k = 0,1,2} \left(S^{(k)} \cdot Q^{(k)}\right) = \sum\limits_{k = 0,1,2} V^{(k)} , 
\end{equation}
where $S^{(k)}$ are spin-tensors constructed from nucleon spin-1/2 operators and $Q^{(k)}$ are tensors of
corresponding rank in coordinate space. For the spin part, the scalar
operators $\big(1$ and $(\vec{\sigma}_1\cdot\vec{\sigma}_2)\big)$ contribute to the central part of the effective interaction, 
the rank-2 term $\big(\!\left[\vec{\sigma}_1 \times \vec{\sigma}_2\right]^{(2)}\!\big)$ corresponds to the so-called  
tensor force.
The vector part ($k=1$) includes a term $\vec{\sigma }_1 + \vec{\sigma }_2$, contributing to the
usual two-body spin-orbit force. The vector part also includes  
two more operators which exchange the intrinsic spin in $LS$-coupled basis and which are called 
the {\it antisymmetric spin-orbit} (ALS) operators
$\big(\!\left[ \vec{\sigma }_1 \times \vec{\sigma }_2 \right]^{(1)}$ and $\vec{\sigma }_1 - \vec{\sigma }_2\big)$. 
Using the $LS$-scheme, it is possible to calculate the matrix elements of each $V^{(k)}$ component of
the interaction from the matrix elements of $V$:
\begin{widetext}\vspace{-4ex}
\begin{multline}
\langle nl,n'l': L S, J M T M_T | V^{(k)} | n''l'',n'''l''' : L' S', J M T M_T \rangle = 
(2k + 1) (-1)^J 
\left\{ \begin{array}{ccc} L & S & J \\ S' & L' & k 
  \end{array} \right\} \\[3mm]
\times \sum_{J' } (-1)^{J' } (2J' +1)
\left\{ \begin{array}{ccc} L & S & J'  \\ S' & L' & k 
  \end{array} \right \} 
\langle nl,n'l': L S,J' M T M_T | V | n''l'',n'''l''' : L' S', J' M T M_T \rangle .
\label{decomp}
\end{multline}
\end{widetext}
In this equation, $|nl,n'l': L S, J M T M_T \rangle $ denotes a two-body state in $LS$ coupling,
where $L$ and $S$ are quantum numbers associated with the total orbital angular momentum operator, $\vec{L}$,
and total spin $\vec{S}$, while $J$ is the quantum number associated with the total \mbox{angular} \mbox{momentum} operator,
$\vec{J}=\vec{L}+\vec{S}$.
Based on the selection rules in $LS$ coupling, it is possible to discriminate between triplet-even (TE), 
triplet-odd (TO), singlet-even~(SE) and single-odd (SO) terms of the central part, between even and odd 
components of the spin-orbit term (LS and ALS, respectively) and 
even and odd components of the tensor term (TNE and TNO, respectively). 
The contribution of these terms to the centroids of the interaction and, 
thus, to the ESPE variations is additive.

The spin-tensor decomposition was used to study the tensorial structure of effective interactions 
in a number of publications~\cite{ElJa68,Kir73,KlKn77,Yoro80,BrRi85,OsSt92}.
The analyses focused on the two-body matrix elements and/or centroids of the interaction, 
even in the context of the ESPE variations~\cite{UmMu04,UmMu06,Smi10,Smi12}.
In this part we discuss the spin-tensor structure of the various $T=1$ and proton-neutron centroids of 
the microscopic interactions and compare them with those from USDB.

The $N=14$ shell gap in $^{22}$O is governed by
the difference between the $V^{T=1}_{s_{1/2} d_{5/2}}$ centroid and the $V^{T=1}_{d_{5/2} d_{5/2}}$ centroid.
Evolution of the $N=14$ shell gap from $^{16}$O to $^{22}$O and the role of
different spin-tensor components is shown in Table~\ref{tab:N14_O}. 
We observe that the increase of the $N=14$ shell gap by 3.03 MeV produced by USDB is mainly 
due to a coherent action of the central (TO) and vector parts of the effective interaction. 
At the same time, microscopic interactions do not show that increase due to much lower contributions
from the central and vector parts. The best result is observed for DJ16, as we mentioned before.
It is seen, that the spin-tensor structure of DJ16A monopole term is consistent with
the structure of the USDB monopole part. This is important since the monopole modification 
to produce DJ16A is both small and distributed among matrix elements in the simplest possible manner, a uniform shift.

\begin{table}[b!]
\caption{Increase of the $N=14$ shell gap (in MeV) from $^{16}$O to $^{22}$O as obtained from
different effective interactions (monopole part only).}
\label{tab:N14_O}\vspace{1ex}
\begin{ruledtabular}
\begin{tabular}{c|c|c|c|c|c|c} 
$\Delta E$  & USDB & BonnC & N3LO & JISP16 & DJ16 & DJ16A \\  
\hline
Total      & $\mathbf{3.03}$ & $\mathbf{0.35}$ & {\bf 0.43} & $\mathbf{0.94}$ & $\mathbf{1.49}$ & $\mathbf{3.06}$\\
\hline
Central    & $\mathbf{1.89}$ & $\mathbf{0.87}$ & {\bf 0.65} & $\mathbf{0.70}$ & $\mathbf{0.90}$ & $\mathbf{1.68}$ \\ 
	TO &      $1.31$  &      $0.28$  &      0.24  &     0.32 &     0.76 &	1.38 \\
	SE &      $0.56$  &      $0.59$  &      0.41  &     0.38 &     0.14 & 	0.30 \\
\hline
Vector     & {\bf  1.26} & ${\mathbf -0.02}$ & {\bf 0.33} & $\mathbf{0.56}$ &  {\bf 0.78} & $\mathbf{1.36}$ \\
      LS   &       0.27  &      -0.10  &      0.07  &     0.35  &  0.24  &	0.43 \\
      ALS  &       0.99  &       0.08  &      0.26  &     0.21  &  0.53  & 	0.93 \\
\hline
Tensor  & $\mathbf{-0.12}$ & ${\mathbf -0.49}$ & $\mathbf{-0.55}$ & ${\mathbf -0.32}$ & $\mathbf{-0.18}$ & $\mathbf{0.02}$\\
\end{tabular}
\end{ruledtabular}
\end{table}

\begin{table*}[!t]
\begin{small}
\centering
\caption{Spin-tensor content (in MeV) of the $T=1$ centroids of TBMEs,
involved into the evolution of the $N=14$ shell gap in O isotopes.
The table shows the results from USDB and of the microscopic effective interactions.}
\label{tab:spin-tensor}\vspace{1ex}
\begin{ruledtabular}
\begin{tabular}{r|rrr|rrr|rrr} 
 & \multicolumn{3}{c|}{USDB} &\multicolumn{3}{c|}{BonnC} &\multicolumn{3}{c}{N3LO} \\
\hline
& $\vphantom{\int^{t}_{q_{q}}} V^{T=1}_{d_{5/2} d_{5/2}}$ & $V^{T=1}_{s_{1/2} d_{5/2}}$ &  $\Delta V$ 
& $ V^{T=1}_{d_{5/2} d_{5/2}}$ & $V^{T=1}_{s_{1/2} d_{5/2}}$ &  $\Delta V$ 
& $ V^{T=1}_{d_{5/2} d_{5/2}}$ & $V^{T=1}_{s_{1/2} d_{5/2}}$ &  $\Delta V$ \\[1mm]
\hline
Total      & $\mathbf{-0.63}$ &       {\bf 0.01} & {\bf 0.64} 
           & $\mathbf{-0.51}$ & $\mathbf{-0.37}$ & {\bf 0.14} 
           & $\mathbf{-0.79}$ & $\mathbf{-0.60}$ & {\bf 0.19}\\
\hline
Central    & $\mathbf{-0.39}$  &      {\bf 0.01} & {\bf 0.40} 
           & $\mathbf{-0.47}$ & $\mathbf{-0.25}$ & {\bf 0.23} 
           & $\mathbf{-0.77}$ & $\mathbf{-0.54}$ & {\bf 0.23}\\

TO         & 0.39 & 0.57 & 0.18 
           & 0.30 & 0.31 & 0.01 
           & 0.10 & 0.13 & 0.03 \\
SE         & $-0.78$ & $-0.57$ & 0.21 
           & $-0.77$ & $-0.55$ & 0.22 
           & $-0.87$ & $-0.67$ & 0.20  \\
\hline
Vector     & $\mathbf{-0.26}$ &       {\bf 0.00} & {\bf 0.26} 
           & $\mathbf{-0.14}$ & $\mathbf{-0.13}$ & {\bf 0.02} 
           & $\mathbf{-0.14}$ & $\mathbf{-0.06}$ & {\bf 0.08}\\

LS         & $-0.06$ & $-0.00$ & 0.06 
           & $-0.05$ & $-0.06$ & $-0.01$ 
           & $-0.06$ & $-0.04$ & 0.02 \\
ALS        & $-0.20$ &  0.00   & 0.21 
           & $-0.10$ & $-0.07$ & 0.03 
           & $-0.09$ & $-0.03$ & 0.06 \\

\hline
Tensor     & $\mathbf{0.03}$ &       {\bf 0.00} & $\mathbf{-0.03}$ 
 	   & $\mathbf{0.10}$ &       {\bf 0.00} & $\mathbf{-0.10}$ 
           & $\mathbf{0.12}$ &       {\bf 0.00} & $\mathbf{-0.12}$  \\  
\hline
%
%
\hline
 & \multicolumn{3}{c|}{JISP16} &\multicolumn{3}{c|}{DJ16} &\multicolumn{3}{c}{DJ16A} \\
\hline
& $ \vphantom{\int_{q_{q}}^{t}} V^{T=1}_{d_{5/2} d_{5/2}}$ & $V^{T=1}_{s_{1/2} d_{5/2}}$ &  $\Delta V$ 
& $ V^{T=1}_{d_{5/2} d_{5/2}}$ & $V^{T=1}_{s_{1/2} d_{5/2}}$ &  $\Delta V$ 
& $ V^{T=1}_{d_{5/2} d_{5/2}}$ & $V^{T=1}_{s_{1/2} d_{5/2}}$ &  $\Delta V$  \\[1mm]
\hline
Total        & $\mathbf{-0.79}$ & $\mathbf{-0.51}$ & {\bf 0.28} 
           & $\mathbf{-0.71}$ & $\mathbf{-0.34}$ & {\bf 0.37}
	   & $\mathbf{-0.63}$ & $\mathbf{ 0.02}$ & {\bf 0.65}\\
\hline
Central        & $\mathbf{-0.69}$ & $\mathbf{-0.46}$ & {\bf 0.23} 
           & $\mathbf{-0.56}$ & $\mathbf{-0.31}$ & {\bf 0.24}
           & $\mathbf{-0.38}$ & $\mathbf{-0.02}$ & {\bf 0.36}\\

TO         & 0.16 & 0.19 & 0.03
           & 0.21 & 0.31 & 0.11 
           & 0.33 & 0.53 & 0.20 \\
SE         & $-0.84$ & $-0.65$ & 0.19 
           & $-0.76$ & $-0.63$ & 0.14 
           & $-0.71$ & $-0.55$ & 0.16 \\
\hline
Vector        & $\mathbf{-0.18}$ & $\mathbf{-0.05}$ & {\bf 0.13} 
           & $\mathbf{-0.19}$ & $\mathbf{-0.02}$ & {\bf 0.17}
           & $\mathbf{-0.24}$ & $\mathbf{ 0.04}$ & {\bf 0.28}\\
LS         & $-0.08$ & $-0.00$ & 0.07 
           & $-0.02$ & $ 0.03$ & 0.05
           & $-0.02$ & $ 0.06$ & 0.08 \\
ALS        & $-0.10$ & $-0.05$ & 0.05 
           & $-0.17$ & $-0.05$ & 0.12 
           & $-0.22$ & $-0.02$ & 0.20 \\
\hline
Tensor        & $\mathbf{0.07}$ &       {\bf 0.00} & $\mathbf{-0.07}$ 
           & $\mathbf{0.04}$ &       {\bf 0.00} & $\mathbf{-0.04}$ 
           & $\mathbf{-0.003}$ &     {\bf 0.00} & $\mathbf{0.003}$  
\end{tabular}
\end{ruledtabular}
\end{small}
\end{table*}

\begin{table*}[!t]
\begin{small}
\centering
\caption{Spin-tensor content (in MeV) of the proton-neutron centroids of USDB and microscopic effective interactions.}
\label{tab:ST-pn}\vspace{1ex}
\begin{ruledtabular}
\begin{tabular}{r|rrr|rrr|rrr} 
 & \multicolumn{3}{c|}{USDB} &\multicolumn{3}{c|}{BonnC} &\multicolumn{3}{c}{N3LO} \\
\hline
& $\vphantom{\int_{q_{q}}^{t}}  V^{pn}_{d_{5/2} d_{5/2}}$ &  $V^{pn}_{d_{3/2} d_{5/2}}$ & $V^{pn}_{s_{1/2} d_{5/2}}$
& $ V^{pn}_{d_{5/2} d_{5/2}}$ &  $V^{pn}_{d_{3/2} d_{5/2}}$ & $V^{pn}_{s_{1/2} d_{5/2}}$ 
& $ V^{pn}_{d_{5/2} d_{5/2}}$ &  $V^{pn}_{d_{3/2} d_{5/2}}$ & $V^{pn}_{s_{1/2} d_{5/2}}$ \\[1mm]
\hline
Total      & $\mathbf{-2.02}$ &   $\mathbf{-1.99}$ & $\mathbf{-1.28}$ 
           & $\mathbf{-1.44}$ & $\mathbf{-2.02}$ & $\mathbf{-1.20}$
           & $\mathbf{-1.73}$ & $\mathbf{-2.27}$ & $\mathbf{-1.51}$\\
\hline
Central    & $\mathbf{-1.98}$ & $\mathbf{-1.94}$ & $\mathbf{-1.30}$ 
           & $\mathbf{-1.70}$ & $\mathbf{-1.72}$ & $\mathbf{-1.11}$ 
           & $\mathbf{-1.92}$ & $\mathbf{-1.98}$ & $\mathbf{-1.48}$\\
\hline
Vector     & $\mathbf{-0.17}$ & $\mathbf{0.14}$ & {\bf 0.02} 
           & $\mathbf{ 0.05}$ & $\mathbf{0.02}$ & $\mathbf{-0.08}$ 
           & $\mathbf{-0.04}$ & $\mathbf{0.05}$ & $\mathbf{ 0.03}$\\
LS         & $-0.01$ & $ 0.03$ & $-0.03$ 
           & $ 0.08$ & $-0.02$ & $-0.04$ 
           & $-0.02$ & $ 0.00$ & $-0.02$ \\
ALS        & $-0.05$ &  0.11   & 0.05 
           & $-0.02$ & $ 0.04$ & $-0.04$ 
           & $-0.03$ & $ 0.04$ & $-0.01$ \\
\hline
Tensor     & $\mathbf{0.13}$ & $\mathbf{-0.19}$ &       {\bf 0.00} 
 	   & $\mathbf{0.20}$ & $\mathbf{-0.31}$ &       {\bf 0.00} 
           & $\mathbf{0.23}$ & $\mathbf{-0.34}$ &       {\bf 0.00} \\  
\hline
%
%
\hline
 & \multicolumn{3}{c|}{JISP16} & \multicolumn{3}{c|}{DJ16} &\multicolumn{3}{c}{DJ16A} \\
\hline
& $ V^{pn}_{d_{5/2} d_{5/2}}$ &  $V^{pn}_{d_{3/2} d_{5/2}}$ & $V^{pn}_{s_{1/2} d_{5/2}}$
& $ \vphantom{\int_{q_{q}}^{t}} V^{pn}_{d_{5/2} d_{5/2}}$ &  $V^{pn}_{d_{3/2} d_{5/2}}$ & $V^{pn}_{s_{1/2} d_{5/2}}$ 
& $ V^{pn}_{d_{5/2} d_{5/2}}$ &  $V^{pn}_{d_{3/2} d_{5/2}}$ & $V^{pn}_{s_{1/2} d_{5/2}}$ \\[1mm]
\hline
Total      & $\mathbf{-1.76}$ & $\mathbf{-2.10}$ & $\mathbf{-1.39}$ 
           & $\mathbf{-1.90}$ & $\mathbf{-1.90}$ & $\mathbf{-1.44}$
           & $\mathbf{-1.91}$ & $\mathbf{-1.75}$ & $\mathbf{-1.22}$\\
\hline
Central    & $\mathbf{-1.86}$ & $\mathbf{-1.92}$ & $\mathbf{-1.32}$ 
           & $\mathbf{-1.89}$ & $\mathbf{-1.89}$ & $\mathbf{-1.43}$
           & $\mathbf{-1.83}$ & $\mathbf{-1.80}$ & $\mathbf{-1.25}$\\
\hline
Vector     & $\mathbf{-0.06}$ & $\mathbf{0.05}$ & $\mathbf{-0.07}$ 
           & $\mathbf{-0.08}$ & $\mathbf{0.09}$ & $\mathbf{-0.01}$
           & $\mathbf{-0.13}$ & $\mathbf{0.13}$ & $\mathbf{ 0.04}$\\
LS         & $-0.03$ & $ 0.01$ & $-0.04$ 
           & $-0.07$ & $ 0.02$ & 0.00 
           & $-0.08$ & $ 0.02$ & 0.03 \\
ALS        & $-0.03$ & $ 0.05$ & $-0.03$ 
           & $-0.02$ & $ 0.08$ & $-0.02$
           & $-0.05$ & $ 0.11$ & 0.01 \\

\hline
Tensor     & $\mathbf{0.16}$ & $\mathbf{-0.24}$ &       {\bf 0.00} 
           & $\mathbf{0.07}$ & $\mathbf{-0.11}$ &       {\bf 0.00} 
           & $\mathbf{0.05}$ & $\mathbf{-0.08}$ &       {\bf 0.00} 
\end{tabular}
\end{ruledtabular}
\end{small}
\end{table*}

Table~\ref{tab:spin-tensor} presents a spin-tensor analysis of the $V^{T=1}_{s_{1/2} d_{5/2}}$ and 
$V^{T=1}_{d_{5/2} d_{5/2}}$ centroids and of their difference, $\Delta V$, as obtained from USDB and various
microscopic interactions.
It is $\Delta V$ which is responsible for the evolution of the~${N=14}$ gap in O isotopes.
For example, the USDB value of $\Delta V=0.64$~MeV results in a significant increase of the $N=14$ shell gap
from $^{16}$O to $^{22}$O. The microscopic effective interactions are characterized by smaller~$\Delta V$ values,
namely, 0.14~MeV for BonnC, 0.19~MeV for N3LO, 0.28~MeV for JISP16 and 0.137~MeV for DJ16. 
This leads to smaller respective increases of the~${N=14}$ shell gap.
To understand the reason, we present \mbox{various} spin-tensor components of the centroids.
We observe that while the $V^{T=1}_{d_{5/2} d_{5/2}}$ centroid is relatively well reproduced by all the interactions,
the $V^{T=1}_{s_{1/2} d_{5/2}}$ centroid of the microscopic interactions is much too attractive.
At the same time, for USDB, TO and SE contributions to  $V^{T=1}_{s_{1/2} d_{5/2}}$ compensate each other.
This is not the case for the \mbox{microscopic} interactions, for which the repulsive TO component to  
$V^{T=1}_{s_{1/2} d_{5/2}}$ stays much smaller in absolute value than the attractive SE component and
 smaller than the TO component of USDB.

We also note that the spin tensor analysis of DJ16A centroids and their difference is very close to that of USDB.

Table~\ref{tab:ST-pn} summarizes a few proton-neutron centroids and their spin-tensor content.
In particular, the difference between  $ V^{pn}_{d_{5/2} d_{5/2}}$ and $V^{pn}_{s_{1/2} d_{5/2}}$ is responsible 
for the evolution of the $N=14$ shell gap from $^{22}$O and $^{28}$Si, while the difference between
$ V^{pn}_{d_{5/2} d_{5/2}}$ and~$V^{pn}_{d_{3/2} d_{5/2}}$ \mbox{governs} the spin-orbit splitting variation for the same nuclei.
The USDB value for the difference between the proton neutron centroids 
$\Delta V=  V^{pn}_{s_{1/2} d_{5/2}}-V^{pn}_{d_{5/2} d_{5/2}}=0.74$~MeV
which results in a fast increase of the $N=14$ shell gap from $^{22}$O to $^{28}$Si by 3.83~MeV 
(see Table~\ref{tab:N14_shell_gaps}).
At the same time, the microscopic effective interactions show smaller differences between these centroids, ranging from
0.24~MeV (BonnC) to 0.46~MeV (DJ16), resulting in a more moderate increase of the $N=14$ gap.
Looking at the values of the centroids in Table~\ref{tab:ST-pn}, 
we may notice that $ V^{pn}_{d_{5/2} d_{5/2}}$ is predicted 
by the microscopic interactions to be slightly smaller
while $ V^{pn}_{s_{1/2} d_{5/2}}$ is always slightly larger  than the corresponding centroids from USDB (in their absolute values). 
For~$ V^{pn}_{d_{5/2} d_{5/2}}$, this is partly due to an insufficient attractive central component of the centroid
and due to a large tensor component, especially for BonnC and N3LO effective interactions,
when compared with the USDB values.

At the same time, we observe that the central component of  $ V^{pn}_{s_{1/2} d_{5/2}}$ provided by 
the microscopic interactions (with the exception of BonnC) is slightly more attractive than that of USDB.

USDB provides proton-neutron centroids $ V^{pn}_{d_{5/2} d_{5/2}}$ and $V^{pn}_{d_{3/2} d_{5/2}}$ 
that are very similar, so the spin-orbit splitting stays almost the same in $^{28}$Si as in $^{22}$O 
(increase of 0.17~MeV, see Table~\ref{tab:N14_shell_gaps}).
This is also valid for the DJ16 results. 
However, the three other microscopic effective interactions (BonnC, N3LO and JISP16) are characterized by
the more attractive $V^{pn}_{d_{3/2} d_{5/2}}$ centroid, 
which leads to a reduction of the spin-orbit splitting in $^{28}$Si, as seen from Table~\ref{tab:N14_shell_gaps}.

In general, we notice some similarity between the DJ16 and the USDB proton-neutron centroids. 
The main deficiency of DJ16 is the more attractive central part of $ V^{pn}_{s_{1/2} d_{5/2}}$, which
is attenuated in DJ16A.


\section{Conclusions and summary}

In the present work we have compared the general properties of the three new microscopic effective $sd$ shell 
interactions obtained from the NCSM wave functions via the OLS transformation.
The NCSM calculations have been performed using the N3LO, JISP16 and Daejeon16 modern $NN$ potentials.

Since the theoretical single-particle energies show major deficiencies when compared with empirical values, 
we have adopted the empirical single-particle energies for these investigations.
In addition, to accommodate the expected dependence of the mean-field with increasing $A$, 
we used the USDB scaling of the TBMEs.

The monopole components of the microscopic effective interactions are compared to those of the 
phenomenological USDB interaction and to an earlier effective interaction obtained within the many-body perturbation
theory from the corresponding $G$ matrix (from the BonnC $NN$ potential).
We have shown that an effective interaction, obtained from a two-nucleon potential only,
(either from an older BonnC potential or from the modern N3LO potential) via one or the other renormalization procedures
show very similar structure of an underlying spherical mean-field and they produce spectra of rather similar 
quality. This supports a previous conclusion~\cite{SchwenkZuker} on the necessity of the $3N$ forces 
in conjunction with these effective $NN$ interactions. 
Indeed, we notice that there is significant progress in the reproduction of the (sub)shell closures 
governed by the $T=1$  centroids  of the microscopic effective interactions, based
on the JISP16 and especially on the Daejeon16 $NN$ potentials, which were adjusted to the properties of light nuclei
(up to $^{16}$O)  in order to mitigate the contributions of $3N$ forces.  
Moreover, DJ16 has the proton-neutron centroids which are in remarkable agreement with those from USDB.
This may signify that the important many-body effects are more accurately included in that potential, 
possibly due to more completely converged NCSM calculations already at $N_{\rm max} = 4$.

To obtain a better description of the data,
we propose minimal modifications to the centroids of the Daejeon16-based effective interaction,
mainly in the $T=1$ channel, which help, in particular, to restore the $N=14$ shell gap, important for the O isotopes.
The modified interaction provides excellent agreement for binding energies of the O isotopes and
greatly improves the excitation spectra of the Oxygen chain and the odd-$A$ F isotopes.

We notice however, that the monopole modifications are not able to remove  completely
the discrepancies seen in $^{28,29}$Si and $^{32}$S and we
speculated that they are related to some non-monopole components of  the interactions. 
Such speculation motivates future investigation.

In conclusion, we note that among three microscopic effective 
$sd$-shell interactions obtained from NCSM calculations
based on the N3LO, JISP16 and Daejeon16 $NN$ potentials, it is DJ16 which agrees significantly better with experiment.
This confirms that Daejeon16 is a promising starting point for various many-body approaches.
The 
valence-space $sd$-shell  interaction DJ16A, obtained by a small phenomenological monopole modification of DJ16, 
yields the best results for considered $sd$-shell spectra as compared with other 
microscopic effective interactions studied here.




\acknowledgments{
We thank Heiko Hergert for providing us with the \mbox{IMSRG} Hamiltonians.
N.~A.~Smirnova acknowledges the financial support of CNRS/IN2P3, France. 
The work of  Y.~Kim and I. J. Shin was
supported by the Rare Isotope Science Project of Institute for Basic Science funded by Ministry of Science
and ICT and National Research Foundation of Korea (2013M7A1A1075764).
This work was supported in part by the US Department of Energy (DOE) under Grant Nos. DE-FG02-87ER40371,
DE-SC0018223 (SciDAC-4/NUCLEI) and DE-SC0015376 (DOE Topical Collaboration in Nuclear Theory for Double-Beta Decay and Fundamental Symmetries). 
Computational resources were provided by the National Energy Research Scientific Computing Center (NERSC), 
which is supported by the US DOE Office of Science under Contract No. DE-AC02-05CH11231.
This work was supported by Higher Education Council of Turkey (YOK),
by The Scientific and Technological Research Council of Turkey (TUBITAK-BIDEB).
N.~A.~Smirnova thanks the Institute for Basic Science, Daejeon, for a hospitality and financial support of her visits.
The work of A.~M.~Shirokov is supported by the Russian Science Foundation under Grant 
No.~16-12-10048.
A.~M.~Shirokov also thanks the University of Bordeaux for a hospitality and financial support of his visit 
to CENBG where a part of this work was done.
}


\appendix 
\section{Tabulation of derived two-body matrix elements}
\label{app-TBMEs}

\begin{longtable}{rrrrrrrrrr}
\mbox{}\\[-4ex]
\caption{\label{tab:TBMEs_DJ16}\;The TBMEs (in MeV) of 
the secondary $sd$-shell}\\[-2.5ex]
\multicolumn{10}{c}{\parbox{1\columnwidth}{ effective Hamiltonian $H_{18}^{P'P}$
obtained from the NCSM calculation with $N_{\rm max}=4$, $\hbar \Omega=14$ MeV, 
and Daejeon16 potential for $^{18}$F are shown as well as the TBMEs of its residual 
valence effective interaction, $V_2^{P'P}$. If the bare potential is used for NCSM,
the corresponding TBMEs are denoted as $H_{18}^{P'}$ and $V_2^{P'}$, respectively.
See Ref.~\protect\cite{Dikmen2015} for details of the formalism.\vphantom{$\int_{\int_{\int}}$}
}}\\ 
\hline \hline\\
 & & & & & & \multicolumn{2}{c}{OLS} & \multicolumn{2}{c}{Bare} \\
\cline{7-8} \cline{9-10}  \\
$2j_a$ & $2j_b$ & $2j_c$ & $2j_d$ & $J$ & $T$ & $H_{18}^{P'P}$ &$V_2^{P'P}$ & $H_{18}^{P'}$ & $V_2^{P'}$ \\
\hline
   1 &   1 &   1 &   1 &   0 &   1 & $ -127.487$ & $   -2.017$ & $ -123.050$ & $   -1.989$ \\
   1 &   1 &   3 &   3 &   0 &   1 & $   -0.631$ & $   -0.631$ & $   -0.769$ & $   -0.769$ \\
   1 &   1 &   5 &   5 &   0 &   1 & $   -1.453$ & $   -1.453$ & $   -1.422$ & $   -1.422$ \\
   3 &   3 &   3 &   3 &   0 &   1 & $ -106.505$ & $   -1.338$ & $ -102.561$ & $   -1.317$ \\
   3 &   3 &   5 &   5 &   0 &   1 & $   -2.240$ & $   -2.240$ & $   -2.500$ & $   -2.500$ \\
   5 &   5 &   5 &   5 &   0 &   1 & $ -127.481$ & $   -2.653$ & $ -123.229$ & $   -2.590$ \\
   1 &   1 &   1 &   1 &   1 &   0 & $ -128.408$ & $   -2.938$ & $ -123.970$ & $   -2.909$ \\
   1 &   1 &   1 &   3 &   1 &   0 & $    0.384$ & $    0.384$ & $    0.381$ & $    0.381$ \\
   1 &   1 &   3 &   3 &   1 &   0 & $    0.752$ & $    0.752$ & $    0.573$ & $    0.573$ \\
   1 &   1 &   3 &   5 &   1 &   0 & $   -2.218$ & $   -2.218$ & $   -2.196$ & $   -2.196$ \\
   1 &   1 &   5 &   5 &   1 &   0 & $   -1.221$ & $   -1.221$ & $   -1.224$ & $   -1.224$ \\
   1 &   3 &   1 &   3 &   1 &   0 & $ -118.939$ & $   -3.620$ & $ -114.717$ & $   -3.565$ \\
   1 &   3 &   3 &   3 &   1 &   0 & $    2.133$ & $    2.133$ & $    2.176$ & $    2.176$ \\
   1 &   3 &   3 &   5 &   1 &   0 & $    1.468$ & $    1.468$ & $    1.458$ & $    1.458$ \\
   1 &   3 &   5 &   5 &   1 &   0 & $   -0.780$ & $   -0.780$ & $   -0.761$ & $   -0.761$ \\
   3 &   3 &   3 &   3 &   1 &   0 & $ -106.618$ & $   -1.451$ & $ -102.587$ & $   -1.343$ \\
   3 &   3 &   3 &   5 &   1 &   0 & $    0.382$ & $    0.382$ & $    0.305$ & $    0.305$ \\
   3 &   3 &   5 &   5 &   1 &   0 & $    2.475$ & $    2.475$ & $    2.465$ & $    2.465$ \\
   3 &   5 &   3 &   5 &   1 &   0 & $ -120.710$ & $   -5.712$ & $ -116.644$ & $   -5.702$ \\
   3 &   5 &   5 &   5 &   1 &   0 & $   -3.379$ & $   -3.379$ & $   -3.375$ & $   -3.375$ \\
   5 &   5 &   5 &   5 &   1 &   0 & $ -125.925$ & $   -1.097$ & $ -121.682$ & $   -1.043$ \\
   1 &   3 &   1 &   3 &   1 &   1 & $ -115.207$ & $    0.111$ & $ -111.017$ & $    0.136$ \\
   1 &   3 &   3 &   5 &   1 &   1 & $   -0.156$ & $   -0.156$ & $   -0.151$ & $   -0.151$ \\
   3 &   5 &   3 &   5 &   1 &   1 & $ -114.966$ & $    0.032$ & $ -110.879$ & $    0.062$ \\
   1 &   3 &   1 &   3 &   2 &   0 & $ -117.201$ & $   -1.882$ & $ -112.992$ & $   -1.839$ \\
   1 &   3 &   1 &   5 &   2 &   0 & $    3.067$ & $    3.067$ & $    3.045$ & $    3.045$ \\
   1 &   3 &   3 &   5 &   2 &   0 & $   -2.336$ & $   -2.336$ & $   -2.290$ & $   -2.290$ \\
   1 &   5 &   1 &   5 &   2 &   0 & $ -125.694$ & $   -0.545$ & $ -121.347$ & $   -0.497$ \\
   1 &   5 &   3 &   5 &   2 &   0 & $    1.931$ & $    1.931$ & $    1.887$ & $    1.887$ \\
   3 &   5 &   3 &   5 &   2 &   0 & $ -118.399$ & $   -3.401$ & $ -114.312$ & $   -3.370$ \\
 {1} & {3} & {1} & {3} & {2} & {1} & {$-115.837$}& {$  -0.518$}& {$-111.652$}& {$  -0.500$}\\
   1 &   3 &   1 &   5 &   2 &   1 & $   -1.500$ & $   -1.500$ & $   -1.498$ & $   -1.498$ \\
   1 &   3 &   3 &   3 &   2 &   1 & $   -0.091$ & $   -0.091$ & $   -0.099$ & $   -0.099$ \\
   1 &   3 &   3 &   5 &   2 &   1 & $    0.450$ & $    0.450$ & $    0.452$ & $    0.452$ \\
\multicolumn{7}{c}{TABLE XI continued.}\\[1ex]
\hline \hline\\
             & & & & & & \multicolumn{2}{c}{OLS} & \multicolumn{2}{c}{Bare} \\
\cline{7-8} \cline{9-10}  \\
$2j_a$ & $2j_b$ & $2j_c$ & $2j_d$ & $J$ & $T$ &
$H_{18}^{P'P}$ &$V_2^{P'P}$ & $H_{18}^{P'}$ & $V_2^{P'}$ \\
\hline\\[-.5ex] 
   1 &   3 &   5 &   5 &   2 &   1 & $   -1.177$ & $   -1.177$ & $   -1.164$ & $   -1.164$ \\
   1 &   5 &   1 &   5 &   2 &   1 & $ -126.524$ & $   -1.375$ & $ -122.191$ & $   -1.341$ \\
   1 &   5 &   3 &   3 &   2 &   1 & $   -0.836$ & $   -0.836$ & $   -0.853$ & $   -0.853$ \\
   1 &   5 &   3 &   5 &   2 &   1 & $    0.374$ & $    0.374$ & $    0.360$ & $    0.360$ \\
   1 &   5 &   5 &   5 &   2 &   1 & $   -0.551$ & $   -0.551$ & $   -0.542$ & $   -0.542$ \\
 {3} & {3} & {3} & {3} & {2} & {1} & {$-105.392$}& {$ -0.225$} & {$-101.429$}& {$  -0.185$} \\
   3 &   3 &   3 &   5 &   2 &   1 & $    0.856$ & $    0.856$ & $    0.855$ & $    0.855$ \\
   3 &   3 &   5 &   5 &   2 &   1 & $   -0.771$ & $   -0.771$ & $   -0.837$ & $   -0.837$ \\
   3 &   5 &   3 &   5 &   2 &   1 & $ -115.134$ & $   -0.136$ & $ -111.057$ & $   -0.116$ \\
   3 &   5 &   5 &   5 &   2 &   1 & $    0.251$ & $    0.251$ & $    0.259$ & $    0.259$ \\
   5 &   5 &   5 &   5 &   2 &   1 & $ -125.990$ & $   -1.162$ & $ -121.770$ & $   -1.132$ \\
   1 &   5 &   1 &   5 &   3 &   0 & $ -129.133$ & $   -3.984$ & $ -124.785$ & $   -3.935$ \\
   1 &   5 &   3 &   3 &   3 &   0 & $   -0.049$ & $   -0.049$ & $   -0.004$ & $   -0.004$ \\
   1 &   5 &   3 &   5 &   3 &   0 & $   -1.535$ & $   -1.535$ & $   -1.527$ & $   -1.527$ \\
   1 &   5 &   5 &   5 &   3 &   0 & $   -2.042$ & $   -2.042$ & $   -2.011$ & $   -2.011$ \\
   3 &   3 &   3 &   3 &   3 &   0 & $ -108.366$ & $   -3.199$ & $ -104.388$ & $   -3.144$ \\
   3 &   3 &   3 &   5 &   3 &   0 & $   -1.756$ & $   -1.756$ & $   -1.743$ & $   -1.743$ \\
\multicolumn{7}{c}{TABLE XI continued.}\\[1ex]
\hline \hline\\
             & & & & & & \multicolumn{2}{c}{OLS} & \multicolumn{2}{c}{Bare} \\
\cline{7-8} \cline{9-10}  \\
$2j_a$ & $2j_b$ & $2j_c$ & $2j_d$ & $J$ & $T$ & $H_{18}^{P'P}$ &$V_2^{P'P}$ & $H_{18}^{P'}$ & $V_2^{P'}$ \\
\hline\\[-.5ex] 
   3 &   3 &   5 &   5 &   3 &   0 & $    0.938$ & $    0.938$ & $    0.977$ & $    0.977$ \\
   3 &   5 &   3 &   5 &   3 &   0 & $ -115.652$ & $   -0.654$ & $ -111.587$ & $   -0.646$ \\
   3 &   5 &   5 &   5 &   3 &   0 & $   -2.062$ & $   -2.062$ & $   -2.071$ & $   -2.071$ \\
   5 &   5 &   5 &   5 &   3 &   0 & $ -125.544$ & $   -0.716$ & $ -121.329$ & $   -0.690$ \\
   1 &   5 &   1 &   5 &   3 &   1 & $ -124.741$ & $    0.408$ & $ -120.416$ & $    0.433$ \\
   1 &   5 &   3 &   5 &   3 &   1 & $    0.304$ & $    0.304$ & $    0.302$ & $    0.302$ \\
   3 &   5 &   3 &   5 &   3 &   1 & $ -114.845$ & $    0.153$ & $ -110.772$ & $    0.169$ \\
   3 &   5 &   3 &   5 &   4 &   0 & $ -119.264$ & $   -4.267$ & $ -115.160$ & $   -4.218$ \\
   3 &   5 &   3 &   5 &   4 &   1 & $ -116.638$ & $   -1.641$ & $ -112.562$ & $   -1.621$ \\
   3 &   5 &   5 &   5 &   4 &   1 & $    1.373$ & $    1.373$ & $    1.371$ & $    1.371$ \\
   5 &   5 &   5 &   5 &   4 &   1 & $ -125.063$ & $   -0.235$ & $ -120.850$ & $   -0.211$ \\
   5 &   5 &   5 &   5 &   5 &   0 & $ -129.314$ & $   -4.486$ & $ -125.087$ & $   -4.448$ \\
\hline\hline \\ \\ 
\end{longtable}
}

%
%
\bibliography{int}

\end{document}